\documentclass[12pt,letterpaper]{article}
\pdfoutput=1
\usepackage[colorlinks=false,
   linkcolor=red, 
   citecolor=blue,
    filecolor=red,
    urlcolor=red,
    linktoc=all, %%%
    pdfstartview=FitV,
    bookmarksopen=true]{hyperref}
\usepackage[left=2cm,top=1cm,right=3cm,nohead]{geometry}

\usepackage[utf8]{inputenc} 
\usepackage{epsfig,latexsym,amsfonts,amsmath,amsthm,amssymb,amsbsy,multirow,slashed,color,
mathrsfs,wasysym,textcomp,subfigure,wrapfig,comment,bbold,array,longtable,multirow}
\usepackage{tabularx}
\usepackage{graphicx}
\usepackage{setspace}
\usepackage{subfigure}
\usepackage[boxsize=0.5em,aligntableaux=center]{ytableau}
\usepackage[bf]{caption}
\usepackage{braket}
\usepackage{comment}
\usepackage{float}
\usepackage{tikz}
\usetikzlibrary{positioning}
\usetikzlibrary{intersections} 
\newlength{\PicScale}
\usepackage{breqn}\usepackage[UKenglish]{babel}
\usepackage[toc,page]{appendix}
\usepackage{hhline}
\usepackage[vcentermath]{youngtab}
\usepackage{geometry}
\usepackage{cite}
\usepackage{datetime}
\usepackage{bbm} 
\usepackage{bm} 
\hypersetup{
    pdftitle={},
    pdfauthor={},
    pdfsubject={}
} 
\newcommand{\red}[1]{\textcolor{red}{#1}}

\newcommand{\blue}[1]{\textcolor{blue}{#1}}
\newcommand{\green}[1]{\textcolor{green}{#1}}

\newcommand{\yellow}[1]{\textcolor{yellow}{#1}}

\newcommand{\be}{\begin{equation}}
\newcommand{\ee}{\end{equation}}

\newcommand{\bea}{\begin{eqnarray}}
\newcommand{\eea}{\end{eqnarray}}
\newcommand{\bmat}{\left(\!\!\begin{array}}
\newcommand{\emat}{\end{array}\!\!\right)}

\newcommand{\RR}{{\mathbb{R}}}

\newcommand{\ZZ}{{\mathbb{Z}}}

\newcommand{\QQ}{{\mathbb{Q}}}
\newcommand{\tm}{{\text{-}}}
\newcommand{\mra}{{\mathrm{A}}}
\newcommand{\mrd}{{\mathrm{D}}}
\newcommand{\mre}{{\mathrm{E}}}
\newcommand{\mrg}{{\mathrm{G}}}
\newcommand{\mrh}{{\mathrm{H}}}

\newcommand{\rO}{{\mathrm{O}}}
\newcommand{\rG}{{\mathrm{G}}}
\newcommand{\tw}{{\mathrm{w}}}

\newcommand{\nltt}{{\mathrm{II}_{2,2}}}
\newcommand{\nlc}{{\mathrm{II}_{1,17}}}
\newcommand{\nlt}{{\mathrm{II}_{2,18}}}
\newcommand{\nld}{{\mathrm{II}_{d,d+16}}}
\newcommand{\uo}{{\mathrm{U}(1)}}
\newcommand{\gr}{{\mathrm{G}_r}}
\newcommand{\sug}{{\mathrm{SU}}}
\newcommand{\sog}{{\mathrm{SO}}}
\newcommand{\hog}{{\mathrm{SO}(32)}}
\newcommand{\hosp}{{\mathrm{Spin}(32)/\mathbb{Z}_2}}
\newcommand{\tC}{{\texttt{C}}}
\newcommand{\tN}{{\texttt{N}}}
\newcolumntype{M}[1]{>{\centering\arraybackslash}m{#1}}
\newcolumntype{N}{@{}m{0pt}@{}}

\newcommand{\mfn}[1]{\mbox{\footnotesize$#1$}}

\newcommand{\msc}[1]{\mbox{\scriptsize$#1$}}
\newcommand{\msm}[1]{\mbox{\small$#1$}}
\newcommand{\op}{\hspace{1pt}}

\def\nn{\nonumber}
\newcommand{\REE}{R_{\small{\rm E}}}
\newcommand{\AEE}{A_{\small{\rm E}}}
\newcommand{\NEE}{N_{\small{\rm E}}}
\newcommand{\RO}{R_{\small{\rm O}}}
\newcommand{\AO}{A_{\small{\rm O}}}
\newcommand{\NO}{N_{\small{\rm O}}}

\numberwithin{equation}{section}

\makeatletter
\def\@cline#1-#2\@nil{%%%
  \omit
  \@multicnt#1%%%
  \advance\@multispan\m@ne
  \ifnum\@multicnt=\@ne\@firstofone{&\omit}\fi
  \@multicnt#2%%%
  \advance\@multicnt-#1%%%
  \advance\@multispan\@ne
  \leaders\hrule\@height\arrayrulewidth\hfill
  \cr
  \noalign{\nobreak\vskip-\arrayrulewidth}}
\makeatother
\graphicspath{{./Images/}}

\begin{document}
\pagestyle{empty}
\begin{center}       
  {\bf\LARGE Exploring the landscape of\\[-1mm]  heterotic strings on $T^d$ \\ [5mm]}

\large{ Anamar\'{\i}a Font$^{\sharp\, *}$, Bernardo Fraiman$^{\dag\, **\,\flat}$, Mariana Gra\~na$^{\flat}$, \\[-1mm]
Carmen A. N\'u\~nez$^{\dag\, **}$ and H\'ector Parra De Freitas$^{\sharp}$
 \\[2mm]}
{\small  $^\sharp$ Facultad de Ciencias, Universidad Central de Venezuela\\[-2mm] }
{\small\it A.P.20513, Caracas 1020-A,  Venezuela\\ }
{\small  $^*$ Max-Planck-Institut f\"ur Gravitationsphysik, Albert-Einstein-Institut\\ [-2mm]}
{\small\it 14476 Golm, Germany\\ }
{\small  $^\dag$ Instituto de Astronom\'ia y F\'isica del Espacio (IAFE-CONICET-UBA)\\ [-2mm]}
{\small  $**$ Departamento de F\'isica, FCEyN, Universidad de Buenos Aires (UBA) \\ [-2mm]}
{\small\it Ciudad Universitaria, Pabell\'on 1, 1428 Buenos Aires, Argentina\\ }
{\small $\flat$  Institut de Physique Th\'eorique, Universit\'e Paris Saclay, CEA, CNRS\\ [-2mm]}
{\small\it  Orme des Merisiers, 91191 Gif-sur-Yvette CEDEX, France.\\[0.2cm] } 

{\small \verb"afont@fisica.ciens.ucv.ve, bfraiman@iafe.uba.ar, mariana.grana@ipht.fr,"\\[-3mm]}
{\small \verb"carmen@iafe.uba.ar, hectorpdf@pm.me"} \\[.2cm]

\small{\bf Abstract} \\[5mm]\end{center}
 
Compactifications of the heterotic string on $T^d$ are the simplest, yet rich enough playgrounds to uncover swampland ideas: 
the $\uo^{d+16}$ left-moving gauge symmetry  gets enhanced at special points in moduli space only to certain groups. We state criteria, based on lattice embedding techniques,  to establish whether a gauge group is realized or not.
For generic $d$, we further show how to obtain the moduli that lead to a given gauge group    
by modifying the method of deleting nodes in the extended Dynkin diagram of the Narain lattice II$_{1,17}$. 
More general algorithms to explore the moduli space are also developed. For $d=1$ and $2$ we list all the maximally enhanced gauge groups, moduli, and other relevant information about the embedding in $\nld$. In agreement with the duality between heterotic on $T^2$ and F-theory 
on K3, all possible gauge groups on $T^2$ match all possible ADE types of singular fibers of elliptic K3 surfaces.
We also present a simple method to transform the moduli under the duality group, and we build the map that relates the charge lattices and moduli of the compactification of the $\mre_8 \times \mre_8$  and $\hosp$ heterotic theories.

\newpage

\setcounter{page}{1}
\pagestyle{plain}
\renewcommand{\thefootnote}{\arabic{footnote}}
\setcounter{footnote}{0}

\tableofcontents
\newpage

\section{Introduction}
\label{sec:intro}

The compactification of perturbative heterotic strings on $d$-dimensional tori  has a long history, starting with the seminal works 
by Narain \cite{Narain:1985jj},  and  Narain, Sarmadi and Witten \cite{narain2}.  
Renewed interest in this subject arose
as a consequence of the many dualities of toroidal compactifications. The  conjectured dualities between the
heterotic on $T^4$ and type IIA on K3 \cite{Seiberg:1988pf}, the heterotic on $T^3$ and M-theory on 
K3 \cite{Hull:1994ys,Witten:1995ex} (for reviews, see \cite{Aspinwall:1996mn, Sen:1998kr}), and the heterotic on $T^2$ and 
F-theory compactified on an elliptic K3 manifold \cite{Vafa:1996xn},
provide ideal frameworks for exploring non-perturbative aspects of string theory. 
Another recent  application is the holographic duality between the average over  toroidal compactifications of  Narain's family 
of two-dimensional CFT's and three-dimensional gravity 
\cite{Maloney:2020nni,Afkhami-Jeddi:2020ezh}. 
Additional  motivations to further investigate this theory include the construction of phenomenologically viable models of 
string compactifications, since  heterotic and F-theory vacua are two of the most promising scenarios to build realistic examples 
\cite{iu, Fmodels}, as well as the test of swampland criteria (see \cite{swampland} for reviews and
\cite{Kim:2019vuc, Kim:2019ths} for recent related work).  

As it is well known,  modular invariance of the heterotic string on $T^d$ requires that the momenta of the worldsheet fields take values on the even self-dual lattice  $\nld$ \cite{Narain:1985jj}. This lattice is unique up to $\mathrm{SO}(d,d+16, \RR)$ transformations, and the precise way in which it is related to the moduli of the theory was  determined in \cite{narain2}. In particular,  the presence of suitable Wilson lines may result in the enlargement of the gauge group of the theory, while further adjusting the metric and Kalb-Ramond background fields, one could continuously interpolate between toroidally compactified versions of the $\mre_8 \times \mre_8$ and  
$\text{Spin}(32)/{\mathbb Z}_2$ heterotic theories. This interpolation was made explicit for the circle in 
\cite{Ginsparg:1986bx,Keurentjes:2006cw}.

Our aim is to examine the structure of  the moduli space  and the pattern of  associated gauge symmetries.
Various interesting related issues that deserve further analysis can be identified. 
One is to find the moduli (up to dualities) that produce a particular group. 
For example, as already noticed in \cite{Narain:1985jj}, the group of maximal dimension allowed is
$\text{SO}(32+2d)$ and values of the moduli for which this group arises were found in \cite{narain2} for $d=1$ and in
\cite{Ginsparg:1986bx} for other $d$. More generally, we might ask for all possible groups and their corresponding background.
The allowed groups are such that their
even positive definite root lattice  can be embedded in the Narain lattice $\mathrm{II}_{d, d+16}$ \cite{Narain:1985jj}. 
Thus, they can  in principle be found using lattice embedding techniques, in
particular the machinery developed by Nikulin \cite{Nikulin80}, as advocated in \cite{Taylor:2011wt}. 
For instance,  Theorem 1.12.4 in \cite{Nikulin80} implies that any ADE group of rank less or equal than $(d+8)$ can be 
embedded in $\nld$, and is thus realized in compactifications of the heterotic theory on $T^d$.

For $d=2$ all allowed gauge groups are known from the work of Shimada and Zhang who 
classified all possible ADE types of singular fibers in elliptic K3 surfaces \cite{SZ, ShimadaK3}. 
As we will explain,
the classification provides all possible heterotic gauge groups because the lattice embedding conditions are 
identical in the K3 and heterotic frameworks. This is consistent with duality between heterotic on $T^2$ and 
F-theory on K3.

Another problem is to obtain the resulting gauge group for specific moduli. 
It can  be solved by organizing the left-moving components of the momenta into roots of an ADE group 
(see \cite{mohaupt,Fraiman:2018ebo} for examples).
However, since this method is cumbersome, it is desirable to develop a more powerful approach which could
also be applied to the question of  finding all possible groups. 
When $d=1$ both problems can be solved using the extended Dynkin diagram (EDD) associated to the
Narain lattice $\mathrm{II}_{1,17}$. For instance, the 44 allowed groups with maximal rank $17$
and the corresponding moduli were determined in \cite{Fraiman:2018ebo} starting from the EDD. 

The generalization  of the powerful EDD algorithm to higher dimensional compactifications clashes with the fact that, unlike $\rO(1,17;\mathbb Z)$,  the T-duality group $\rO(d,d+16;\mathbb Z)$ is no longer generated by simple reflections.  In the absence of  a Dynkin diagram to describe  $\mathrm{II}_{d,d+16}$, what we can do to explore the landscape of heterotic strings on $T^d$, for generic dimension $d>1$, is to develop alternative methods.  

To begin we will revisit Nikulin's criteria, and apply them to compactifications of the heterotic string on $T^d$. 
The study of embeddings in $\nld$ will enable us to characterize the allowed gauge groups in terms of lattice data
consisting of the pair $(L,T)$, where $L$ is the root lattice of the group, and $T$ is the dual lattice of the right-moving
momenta. Conversely, $(L,T)$ can be determined from the moduli that originate the group.

We  also present three other methods to examine the toroidal landscape. We focus mainly on maximal enhancing in $T^2$ 
compactifications of the $\mre_8 \times \mre_8$ theory, but the algorithms work in higher dimensions. In particular, 
we will obtain all semisimple groups with maximal rank $d+16$,
occurring in $d=1,2$.  Moreover, the  moduli in the  $\text{Spin}(32)/{\mathbb Z}_2$ theory can be deduced from 
those of the $\mre_8 \times \mre_8$ theory by making use of
an $\rO(d,d+16)$ transformation that  generalizes the map constructed in \cite{Keurentjes:2006cw}  for $d=1$. 

One of the methods developed  mimics the EDD approach by employing the shift vector algorithm, based on original work by 
Kac \cite{Kac}. The algorithm gives in particular pairs of Wilson lines that break $\mre_8 \times \mre_8$ to a subgroup of rank 16.  
By choosing special values of the torus metric and the Kalb-Ramond field one can construct extended diagrams containing $d+18$ nodes, where the $d$ nodes coming from the torus connect the two $9$-node diagrams of two extended $\mre_8$ groups. Deleting two nodes leads to a semisimple group of rank $16+d$ that is realized in the heterotic string, and the construction gives the point in moduli space where it is realized. 
This method gives all groups of maximal enhancement in circle compactifications, but already for $d=2$ fails to give some groups that are known to appear from the results of \cite{SZ}.
 To search for additional groups we elaborate an algorithm that determines maximal enhancements for other values of the moduli, 
but still starting from Wilson lines that leave unbroken a subgroup of rank 16. More groups can then be found,   
but still for $d=2$ there are 2 of the 325 groups of the list in \cite{SZ} that do not appear. We argue that these groups cannot be obtained from enhancing a rank 16 subgroup of $\mre_8 \times \mre_8$, which is the departing point of this method.

To recover all allowed groups we use a more general technique. The idea is to start from a point of maximal enhancement, i.e.
a rank $d+16$ group with no $\uo$ factors, move along lines in moduli space where there is a breaking to a group with one $\uo$ factor, 
and then find all maximal enhancings that can be reached from the neighborhood of the initial point. We have fully exploited this technique in $d=2$, finding all enhancements reported in \cite{SZ}. We have also done a quick exploration in $d=3$.

The paper is organized as follows.  In section \ref{sec:review}, we  briefly review the basics of heterotic compactification on $T^d$ and 
present a simple method to find the transformation of the background fields under the action of  $\rO(d,d+16)$. We also review the map 
relating  the charge vectors and moduli of the $\mre_8 \times \mre_8$ and  $\text{Spin}(32)/{\mathbb Z}_2$ theories on the circle, 
and formulate it for generic $d$.
In section \ref{sec:lattices} we state criteria, based on lattice embedding techniques,
that can be used to detect whether a group is allowed or not. 
We additionally explain how to translate between heterotic moduli and lattice data.
The notation and essential concepts about lattices that supplement this 
section are contained in appendices \ref{ap:not} and \ref{ap:extralat}.
The method of the EDD and the results that were obtained in the circle are recalled in section \ref{sec:circle}, 
where we also show how they perfectly fit within the formalism of lattice embeddings. 
Compactifications on $T^2$ are the subject of section \ref{sec:d2}. In  section \ref{sec:complex}, we  introduce the complex moduli and their duality transformations, and 
review the action of $\rO(2,3;\mathbb Z)$ on a particular slice of the moduli space.
In section \ref{sec:shiftd2} we describe a method based on extended diagrams and apply it to analyze maximal enhancings.
In section \ref{sec:explore}, we present two computational algorithms to obtain the moduli underlying semisimple groups of
maximal rank; one generalizes the method of extended diagrams while the other explores the neighborhood of points of maximal enhancement. 
In section \ref{ss:alld2} we discuss several features of the models appearing in $d=2$. 
Some results about compactifications on $T^d$, $d >2$, are summarized in section \ref{sec:dd}. 
We further discuss the results and open problems in section \ref{sec:conclusions}.
Tables containing all the groups of maximal enhancement in one and two dimensions, and the points in moduli space where they arise, are presented in appendix \ref{app:tablesmaximal}.

\section{Toroidal compactification of the heterotic string}
\label{sec:review}

In this section we briefly review the basics of heterotic compactification on $T^d$ and outline our notation.
The torus is defined by identifications in a lattice $\tilde \Lambda_d$ generated by vectors $e_i$, $i=1,\ldots, d$.
The constant torus metric is $g_{ij} = e^a{}_i\delta_{ab}e^b{}_j$, $a=1,\ldots, d$. 
The vectors $\hat e^{*i}=g^{ij} e_i$, $g^{ij}=g^{-1}_{ij}$, span the dual lattice $\tilde\Lambda^*_d$.
The background is further specified by the constant antisymmetric two-form field $b_{ij}$ and $d$ independent
Wilson lines $A_i^I$, $I = 1,...,16$. The latter are constant components of the 10-dimensional
gauge field in the Cartan sub-algebra of $\mre_8 \times \mre_8$ or $\text{SO}(32)$.
It is convenient to introduce the tensor $E_{ij}$ given by
\be
E_{ij}=g_{ij}+\frac12A_i\cdot A_j+b_{ij} \, , 
\label{genmetric}
\ee
where $A_i\cdot A_j=A_i^I A_j^I$. We use conventions $\alpha^\prime=1$.

The momenta of the worldsheet fields of  perturbative heterotic string theory compactified on a $d$-dimensional torus $T^d $ 
must take values on an even self-dual lattice $\mathrm{II}_{d,d+16}$ \cite{Narain:1985jj}. 
As shown in \cite{narain2}, the left and right components of the canonical center of mass momenta can be expressed in terms 
of the compactification moduli, $g_{ij}$, $b_{ij}$ and $A_i^I$, as
\begin{subequations}
\begin{align}
p_{R}&= \frac1{\sqrt2}\left[n_i-E_{ij}w^j-\pi \cdot A_i\right]\hat e^{*i}\, , \label{pR} \\
p_{L}&=\frac1{\sqrt2}\left[n_i+\left(2g_{ij}-E_{ij}\right)w^j-\pi\cdot A_i\right]\hat e^{*i}
=\sqrt2 \, w^i e_i + p_R\, , \label{pL} \\[2mm]
p^I &= \pi^I+A^I_iw^i\, . \label{pI}
\end{align}
\label{momenta}
\end{subequations}
Here $n_i$ and $w^i$ are the integer momenta and winding numbers on the torus.
The $\pi^I$ are the components of a vector belonging to the gauge lattice denoted $\Upsilon_{16}$, given by
\be
\Upsilon_{16}=\left\{\begin{array}{ll}
\Gamma_8 \times \Gamma_8\, , & {\rm for \  the \  } \mre_8 \times \mre_8  \  {\rm theory \ \, (HE)} \\
\Gamma_{16}\, , & {\rm for \ the \ } \hosp  \ {\rm theory \ \, (HO)} 
\end{array}
\right.
\, ,
\label{up16def}
\ee
where $\Gamma_{8q}$ is the even self-dual lattice consisting of vectors $(m_1,...,m_{8q})$ and $(m_1+\frac12,...,m_{8q}+\frac12)$,  with $m_k\in{\mathbb Z}$ and $\sum_{k=1}^{8q}m_k=\text{even}$. Then $\pi^I \pi^I=\text{even}$.
The $\pi^I$ can also be written 
as $\pi^I=\pi^A \alpha_A^I$, with $A=1,\ldots, 16$, where $\alpha_A^I$ is a basis of $\Upsilon_{16}$ 
such that $\alpha_A^I \alpha_B^I = \kappa_{AB}$ is the lattice metric.

The total momentum ${\bf {p = (p_R;p_L)}}$, with ${\bf p_R} = p_{Ra}$, ${\bf p_L} = (p_{La},p^I)$,
transforms as a vector under $\rO(d,d+16;\mathbb R)$. It spans the $2d\text{+}16$-dimensional  
momentum lattice $\mathrm{II}_{d,d+16} \subset \mathbb R^{2d+16}$, satisfying 
\be
{\bf p\cdot p} = {\bf p_L}^2 -{\bf p_R}^2= 2w^in_i + \pi^I\pi^I  \,  \in 2 \ZZ 
\label{evencond}
\ee
Thus, $\mathrm{II}_{d,d+16}$ is even and  
it can be shown that it is self-dual, i.e. $\mathrm{II}_{d,d+16} =\mathrm{II}^*_{d,d+16}$.
Notice that we are using signature $((-)^d;(+)^{d+16})$
for the Lorentzian metric.

The space of inequivalent lattices and inequivalent backgrounds is described by
\be\label{modulispace}
\frac{\rO(d,d+16;\mathbb R)}{\rO(d;\mathbb R)\times \rO(d+16;\mathbb R)\times \rO(d,d+16;\mathbb Z)}
\ee  
where $\rO(d,d+16;\mathbb Z)$ is the T-duality group that leaves invariant the spectrum
of the theory. 
We refer to \cite{Fraiman:2018ebo} for a complete discussion of the $\rO(d,d+16;\mathbb Z)$ generators, 
see also \cite{Giveon:1994fu}.
Typical elements are a change of basis of the torus lattice $\tilde \Lambda_d$, shifts of the $B$-field by an antisymmetric 
integer matrix, and transformations of the Wilson lines by translations or automorphisms in $\Upsilon_{16}$.
There are also factorized dualities that correspond to exchanging winding and momenta in one internal direction. 
In section \ref{ss:duality} we will discuss duality transformations in more detail.

The spectrum of states depends on the background fields. It can be obtained from the mass formula and level-matching condition 
given by
\bea 
m^2& =& {\bf p_L}^2+{\bf p_R}^2+2 \left(N_L + N_R -\left\{\begin{matrix}1 \ \ {\rm  R \ sector} \\
\frac32 \ \ {\rm NS \ sector}\end{matrix}\right.\right) , \label{mass}\\
 0 &=&{\bf p_L}^2-{\bf p_R}^2+ 2 \left(N_L -N_R -\left\{\begin{matrix}1 \  \ {\rm R \ sector}\\ \frac12  \ \ {\rm NS \ sector} \end{matrix}\right.\right)\, , \label{levelmatching}
\eea
where $N_L$ and $N_R$ are left- and right-moving oscillator numbers. These equations are invariant under
the duality group $\rO(d,d+16;\ZZ)$.

In the NS sector the lowest lying states have $N_R=\dfrac12$ and their supersymmetric partners in R have $N_R=0$.
These states can be massless only if 
 \be
{\bf p_R} = 0, \quad {\bf p_L}^2+2(N_L-1)=0 \, .
\label{extra}
\ee
The condition ${\bf p_R}=0$ requires that the the momentum numbers $n_i$ satisfy (see \eqref{momenta})
\be
\label{pR0}
n_i = E_{ij} w^j + \pi \cdot A_i \in \ZZ \, .
\ee
Moreover, from \eqref{evencond} it follows that
\be
\label{pL2}
{\bf p_L}^2 = 2w^in_i + \pi\cdot \pi \, .
\ee
For generic values of the moduli the only solution 
is $w^i=0$, $n_i=0$, $\pi^I=0$, 
implying ${\bf p_L}=0$, and $N_L=1$ in \eqref{extra}.
It gives rise to the gravity multiplet plus gauge multiplets of $\uo^{d+16}$.
On the other hand, for special values of the moduli there can exist solutions with $N_L=0$, and ${\bf p_L}^2=2$.
The set of ${\bf p_L}$ then gives the roots of a Lie group $\gr$ of rank $r\le d+16$. In this case there will be gauge 
multiplets of a group  $\gr\times \uo^{d+16-r}$. The non-Abelian piece $\gr$ is in turn a product of ADE factors of 
total rank $r$. Our main task for the next sections is to study which groups can occur and to determine
the underlying moduli. 

We will mostly work with the HE theory. The results for the HO can be deduced from the
the map discussed in section \ref{ss:hehomap}.

\subsection{Duality transformations of the moduli}
\label{ss:duality}

In this section we present a simple way of finding the action of $\rO(d,d+16)$ transformations on the background fields $(g_{ij}, b_{ij}, A_i^I)$. 

We first start by the transformation of the $2d+16$ charge vectors, defined as
\be
|Z \rangle =|w^i,n_i; \pi^I \rangle \, .
\label{Zdef}
\ee
The inner product between charge vectors is computed using the $\rO(d,d+16)$ invariant metric 
\be
\eta=\left(\begin{matrix}0&1_{d\times d}&0\\
1_{d\times d}&0&0\\
0&0&\delta_{IJ}\end{matrix}\right)\, .
\label{etadef}
\ee
and is given by
\be \label{innerprod}
\langle Z'|Z \rangle = {w'}^{i} n_i + n'_i w^i + {\pi'}^{I} \pi^I \ .
\ee

Given the generators $O\in \rO(d,d+16; \mathbb Z)$ presented in \cite{Fraiman:2018ebo}, the transformation of 
$\ket{\tilde Z} \equiv \eta |Z\rangle$  is simply{\footnote{For instance, when $b_{ij} \to b_{ij} + \Theta_{ij}$, 
with $ \Theta_{ij}=-\Theta_{ji} \in \ZZ$, $|Z \rangle \to |w^i, n_i + \Theta_{ij} w^j; \pi^I\rangle$. \label{f1}}
\be
|\tilde Z \rangle \, \rightarrow  O \,  |\tilde Z \rangle\, ,  
\label{etaaction}
\ee
The transformation of the moduli can be obtained from the transformation of the generalized metric, discussed for example in 
\cite{Fraiman:2018ebo}.
It is generically simpler though to find the transformation of the moduli using the vielbein ${\mathcal E}$ for the generalized
metric. This vielbein can be built using that the left and right moving momenta \eqref{momenta} are
\be
{\bf p}={\cal E} |\tilde Z \rangle \ .
\ee
Under $\rO(d,d+16)$, the vielbein transforms as
\be\label{vielbeintransf}
{\cal E} \to {\cal E} \, \eta O^T \eta \ .
\ee
From this transformation law it follows that the first $d$ rows of $\eta\mathcal{E}$, which we write as
\begin{equation}\label{defvecs}
	|{\mathcal{\tilde E}_a} \rangle\equiv \frac{1}{\sqrt{2}}{\hat e^{i*}}_a|{E_{ik},- \delta_{i}{}^j;{A_i}^I}\rangle, ~~~~~ a = 1,...,d,
\end{equation}
 are $\rO(d,d+16)$ vectors. Taking the transpose of \eqref{vielbeintransf} we find 
\begin{equation}
	\ket{\mathcal{\tilde E}_a} \to O\ket{\mathcal{\tilde E}_a}.
\end{equation} 
These vectors also form a negative definite orthonormal set:
\begin{equation}
\braket{\mathcal{\tilde E}_a | \mathcal{\tilde E}_b} = \frac{1}{2} {\hat e^{i*}}_a{\hat e^{j*}}_b (-2E_{ij} + A_i \cdot A_j) =  \frac{1}{2} {\hat e^{i*}}_a{\hat e^{j*}}_b (-2g_{ij}) = -\delta_{ab}.
\end{equation}

To get the transformation laws for the moduli under an $\rO(d,d+16;{\mathbb R})$ element we simply construct the vectors $\ket{\mathcal{\tilde E}_a}$, transform them to  $\ket{\mathcal{\tilde E}_a'} = O\ket{\mathcal{\tilde E}_a}$, and extract  the transformed moduli $E^{\prime}_{ij}, A^{\prime}_i$. In practice, however, this procedure can be simplified as follows. Construct the $d\times (2d+16)$ matrix 
\begin{equation}\mathcal{A} \equiv \begin{pmatrix}  E_{ij} & -\delta_{i}{}^j & {A_i}^I\end{pmatrix},
\end{equation}
with rows labeled $\mathcal{A}_i$. These differ from the vectors $\ket{\mathcal{\tilde E}_a}$ in that the factor $(1/\sqrt{2}){\hat e^{*i}}_a$ is missing (cf. eq. \eqref{defvecs}). We may however interpret this as taking ${\hat e^{*i}}_a = \sqrt{2}\delta^i_a$, so that the rows $\mathcal{A}_i$ can also be transformed as $\rO(d,d+16)$ vectors, $\mathcal{A}_i\to \mathcal{A}_i' = O\mathcal{A}_i$. From the new matrix $\mathcal{A}'$ one then extracts the moduli with the formula
\begin{equation} \label{modprime}
\begin{pmatrix}
E_{ij}' & -\delta_{i}{}^{j} & A_i^{I'}
\end{pmatrix} = -
	\begin{pmatrix}
	\mathcal{A}'_{1,d+1} & \cdots & \mathcal{A}'_{1,2d}\\
	\vdots & \ddots & \vdots\\
	\mathcal{A}'_{d,d+1}  & \cdots & \mathcal{A}'_{d,2d}
	\end{pmatrix}^{-1}
	\mathcal{A}'\, ,
\end{equation}
where on the right hand side we multiply by minus the inverse of the $d \times d$ middle block of $\cal A'$, which is the vielbein for the transformed metric $e'_{ai}$. 

We now proceed to illustrate this method with a pair of examples where we restrict to the T-duality group $\rO (d,d+16,\mathbb{Z})$. Consider first the case $d=2$, and apply the transformation given by the matrix
\begin{equation}
O_{\Lambda_1} = 
\begin{pmatrix}
1 & 0  & -\frac{1}{2}\Lambda^2_1  & 0 &\Lambda \\
0 & 1 & 0 & 0 & 0\\
0  & 0 & 1 &  0 & 0\\
0 & 0 & 0 & 1 & 0\\
0  & 0 &-\Lambda^t_1  & 0 &1_{16\times 16}
\end{pmatrix}, ~~~~~ \Lambda_1 \in \Upsilon_{16}
\end{equation}\, ,
which shifts $A_1$ by $\Lambda_1$. After transforming the rows of $\mathcal{A}$ with $O_{\Lambda_1}$, we obtain
\begin{equation}\label{2dWLshift}
\mathcal{A}' =  \begin{pmatrix}
E_{11} + \tfrac{1}{2}\Lambda^2 + \Lambda_1 \cdot A_1 & E_{12} & -1 & 0 & A_1 + \Lambda_1\\
E_{21} + \Lambda_1 \cdot A_2 & E_{21} & 0 & -1 & A_2 
\end{pmatrix}.
\end{equation}
Since the second $2\times 2$ block of $\mathcal{A}$ remains invariant, minus its inverse, which appears in \eqref{modprime}, is the identity. The transformed $E_{ij}$ and $A_i$ can then be read off from eq. \eqref{2dWLshift}. In terms of the background fields $g_{ij}, b_{12}, A_i$, we see that
\begin{equation}
O_{\Lambda_1}:~~~~~g_{ij}' = g_{ij}, ~~~~~ b'_{12} = b_{12} - \frac{1}{2}\Lambda_1\cdot A_2, ~~~~~ A_1' = A_1 + \Lambda, ~~~~~ A_2' = A_2.
\end{equation}
This result highlights the fact that, generically, a shift of one Wilson  line $A_i$  by a vector $\Lambda_i \in \Upsilon_{16}$ must be accompanied by a $b$-field shift $b_{ij}' = b_{ij} - \tfrac{1}{2}\Lambda_i \cdot A_j$. The components of the charge vector $\ket{Z}$ transform as
\begin{equation}
O_{\Lambda_i}:~~\pi^{I} \to \pi^I - \Lambda_i^I w^i, ~~~ n_i \to n_i -\frac{1}{2}\Lambda_i^2 w^i + \pi \cdot \Lambda_i, ~~~ n_j \to n_j \ (j \neq i) \ ,~~~w^i \to w^i \, .~~~
\end{equation}

Now let us use this method to obtain the factorized duality $O_{D_1}$, which exchanges $n_1 \leftrightarrow w^1$ in generic dimension $d$. The action of $O_{D_1}$ on the matrix $\mathcal{A}$ exchanges the first and the $(d+1)$th columns, and so 
\begin{equation}
\begin{pmatrix}
E_{ij}' & -\delta_{i}{}^{j} & A_i^{I'}
\end{pmatrix}= 
	\begin{pmatrix}
	E_{11} & \delta^i{}_1\\
	\vdots & \vdots \\
	E_{d1} & \delta_d{}^i
	\end{pmatrix}^{-1}
	\begin{pmatrix}
	 \delta^1_1 & E_{1i} & -E_{11} & -\delta^i{}_1& {A_1}^I\\
	\vdots & \vdots&\vdots & \vdots& \vdots\\
	\delta^1_d & E_{di} & -E_{d1} & - \delta_d{}^i &{A_d}^I
	\end{pmatrix}, ~~~~~ i = 2,...,d.
\end{equation}
After performing this matrix operation, we obtain the transformation rules 
\begin{equation}\small
E' =
\frac{1}{E_{11}}
\begin{pmatrix}
1 & -E_{1j}\\
-E_{i1} & E_{11}E_{ij} - E_{i1}E_{1j}
\end{pmatrix}, ~~~
A'_i = \frac{1}{E_{11}}
\begin{pmatrix}
-A_1\\
E_{11}A_{i} - E_{i1}A_{1}
\end{pmatrix} \, , ~~i,j = 2,...,d\, .
\end{equation}
This result generalizes to a factorized duality in an arbitrary direction $\theta$, 
\begin{equation}\label{buscher}
\begin{split}
	O_{D_\theta}: ~~~E_{\theta\theta}' &= \frac{1}{E_{\theta\theta}}, ~~~ E_{\theta j}' = -\frac{E_{\theta j}}{E_{\theta\theta}}, ~~~ E_{i\theta}' = -\frac{E_{i\theta}}{E_{\theta\theta}}, ~~~ E_{ij}' = \frac{E_{\theta\theta}E_{ij} - E_{i\theta} E_{\theta j}}{E_{\theta\theta}},\\
	A_\theta' &= -\frac{A_\theta}{E_{\theta\theta}}, ~~~~~ A_i' = \frac{E_{\theta\theta} A_i - E_{i\theta}A_\theta}{E_{\theta\theta}}, ~~~~~ ~~~~~ ~~~ i,j = 1,...,d \neq \theta
\end{split}
\end{equation}
in agreement with the heterotic Buscher rules found originally in \cite{buscher1} and discussed also in \cite{buscher}. 

\subsection{The HE $\leftrightarrow$ HO map}
\label{ss:hehomap}

Due to the uniqueness of the Narain lattices, the HO and HE theories compactified on $T^d$ share the same moduli space. For the circle, an explicit map relating the charge lattices of both theories  was given in \cite{Ginsparg:1986bx} and the precise relation between the  moduli was worked out in \cite{Keurentjes:2006cw}. 

The $\rO(1,17)$ transformation relating a basis of vectors of the $\Gamma_8\times\Gamma_8$ embedding into II$_{1,17}$ to another one of the $\Gamma_{16}$ embedding  is given by \cite{Ginsparg:1986bx}
\be
\Theta_{{\small{\rm E}}\rightarrow {\small{\rm O}}}=O_{\Lambda_{\small{\rm O}}}O_\Omega O_{P_1}O_{D_1}O_{-\Lambda_{\small{\rm E}}}\, , \label{he-ho1}
\ee
where $O_{\Lambda_{\small{\rm E}}}$,   $O_{\Lambda_{\small{\rm O}}}$ are
 shifts of the Wilson line by 
\be
\Lambda_{\small{\rm E}}=(0^7,1,-1,0^7)\, , \qquad
\Lambda_{\small{\rm O}}=\left({\tfrac12}^8,0^8\right)\label{LHOapp}\, ,
\ee
$O_{D_1}$ is a T-duality  in the circle direction, $O_{P_1}$ an inversion and $O_{\Omega}$ a rescaling. 
Their action on the charge vectors and moduli is given by
\begin{equation}
	\begin{split}
	O_\Lambda:&~~~ \ket{w,n;\pi} \to \ket{w,n+\pi\cdot \Lambda - \tfrac{1}{2}w\Lambda^2;\pi - w\Lambda}, ~~~~~~~~~ (R,A) \to (R,A+ \Lambda),\\
	O_{D_1}:&~~~ \ket{w,n;\pi} \to \ket{n,w;\pi}, ~~~~~~~~~~~~~~~~~ (R,A) \to \left(\frac{R}{R^2 + \frac{1}{2}A^2},-\frac{A}{R^2 + \frac{1}{2}A^2}\right),\\
	O_{P_1}:&~~~ \ket{w,n;\pi}\to\ket{-w,-n;\pi}, ~~~~~~~~~~~~~~~~~~~~~~~~~~~~~~~~~~~~~ (R,A) \to (R,-A),\\
	O_\Omega:&~~~\ket{w,n;\pi} \to \ket{2w,\tfrac{1}{2}n;\pi}, ~~~~~~~~~~~~~~~~~~~~~~~~~~~~~~~~~~~~~~(R,A) \to (\tfrac{1}{2}R,\tfrac{1}{2}A).
	\end{split}
\end{equation}
Hence the total transformation \eqref{he-ho1} gives
\begin{equation}\label{hehos1}
	\begin{split}
		\Theta_{\small{\text{E}\to\text{O}}}:&~~~w \to 2w - 2n + 2\pi\cdot \Lambda_{\small{\rm E}},~~~ n \to -2w + 2n + \pi\cdot(\Lambda_{\small{\rm O}} - 2\Lambda_{\small{\rm E}}),\\
		&~~~\pi \to w(\Lambda_{\small{\rm E}} - 2\Lambda_{\small{\rm O}}) + 2n\Lambda_{\small{\rm O}} + \pi -2\Lambda_{\small{\rm O}} (\Lambda_{\small{\rm E}} \cdot \pi),\\
		&~~~R \to \frac{R}{2R^2 + (A - \Lambda_{\small{\rm E}})^2}, ~~~A \to \frac{A - \Lambda_{\small{\rm E}}}{2R^2 + (A - \Lambda_{\small{\rm E}})^2} + \Lambda_{\small{\rm O}},
	\end{split}
\end{equation}
corresponding to the $O(1,17,\mathbb{R})$ matrix
\bea
\Theta_{{\small{\rm E}}\rightarrow {\small{\rm O}}}=\left(\begin{matrix}2&-2&\Lambda_{\small{\rm O}}-2\Lambda_{\small{\rm E}}\\
-2&2&2\Lambda_{\small{\rm E}}\\
2\Lambda_{\small{\rm O}}^t&\Lambda_{\small{\rm E}}^t-2\Lambda_{\small{\rm O}}^t&
\mathbb{1}_{16}-2\Lambda_{\small{\rm O}}\otimes \Lambda_{\small{\rm E}}\end{matrix}\right)\, , \label{he-ho11}
\eea
where $\otimes$ is an outer product. 

Labeling $E_{\small{\rm E}}=R_{\small{\rm E}}^2+\frac12A_{\small{\rm E}}^2$ and the Wilson line $A_{\small{\rm E}}$ in the HE theory, the transformation \eqref{hehos1} gives the HO moduli as \cite{Keurentjes:2006cw}
\be
\left(E_{\small{\rm O}},A_{\small{\rm O}}\right)=
\left(1+\frac{A_{\small{\rm E}}\cdot\Lambda_{\small{\rm O}}}
{2(E_{\small{\rm E}}+1-A_{\small{\rm E}}\cdot\Lambda_{\small{\rm E}})} \ , \ \frac{A_{\small{\rm E}}-\Lambda_{\small{\rm E}}}
{2(E_{\small{\rm E}}+1-A_{\small{\rm E}}\cdot\Lambda_{\small{\rm E}})}+\Lambda_{\small{\rm O}}\right)\, . 
\ee \label{duality1app} \, 
The map from HO to HE is simply obtained by exchanging $\left(E_{\small{\rm O}},A_{\small{\rm O},},\Lambda_{\small{\rm O}}\right)\leftrightarrow \left(E_{\small{\rm E}},A_{\small{\rm E}},\Lambda_{\small{\rm E}}\right)$.

To extend  \eqref{he-ho11} from the circle to  $T^d$, it is sufficient to consider a decomposition of the  Narain lattice of the form
\begin{equation}
	\text{II}_{d,d+16} = \text{II}_{1,1} \oplus \cdots \oplus \text{II}_{1,1} \oplus \Gamma_{8} \oplus \Gamma_{8},
\end{equation}
where the number of $\text{II}_{1,1}$ lattices is $d$. We use $\Theta_{\text{E}\to \text{O}}$ to transform 
\begin{equation}
	\Theta_{\text{E}\to \text{O}}:~~~\text{II}_{1,1} \oplus \Gamma_{8} \oplus \Gamma_{8}~\to~ \text{II}_{1,1} \oplus \Gamma_{16},
\end{equation}
choosing $\text{II}_{1,1}$ to be in the direction given by the torus lattice vector $e_1$, without loss of generality. 
This brings the Narain lattice into the form
\begin{equation}
\text{II}_{d,d+16} = \text{II}_{1,1}\oplus \cdots \oplus \text{II}_{1,1} \oplus \Gamma_{16}.
\end{equation}
It follows that the desired extension is
\begin{equation}
	\Theta^{(d)}_{\text{E}\to \text{O}} \to {\mathbb 1}_{(2d-2)\times (2d-2)} \oplus \Theta_{\text{E}\to \text{O}}=\left(\begin{matrix}\mathbb{1}_{(2d-2)\times (2d-2)}&0\\0&\Theta_{\text{E}\to \text{O}} \end{matrix}\right)\, ,
\end{equation}
which holds provided the ordering $\ket{Z} = \ket{w^2,n_2,...,w^d,n_d,w^1,n_1;\pi}$ is used.
In practice one may wish to keep the order in \eqref{Zdef} and
rearrange the entries of $\Theta^{(d)}_{\text{E}\to \text{O}}$ instead, which is reasonable for 
low values of $d$. 

To get the transformation rules for the moduli, we proceed constructively using the factorized form of  
$\Theta_{{\small{\rm E}}\rightarrow {\small{\rm O}}}$ in \eqref{he-ho1}, and
generalizing each intermediate transformation. Each of the generalized transformation rules can be obtained by the method detailed in section \ref{ss:duality}, which is valid not only for T-dualities but for generic $\rO(d,d+16)$ transformations such as $O_{\Lambda_{\small{\rm O}}}$ (in HE) and $O_{D_1}$. 

Let us first take a detailed look at the map $\Theta_{\text{E} \to \text{O}}$ for $d = 2$. The generalization to arbitrary $d$ is straightforward. Preserving the usual ordering of the components of $\ket{Z}$, namely $\ket{w^1,w^2,n_1,n_2;\pi}$, we write 
\begin{equation}\label{hehot2}
	\Theta^{(2)}_{\text{E}\to\text{O}} = 
	\begin{pmatrix}2&0&-2&0&\Lambda_{\small{\rm O}}-2\Lambda_{\small{\rm E}}\\
	0&1&0&0&0\\
	-2&0&2&0&2\Lambda_{\small{\rm E}}\\
	0&0&0&1&0\\
	2\Lambda^t_{\small{\rm O}}&0&2\Lambda^t_{\small{\rm E}}-\Lambda^t_{\small{\rm O}}&0&
	\mathbb{1}_{16\times 16}-2\Lambda_{\small{\rm O}}\otimes\Lambda_{\small{\rm E}}\end{pmatrix},
\end{equation}
The transformation rules for the quantum numbers are exactly the same as in the $d = 1$ case for $w^1$,$n_1$ and $\pi$, while $w^2$ and $n_2$ are invariant, as expected. 

To work out the map, we proceed by applying the transformations in the r.h.s. of \eqref{he-ho1}   in succession. The Wilson line shift in direction 1 acts as
\be
O_{\Lambda}:\ \ \ E\rightarrow\left(\begin{matrix}E_{11}-\Lambda\cdot A_1+1&E_{12}\\
E_{21}-\Lambda\cdot A_2&E_{22}\end{matrix}\right)\, , \quad A_1\rightarrow A_1-\Lambda\, , \quad A_2\rightarrow A_2.
\ee
Note that $E_{12}$ is invariant since the $b$-field is also shifted (see the footnote \ref{f1}). The factorized duality acts as
\be\label{fac1d2}
O_{D_1}: \ \ E\rightarrow\frac1{E_{11}}\left(\begin{matrix}1&-E_{12}\\
E_{21}&\det E\end{matrix}\right)\, , \quad A_1\rightarrow -\frac{A_1}{E_{11}}\, , \quad A_2\rightarrow A_2-\frac{E_{21}}{E_{11}}A_1\, ,
\ee
and finally  $O_{P_1}$ and $O_{\Omega}$ produce the transformations
\bea
O_{P_1}&:&\quad E\rightarrow\left(\begin{matrix}E_{11}&-E_{12}\\
-E_{21}&E_{22}\end{matrix}\right)\, , \quad A_1\rightarrow -{A_1}\, , \quad A_2\rightarrow A_2\, ,\ \\
O_{\Omega}&:&\quad E\rightarrow\left(\begin{matrix}\frac14E_{11}&\frac12E_{12}\\
\frac12E_{21}&E_{22}\end{matrix}\right)\, , \quad A_1\rightarrow \frac12{A_1}\, , \quad A_2\rightarrow A_2\, .
\eea
Putting all together, we get
\begin{multline}
 {\small{\left(\begin{matrix}E_{11}& E_{12} & A_1 \\
E_{21} & E_{22} & A_2 \end{matrix}\right) 
\rightarrow 
\left(\begin{matrix}1& 0 & \Lambda_{\small{\rm O}} \\
\Lambda_{\small{\rm O}}\cdot A_2 & E_{22} & A_2\end{matrix}\right) }}\\
{\small{+\ \frac1{E_{11}-\Lambda_{\small{\rm E}}\cdot A_1+1}\left(\begin{matrix} \tfrac12 \\
\Lambda_{\small{\rm E}}\cdot A_2-E_{21}\end{matrix}\right)
\left(\begin{matrix}\Lambda_{\small{\rm O}}\cdot A_1& E_{12}&A_1-\Lambda_{\small{\rm E}}\end{matrix}\right)}} \, . 
\label{OES}
\end{multline}

The map for generic $d$ can be worked out in a similar fashion. The final result reads 
\begin{multline}
 {\small{\left(\begin{matrix}E_{11}& E_{12} & \cdots &  E_{1d} & A_1 \\
E_{21} & E_{22} & \cdots & E_{2d} & A_2 \\
\vdots & \vdots & \ddots & \vdots & \vdots \\ 
E_{d1} & E_{d2} & \cdots & E_{dd} & A_d 
\end{matrix}\right) 
\rightarrow 
\left(\begin{matrix}1& 0 &  \cdots &  0 &\Lambda_{\small{\rm O}} \\
\Lambda_{\small{\rm O}}\cdot A_2 & E_{22} & \cdots & E_{2d} & A_2 \\
\vdots & \vdots & \ddots & \vdots & \vdots \\ 
\Lambda_{\small{\rm O}}\cdot A_d  & E_{d2} & \cdots & E_{dd} & A_d 
\end{matrix}\right) }}\\
{\small{+\ \frac1{E_{11}-\Lambda_{\small{\rm E}}\cdot A_1+1}
\left(\begin{matrix} \tfrac12 \\
\Lambda_{\small{\rm E}}\cdot A_2-E_{21}\\
\vdots \\
\Lambda_{\small{\rm E}}\cdot A_d-E_{d1}
\end{matrix}\right)
\left(\begin{matrix}
\Lambda_{\small{\rm O}}\cdot A_1& E_{12}& \cdots &  E_{1d} & A_1-\Lambda_{\small{\rm E}}
\end{matrix}\right)}} \, . 
\label{OESd}
\end{multline}
In the forthcoming sections we will  apply the HE-HO map in compactifications to $d=1$ and 2 and give some examples for 
other values of $d$.

\section{Embedding in Narain lattices}
\label{sec:lattices}

In this section we discuss how to determine which gauge groups $\gr\times \uo^{d+16-r}$  occur in the compactification of 
perturbative heterotic strings on $T^d$. 
We are mostly interested in heterotic compactification on $T^2$, which is dual to
F-theory compactifications on elliptic K3 surfaces \cite{Vafa:1996xn}. Not surprisingly,
for $d=2$ the problem of finding all allowed $\gr$ happens to be
related to the classification of possible singular fibers of ADE type in elliptic K3 surfaces.
The explicit solution has been obtained in the K3 framework in \cite{SZ, ShimadaK3}, using Nikulin's formalism.
The results are expected to hold in the heterotic context too. The reason is that in the K3 context, the condition
on the allowed $\gr$ is that its even positive definite root lattice can be embedded in $\mathrm{II}_{2,18}$ which is precisely the
Narain lattice.

According to Theorem 1.12.4 in \cite{Nikulin80}, any $\gr$ of type ADE with $r \le 10$ is allowed for $d=2$,
as indeed found in \cite{ShimadaK3}. 
For larger $r$ more complicated conditions have to be verified as we will explain shortly. This program
has been carried out in \cite{ShimadaK3}. It turns out that for $r=11,12$, also all ADE $\gr$ can be
embedded in $\mathrm{II}_{2,18}$. For $r=13$, only $13\mra_1$ and $11\mra_1 + \mra_2$ are precluded.
Henceforth $\gr$ will be denoted by the chain of ADE factors of its algebra. For $r=14$, except 
$8\mra_1 + \mre_6$, all other forbidden groups, e.g. $14\mra_1$, were predicted to
be prohibited because singular fibers with such $\gr$ could not fit in a K3 where the vanishing degree of the discriminant 
must be 24. For $r\ge 15$ there are many more forbidden groups. In particular, 
there are 1599 ADE groups of rank 18 \cite{ShimadaK3} but according to the analysis of \cite{SZ, ShimadaK3}, only
325 are expected to be realized in compactifications of the heterotic string on $T^2$. 
A natural question is why some groups are forbidden. To answer it, we will present
some tools that can be applied to decide when a group is allowed or not. 
Our purpose is to illustrate the main ideas, not to do a systematic search as in \cite{SZ, ShimadaK3} for $d=2$.

We will mostly focus on the case of maximal enhancing, i.e. $\gr$ with $r=d+16$.
In \ref{s:max}, we will first discuss three criteria that can be applied for generic $d$. We then specialize to $d=1,2$, and 
in less detail to $d=8$. The criteria for groups with $r < 16 +d$ are presented in appendix \ref{ap:gen}.
The connection of the criteria to heterotic compactifications is addressed in section \ref{ss:nikhet}.
We refer to \cite{K3book, Kondo, Braun:2013yya} for short expositions of the main results of Nikulin's \cite{Nikulin80} relevant for
our analysis, see also \cite{Morrison, Moore:1998pn, Gaberdiel:2011fg, Cheng:2016org}. 
Before jumping into matters the reader is advised to consult appendix \ref{ap:not} where the notation and some 
basic concepts are introduced.

\subsection{Embeddings of groups with maximal rank $r=d+16$}
\label{s:max}

The problem is to embed a lattice $L$ of signature $(0,d+16)$ in the even unimodular Narain lattice
$\mathrm{II}_{d,d+16}$. In the heterotic context $L$ is the root lattice of a group of maximal rank 
arising upon compactification on $T^d$.
Nikulin \cite{Nikulin80} provides powerful results that serve to determine whether or not such embedding  exists. 
In particular,  adapting respectively Corollary 1.12.3 and Theorem 1.12.4(c) of \cite{Nikulin80} to the case at hand
leads to the criteria

\begin{quotation}
\noindent
{\bf {Criterion 1}} \hfill\\
\noindent
{\em If $\ell(A_L)<d$  then $L$ has a primitive embedding in $\mathrm{II}_{d,d+16}$.}
\end{quotation}
\begin{quotation}
\noindent
{\bf {Criterion 2} }\hfill\\
\noindent
{\em $L$ has a primitive embedding in $\mathrm{II}_{d,d+16}$ if and only if there exists a lattice $T$ of signature $(0,d)$
such that $(A_T, q_T)$ is isomorphic to $(A_L,q_L)$. } 
\end{quotation}
\noindent
Here $A_L$ and $q_L$ are respectively the discriminant group and the quadratic discriminant form of $L$, whereas
$\ell(A_L)$ is the minimal number of generators of $A_L$, and analogously for $T$ (see appendix \ref{ap:not} for details).
Since $\ell(A_T) \le d$, groups with $\ell(A_L)=d$ could pass criterion 2 which actually requires $d(L)=d(T)$.
We will shortly explain how the lattice $T$ can be determined when $d=1,2$. There could exist more than one
$T$, as found for some groups in \cite{SZ}. Notice that in our conventions $(0,d)$ means positive
signature.

Now, criteria 1 and 2 cannot be the whole story. We know groups with $\ell(A_L) > d$ that can be realized
in heterotic compactifications on $T^d$. For example, when $d=2$, heterotic moduli that give $L=3 \mre_6$ 
are known. Hence, there should be an embedding of this $L$ in $\mathrm{II}_{2,18}$ even though $\ell(A_L)=3$.
We also know examples with $d=1$. In particular,  $L=\mrd_{16}+\mra_1$
with $\ell(A_L)=3$, would be forbidden by criterion 2 but must admit an embedding in $\mathrm{II}_{1,17}$ because
it certainly arises in the heterotic string on $S^1$.  Actually, for $d=1$ 
the 44 groups with maximal rank found in \cite{Fraiman:2018ebo} have $\ell(A_L) \le 3$. 
Only the groups with $\ell(A_L)=1$, e.g. $L=2\mre_8+\mra_1$, could possibly be allowed by criterion 2. The problem is
that criteria 1 and 2 refer to {\em primitive} embeddings and this need not be the case. From the arguments in
\cite{SZ, ShimadaK3} it transpires that this condition can be relaxed by demanding that $L$ has an overlattice 
$M$ which can be embedded primitively in the Narain lattice. For instance, we know that $\mrd_{16}$ has an
overlattice given by the even unimodular HO lattice $\Gamma_{16}$ with trivial discriminant group. 
Therefore, $L=\mrd_{16} +\mra_1$ has an overlattice $M=\Gamma_{16}+ \mra_1$ with $A_M=\ZZ_2$ and $\ell(A_M)=1$.
The overlattice $M$ could then pass criterion 2 with an even 1 dimensional lattice $T$ equal to the $A_1$ lattice.

The above arguments lead to a third criterion obtained adapting Theorem 7.1 \cite{ShimadaK3}. It reads
\noindent
\begin{quotation}
\noindent
{\bf {Criterion 3}} \hfill\\
\noindent
{\em $L$ has an embedding in $\mathrm{II}_{d,d+16}$ if and only if $L$ has an overlattice $M$ with
the following properties:
\begin{itemize}
\item[\rm{(i)}] there exists an even lattice $T$ of signature $(0,d)$ such that $(A_T, q_T)$ is isomorphic to $(A_M,q_M)$,
\item[\rm{(ii)}] the sublattice $M_{root}$ of $M$ coincides with $L$.
\end{itemize}
}
\end{quotation}
\noindent
Since $L$ is an overlattice of itself, criterion 2 is a subcase of criterion 3. As explained in appendix \ref{ap:not}, for an overlattice
$M$ to exist, there must be an isotropic subgroup $H_L$ of $A_L$ such that $M/L\cong H_L$ and 
$|H_L|^2=d(L)/d(M)$. When criterion 3 is satisfied, $d(M)=d(T)$. We then obtain the useful relation
\be
d(L) = d(T) |H_L|^2 \, .
\label{mwrel}
\ee
We will refer to $T$ as the complementary lattice in the following.

In the K3 framework, in which $d=2$, $H_L$ corresponds to the torsion part of the Mordell-Weil group, called $MW$ in \cite{SZ}. It can be
checked that all pairs $(L,T)$ in Table 2 of \cite{SZ}, reproduced in our Table \ref{tab:alld2}, satisfy the relation \eqref{mwrel}.
We remark that there could exist more than one $M$, as found for some groups in \cite{SZ}.  

In the work of Shimada and Zhang \cite{SZ}, the focus is on the classification of all possible ADE types of singular fibers of 
{\em extremal} elliptic  K3 surfaces. Such a surface, called $X$, is characterized by having Picard number, 
$\rho(X)$, equal to 20, and finite Mordell-Weil group \cite{K3book}. In this case the 
N\'eron-Severi lattice, $NS_X$, and the transcendental lattice, $T_X$, have signatures $(1,19)$ and $(2,0)$
respectively\footnote{By definition,
$NS_X=H^{1,1}(X,\mathbb{R}) \cap H^2(X,\mathbb{Z})$ and has signature $(1,\rho(X)-1)$. 
The transcendental lattice is the orthogonal complement of
$NS_X$ in  $H^2(X,\mathbb{Z})$ and has signature $(2,20-\rho(X))$.  
With the intersection form of $X$, the second cohomology group $H^2(X,\mathbb{Z})$ is 
isometric to $\mathrm{II}_{3,19}$. The N\'eron-Severi lattice can be decomposed as 
$NS_X=\mathrm{II}_{1,1} \oplus  W_X$, where $\mathrm{II}_{1,1}$ is generated by the
zero section and the generic fiber. The lattice $W_X$ is the orthogonal complement of $\mathrm{II}_{1,1}$
in $NS_X$ and has signature $(0,\rho(X)-2)$. Thus, $\mathrm{II}_{1,1} \oplus  W_X \oplus T_X \subset \mathrm{II}_{3,19}$.}. 
The lattice $W_X$ has signature $(0,18)$ and contains the sublattice 
$L(\Sigma)$ of rank 18, where $\Sigma$ is the formal sum of the ADE types of singular fibers (determined by the 
Kodaira classification). 
It follows that $L(\Sigma)$ must admit an embedding in $\nlt$.
Now, in the heterotic compactification on $T^2$, the semisimple ADE groups of maximal rank 18 that can occur
are such that their root lattice can be embedded in the Narain lattice $\nlt$.
Thus, the results of \cite{SZ} for all possible $L(\Sigma)$ translate into
all possible maximal enhancings in the heterotic compactification on $T^2$.
Notice that the complementary lattice of criteria 2 and 3 above is related to the transcendental lattice
by a change of sign of the Gram metric, i.e. $T=T_X\langle -1 \rangle$. 
In section \ref{ss:nikhet} we will discuss to greater extent the connection to heterotic compactifications. 

 We illustrate below the application of criteria 1,2,3 to the cases $d=1,2$. We will also comment briefly on $d=8$.
In practice we first try criterion 1. If $L$ passes it, then it is allowed. If not, we continue with criterion 2. If $L$ satisfies
it, we are done, otherwise we apply criterion 3. If $L$ also fails criterion 3 we conclude that $L$ is not allowed.
A consistency check is that if $L$ passes criterion 1 it must also fulfill criterion 3. Let us mention
that the steps taken by Shimada and Zhang to compile their list, cf. section 3 in \cite{SZ},  
indicate that they run a computer program based on the more general criterion 3.

\subsubsection{$d=1$}
\label{ss:d1}

As a warm up we will study the $d=1$ case which is simple yet instructive.
Moreover, all allowed groups of maximal enhancing appearing in heterotic compactification on $S^1$ 
are already known \cite{Fraiman:2018ebo}. Thus, there are many examples to illustrate the application 
of the lattice embedding techniques.

When $d=1$ the easy criterion 1 gives no information.
When $\ell(A_L)=1$ we then apply criterion 2. In Table \ref{d1l1} we give some examples of
allowed groups. It is easy to propose the corresponding $T$ because 
it must be $d(T)=d(L)$ and the (0,1) even lattices are of type $\mra_1\langle m \rangle$, defined to be the $\mra_1$ 
lattice rescaled so that its basis vector has norm $u_1^2=2m$.  One still has to check that the discriminant forms do match,
more precisely that there is an isomorphism $(A_L, q_L) \cong (A_T, q_T)$. For example, for $L=\mrd_{17}$,
$A_L$ is generated by the spinor class with $s^2=\frac{17}4 = \frac14 \, \text{mod}\, 2$,  so $q_L$ takes
values $\frac{j^2}4 \, \text{mod}\, 2$, $j=0,\ldots 3$. This matches the $q_T$ of $\mra_1\langle 2\rangle$ which takes 
the same values because $(u_1^*)^2=\frac14$. It is more challenging to check $L=\mre_7 + \mra_{10}$. For
the proposed $T$, $A_T$ is generated by $u_1^*$ with $(u_1^*)^2=\frac1{22}$, whereas $A_L$ is generated
by $w_{\bf{56}}\times w_1$ with $w_{\bf{56}}^2=\frac32$ and $w_1^2=\frac{10}{11}$. To see that $q_L$ and $q_T$
match it suffices to verify that \mbox{$\big(\frac32 + \frac{10 j^2}{11} = \frac{1}{22} + \msm{2 k}\big)$}
is satisfied by integers $j$ and $k$, e.g.  $j=4$, $k=8$.

\begin{table}[h!]\begin{center}
\renewcommand{\arraystretch}{1.0}
\begin{tabular}{|c|c|c|}
\hline
$L$ & $A_L$ & $T$   \\[3pt]
\hline
$2\mre_8 + \mra_1$ &  $\ZZ_2$  & $\mra_1$
  \\ \hline
$\mrd_{17} $ &  $\ZZ_4$ & $\mra_1\langle 2\rangle$
   \\ \hline
   $\mre_8 + \mrd_9 $ &  $\ZZ_4$ & $\mra_1\langle 2\rangle$
   \\ \hline
$\mre_7 + \mra_{10} $ &  $\ZZ_2 \times \ZZ_{11} \cong \ZZ_{22}$  & $\mra_1\langle 11\rangle$
 \\ \hline 
 \end{tabular}
\caption{Examples of allowed $L$ with $\ell(A_L)=1$, when $d=1$.}
  \label{d1l1}\end{center}\end{table}
The allowed groups \cite{Fraiman:2018ebo} with maximal enhancing of the form 
$L=\mre_8 + \mre_{9-p} + A_p$, $p=1, \ldots, 9$, $p\not=7$,
all have $\ell(A_L)=1$. Only for $p=8$ there is an isotropic subgroup (actually for the $\mra_8$ component) but
the $M_{\text{root}}$ of the associated $M$ is larger than $L$. Hence, all these groups should be allowed by
criterion 2. We find that the corresponding $T$ is $\mra_1\langle \frac{p(p+1)}2\rangle$, $p=1, \ldots, 6$, and
$\mra_1\langle \frac{(10-p)(p+1)}2\rangle$, $p=8,9$.

It is straightforward but cumbersome to check exhaustively which of the known groups with maximal enhancing and 
$\ell(A_L)=1$ satisfy criterion 2, and if not apply criterion 3. In many cases, e.g. $L=\mre_7 + \mre_6 + \mra_4$, 
$A_L=\ZZ_{30}$, one can quickly see that  an overlattice cannot exist because there is no isotropic subgroup. 
Since this $L$ is known to appear, criterion 2 should allow it, and indeed $T=\mra_1\langle 15\rangle$ fulfills the conditions.

A neat example with $\ell(A_L)=1$ is  $L=\mra_{17}$, $A_L=\ZZ_{18}$. 
The candidate $T$ would be $\mra_1\langle 9 \rangle$ but the discriminant forms do not match because there are no 
integers $j$ and $k$ such that \mbox{$\big(\frac{17 j^2}{18} = \frac{1}{18} + \msm{2 k}\big)$}
is satisfied. Fortunately, $\mra_{17}$ has an overlattice $M$ associated to the isotropic
subgroup $H_L=\ZZ_3$, generated by $w_6$ with $w_6^2= 4 = 0 \, \text{mod}\, 2$. From \eqref{mwrel} we see that
$d(T)=2$ so it must be $T =\mra_1$. Since $d(M)=d(T)$ also $A_M= \ZZ_2$.
It remains to check that the discriminant forms of $A_M$ and $A_T$ coincide.
To this end we need to determine the orthogonal complement $H_L^\perp$ of $H_L$ in $A_L$
and restrict $q_L$ to $H_L^\perp/H_L$. We then look for weights
orthogonal to the generator $w_6$, i.e. weights such that $w_i \cdot w_6=0\, \text{mod}\, 1$. Besides $w_6$ and $w_{12}$
which belong to $H_L$, $w_3$, $w_9$ and $w_{15}$ are orthogonal. 
Now, $w_i^2=\frac12\, \text{mod}\, 2$, for $i=3, 9, 15$.
This confirms that $A_M = \ZZ_2$, with the discriminant form $q_M$ taking values 0 and $\frac12$. These are the same values
taken by $q_T$.
Finally, the root sublattice of $M$ is equal to $L$ because $w_6^2=4$.

We can also study known allowed groups with $\ell(A_L)\ge 2$ where criterion 3 must be
applied. An example is the group with $L=\mre_6 + \mra_{11}$, $A_L=\ZZ_3 \times \ZZ_{12}$. There exists an 
overlattice with $H_L=\ZZ_3$ and it can be shown that criterion 3 is satisfied with $T=\mra_1\langle 2\rangle$.
For a second example take $L=\mra_1 + \mra_2 + \mra_{14}$,
$A_L=\ZZ_2\times \ZZ_3 \times \ZZ_{15} \cong \ZZ_6 \times \ZZ_{15}$. The piece $\tilde L=\mra_2 + \mra_{14}$
has an overlattice $\tilde M$ with $d(\tilde M)=5$ so necessarily $A_{\tilde M}=\ZZ_5$. Thus, 
$L$ has an overlattice $M=\mra_1 + \tilde M$, $A_M = \ZZ_2 \times \ZZ_5 \cong \ZZ_{10}$ and a candidate
$T$ is $\mra_1\langle 5 \rangle$.
With $\ell(A_L)=3$ we already discussed how $L=\mrd_{16} + \mra_1$ passes the test. In Table \ref{tab:alld1}
we give full results.
 
So far we have discussed groups with maximal enhancing which are known to occur. It is reassuring that they 
are allowed by the lattice embedding criteria but our main motivation was to understand why some groups
are forbidden. Let us then finally offer a couple of examples of forbidden groups.
Take $L=\mra_6 + \mrd_{11}$, 
$A_L=\ZZ_{28}$.
A candidate $T$ is $A_1\langle 14 \rangle$, but $q_T \ncong q_L$. 
An overlattice cannot exist because there is no isotropic subgroup of $A_L$. Thus, this $L$ fails criteria 2 and 3.
A less trivial example is $L=2\mrd_8 + \mra_1$, $A_L=\ZZ_2^5$. In appendix \ref{ap:not} we explained that $\mrd_8$
admits $\mre_8$ as an overlattice. For $L$ this leads to a full overlattice given by $M=2\mre_8 + \mra_1$.
Now $A_{M}=\ZZ_2$ and an adequate $T$ would be $\mra_1$. However, condition (ii) in criterion 3 is not
satisfied. As remarked in appendix \ref{ap:not}, the root sublattice of $2\mre_8$ is not equal to $2\mrd_8$.
Actually, $L$ admits also an overlattice $M^\prime =\mre_8 +\mrd_8+ \mra_1$ with $A_{M^\prime}=\ZZ_2^3$
and $\ell(A_{M^\prime})=3$ so there can be no associated $T$.
It would be interesting to study more examples of forbidden groups.

\subsubsection{$d=2$}
\label{ss:d2}

When $d=2$, criterion 1 implies that lattices with $\ell(A_L)=1$ give allowed groups.
In Table \ref{l1} we present a few examples of this type.

\begin{table}[h!]\begin{center}
\renewcommand{\arraystretch}{1.0}
\begin{tabular}{|c|c|c|}
\hline
$L$ & $A_L$ & $T$   \\[3pt]
\hline
$\mra_{18} $ &  $\ZZ_{19}$  & [2, 1, 10]
  \\ \hline
  $\mra_4 + \mre_6 + \mre_8$ &  $\ZZ_5 \times \ZZ_3 \cong \ZZ_{15}$ & [2, 1, 8] 
  \\ \hline
$\mra_2 + \mra_{16} $ &  $\ZZ_3 \times \ZZ_{17} \cong \ZZ_{51}$  & [6, 3, 10]
  \\ \hline
$\mra_8 + \mra_{10} $ &  $\ZZ_9 \times \ZZ_{11} \cong \ZZ_{99}$  & [10, 1, 10]
  \\ \hline
$\mra_6 + \mra_{12} $ &  $\ZZ_7 \times \ZZ_{13} \cong \ZZ_{91}$  & [2, 1, 46]
  \\ \hline
$\mre_6 + \mra_{12} $ &  $\ZZ_3 \times \ZZ_{13} \cong \ZZ_{39}$ & [4, 1, 10]
 \\ \hline 
\end{tabular}
\caption{Examples of allowed $L$ with $\ell(A_L)=1$, when $d=2$. $T$ is denoted by its Gram matrix $[u_1^2,u_1\cdot u_2,u_2^2]$.}
  \label{l1}\end{center}\end{table}

Before considering examples with $\ell(A_L)=2$ let us describe how
to find the lattice $T$. To begin, $d(T)$ is known because it must be equal to
$d(L)$ or $d(M)$. Next, the even 2 dimensional lattices of determinant less than 50 are listed in Table 15.1 of \cite{CS}, and
for larger $d(T)$ they can be found using the {\tt SageMath} module on binary quadratic forms \cite{sage}.
Given $T$, the pair $(A_T, q_T)$ can be deduced as explained in appendix \ref{ap:not}.
We then check if $(A_T, q_T) \cong (A_L, q_L)$.

Criterion 2 must also hold when $\ell(A_L)=1$ since in this case the existence of a primitive embedding
is guaranteed by criterion 1. In Table \ref{l1} we have shown the corresponding matrices $T$.
For example, with $d(T)=19$ there is only the lattice 
with Gram matrix $Q$ given in Table \ref{l1}.
It can be checked that $A_T\cong \ZZ_{19}$ and that the values of $q_T$ are such that indeed $(A_T,q_T)$
is isomorphic to $(A_L, q_L)$ for $L=A_{18}$.
For $L=\mra_4+\mre_6+\mre_8$ we need a $T$ with $d(T)=15$. In this case there are two possible 
lattices, $[2,1,8]$ and $[4,1,4]$, both with $A_T=\ZZ_{15}$. It can be checked that only the discriminant form of 
the first does match $q_L$.

The allowed $L$'s are given  in Table 2 in \cite{SZ}. It is a simple task to find $A_L$ and $\ell(A_L)$.
Groups accepted by criterion 2 have $\ell(A_L)=2$ and $MW=[0]$. In our language trivial $MW$ means trivial $H_L$,
i.e. trivial overlattice $M=L$. There are many examples of this type. In Table \ref{l2} we show a few. To find
$T$ we proceed as explained before, looking first for even lattices of determinant $d(T)=d(L)$ and $A_T=A_L$. 
There might be more than one, the correct ones must have $(A_T,q_T) \cong (A_L,q_L)$. In Table \ref{l2} we
have displayed in red candidates for $T$ that are discarded because $q_T$ is incongruent with $q_L$. 
The incorrect $T$'s are more or less obvious. 
Checking the isomorphism for the correct ones is more laborious. For instance, for $L=\mre_6 +\mrd_{12}$, the
distinct values that can appear in $q_L$ are in the set $\{0, \frac13, \msm{1}, \frac43\}$. Both $T$'s have
$A_T=\ZZ_2 \times \ZZ_6$, but  the values of $q_L$ can only be matched to the values
in the $T$ with $Q^{-1}=[\frac13, -\frac16, \frac13]$.
\begin{table}[h!]\begin{center}
\renewcommand{\arraystretch}{1.0}
\begin{tabular}{|c|c|c|}
\hline
$L$ & $A_L$ & $T$   \\[3pt]
\hline
$2\mrd_{9} $ &  $\ZZ_4 \times \ZZ_4$  & [4, 0, 4] 
  \\ \hline
  $\mra_4 + 2\mre_7$ &  $\ZZ_5 \times \ZZ_2 \times \ZZ_2 \cong \ZZ_{10} \times \ZZ_2$ & [4, 2, 6] \red{[2,0,10]}
  \\ \hline
$\mre_6 + \mrd_{12} $ &  $\ZZ_3 \times \ZZ_2 \times \ZZ_2 \cong \ZZ_6\times \ZZ_2$  & [4, 2, 4] \red{[2,0,6]}
  \\ \hline
$\mra_1 + \mra_{17} $ &  $\ZZ_2 \times \ZZ_{18}$  & [4, 2, 10] \red{[2,0,18]} 
  \\ \hline
\end{tabular}
\caption{Examples of allowed $L$ with $\ell(A_L)=2$, when $d=2$. The candidates for $T$ with $d(T)=d(L)$, but
with  $(A_T,q_T) \ncong (A_L,q_L)$, are displayed in red.}
  \label{l2}\end{center}\end{table}

The example $L=\mra_1 + \mra_{17}$ is interesting because it also admits an overlattice. Indeed, in section
\ref{ss:d1} we saw that $\tilde L=\mra_{17}$ has an overlattice $\tilde M$ with $\tilde M/\tilde L\cong \ZZ_3$,
$A_{\tilde M} = \ZZ_2$ and $q_{\tilde M} = \{\msm{0}, \frac12\}$. Thus, the full $L$ has an overlattice
$M=\mra_1 + \tilde M$ with $A_M=\ZZ_2 \times \ZZ_2$ and $M/L\cong \ZZ_3$. Now criterion 3 can be fulfilled
with $T=[2,0,2]$. This agrees with results of \cite{SZ} for this $L$.

When $\ell(A_L) \ge 3$ we can check that the allowed groups pass criterion 3 with the data given in Table 2 of
\cite{SZ}. One example is $L=3\mra_6$, $A_L=\ZZ_7^3$. There is an isotropic subgroup $H_L=\ZZ_7$ generated
by $\mu=w_1(1)\times w_2(2)\times w_4(3)$, where $w_i(a)$ denotes weights of the $a^{\text{th}}$ $\mra_6$ factor.
Notice that $\mu^2=4=0\, \text{mod}\, 2$. From \eqref{mwrel}, $d(M)=7$ so necessarily $A_M=\ZZ_7$.
Following the procedure to determine $q_M$ shows that it matches the $q_T$ of $T=[2,1,4]$
which is the unique even 2-dimensional lattice with $d(T)=7$.

Finally we come to forbidden groups. Let us discuss the examples in Table \ref{for2}. In all three there are no suitable
lattices $T$. The possible candidates, shown in red, are discarded 
because their $q_T$ does not match $q_L$. We conclude that these groups do not satisfy criterion 2 and continue to check criterion 3.
In example 1 we know that $\mrd_8$ has an overlattice $\mre_8$ so the full $L$ has an overlattice
$M=2\mre_8 + \mra_2$, $M/L\cong \ZZ_2$  so $d(M)=3$, consistent with $A_M=\ZZ_3$. Now $q_M$ matches
the $q_T$ of $T=[2,1,2]$ but still criterion 3 fails because $M_{\text{root}} \not= L$.
In example 2, there is an isotropic subgroup $H_L=\ZZ_2$ generated by $\mu=v \times w_2$, where $v$ is the vector weight
of $\mrd_{15}$ and $w_2$ is the weight of the ${{\bf{10}}}$ of $\mra_3$. Since $v^2=1$ and $w_2^2=1$, $\mu^2=2$.
From \eqref{mwrel}, $d(M)=\frac{16}{2^2}=4$. The only possible $T$ with $d(T)=4$ is $[2,0,2]$ and it could be
that $q_T$ matches $q_M$. However, $M$ has elements $y + n \mu$, $y\in L$, $n=0,1$ and since $\mu^2=2$, 
$M_{\text{root}} \not= L$. Hence, example 2 does not pass criterion 3. 
Concerning example 3, it flops criterion 3 because there is no isotropic subgroup of $A_L$. To see this, first
observe that \eqref{mwrel} implies that only $|H_L|=7$ would be consistent with $d(M)$ being an integer.
Thus, $H_L$ would have to be $\ZZ_7$ and its generator would have to be a product of weights of the $\mra_6$'s, say
$\mu=w_i(1) \times w_j(2)$. 
However it is not possible to obtain $\mu^2=0\, \text{mod}\, 2$.
  
\begin{table}[h!]\begin{center}
\renewcommand{\arraystretch}{1.0}
\begin{tabular}{|c|c|c|c|}
\hline
\#& $L$ & $A_L$ & $T$   \\[3pt]
\hline
1 & $\mre_8 + \mrd_8 + \mra_2 $ &  $\ZZ_2 \times \ZZ_2 \times \ZZ_3 \cong \ZZ_2 \times \ZZ_6$  & \red{[2, 0, 6]} \red{[4,2,4]}
  \\ \hline
2 & $\mrd_{15} + \mra_3$ &  $\ZZ_4 \times \ZZ_4$ & \red{[4,0,4]}
  \\ \hline
3 & $2\mra_6 + \mre_6$ &  $\ZZ_7 \times \ZZ_7 \times \ZZ_3 \cong \ZZ_7\times \ZZ_{21}$  & \red{[14,7,14]} 
    \\ \hline
\end{tabular}
\caption{Examples of forbidden $L$ when $d=2$.}
  \label{for2}\end{center}\end{table}
In summary, we have provided several examples where it was relatively simple to apply by hand the criteria
that serve to determine whether a group of maximal rank is allowed or not. Clearly,
to make a full search, or even to check more complicated examples, would require computer aid.

In Table \ref{tab:alld2} we give the subgroups $H_L$ and the lattice $T$ for all the allowed $L$'s
found in the K3 framework \cite{SZ}. They correspond to all maximal enhancements arising in heterotic compactifications on $T^2$.

\subsubsection{$d=8$}
\label{ss:d8}

The case $d=8$ is peculiar because there exists an even unimodular lattice of signature $(0,8)$, namely $\mre_8$.
To see how this enters the analysis, consider $L=3 \mre_8$ which has trivial $A_L$. Since $\ell(A_L)=0$, this
$L$ easily passes criterion 1. Now, since criterion 2 must also be fulfilled there has to be an even lattice of signature
$(0,8)$ and trivial $A_T$. This requires $d(T)=1$ so $T=\mre_8$.
This indicates that in the heterotic on $T^8$ it is possible to obtain the group $3\mre_8$.
Indeed, it can be found in the HE by setting all the Wilson lines to zero and taking the internal torus with
metric $g_{ij} = \frac12 \tilde g_{ij}$, where $\tilde g_{ij}$ is the Cartan matrix of $\mre_8$. 
The antisymmetric field must be chosen as
\be
b_{ij} = \left\{\!\!\!\begin{array}{ll}
\phantom{-}\tfrac12 \tilde g_{ij}, & i < j, \\
-\tfrac12 \tilde g_{ij}, & i > j, \\
\phantom{-}0, & i=j.
\end{array}
\right.
\, .
\label{bginsparg}
\ee
This is an example of the general type discussed in \cite{Goddard:1983at, Ginsparg:1986bx} in which
$p_L - p_R$ belongs to the root lattice of an ADE group of rank $d$. 

A second interesting example is $L=24 \mra_1$, $A_L=\ZZ_2^{24}$. Since $\ell(A_L)=24$, $L$ fails criterion 1
and criterion 2 as well because $\ell(T) \le 8$. To apply criterion 3 we recall that this $L$ admits an even unimodular overlattice
given by one of the Niemeier lattices, say $N_\psi$, with $N_\psi/L\cong \ZZ_2^{12}$ (see chapter 16 in \cite{CS}).
It is also known that the root lattice of $N_\psi$ and $L$ coincide.
Thus, $L$ fulfills criterion 3 with $M=N_\psi$ and $T=\mre_8$. 
By the same token $L=12 A_2$ is also allowed by criterion 3.
Niemeier lattices in heterotic compactifications on $T^8$ have appeared in \cite{Harrison:2016pmb}.

\subsection{Connection to heterotic compactifications}
\label{ss:nikhet}

We have seen that the groups of maximal rank that can be embedded in $\nld$ 
are characterized
by an ADE lattice $L$ of rank $d+16$, the isotropic subgroup $H_L\subset A_L$, the associated overlattice $M$ and
the complementary even lattice $T$ of rank $d$, satisfying $(A_T,q_T) \cong (A_M,q_M)$.
The isotropic subgroup $H_L$ is the torsion part of the embedding, in the sense that $M/L\cong H_L$.
For an embedding to exist, it must be that $d(M)=d(T)=d(L)/|H_L|^2$.
In the heterotic framework $L$ is the root lattice of some gauge group with maximal enhancing.
We now want to identify $T$, which we call the complementary lattice.

There is a natural candidate for an even lattice of rank $d$, namely the sublattice of $\nld$, denoted $K$, 
obtained by setting ${\bf p_L}=0$. This is
\be
K=\Big\{({\bf p_R};{\bf p_L}) \in \nld \ ||\  {\bf p_L}=0\Big\} \, .
\label{Kheterotic}
\ee
Let us next examine the consequences of setting ${\bf p_L}=0$. First, from \eqref{pI} we find that $p^I=0$ implies
\be
\pi^I = - w^i A_i^I \, . 
\label{pI0}
\ee
Second, imposing $p_L=0$ leads to
\be
n_i = - w^j E_{ji} \, , 
\label{pL0n}
\ee
after substituting \eqref{pI0} in \eqref{pL}. 
From $p_L=0$ it further follows that
\be
p_R =- \sqrt 2 w^i e_i \, . 
\label{pL0}
\ee
Thus, $p_R$ lies in a lattice of rank $d$ as long as all the windings $w^i$ are allowed to be different from zero.
Since $\pi$ is a vector in the gauge lattice $\Upsilon_{16}$, the condition \eqref{pI0} can only be fulfilled 
with $w^i \not=0$ if the Wilson lines $A_i$ are quantized, in the sense that they are given by a vector in
$\Upsilon_{16}$, divided by a positive integer. 
We define the order of the Wilson line $A_i$ as the smallest positive integer $N_i$ such that 
\be
N_i A_i \in \Upsilon_{16} \quad  (\text{no sum in } i) \, .
\label{orderA}
\ee
If $A_i=0$, its order is 1.
In section \ref{sec:shift} we will review an algorithm to find such Wilson lines.
All $A_i$ must be quantized so that \eqref{pI0} does not force some windings $w^i$ to be identically zero.
The quantization condition in \eqref{pL0n} is also very restrictive. It clearly demands the $E_{ij}$ to be rational numbers.
Taking into account quantization of the Wilson lines then requires
the $T^d$ metric components $g_{ij}=e_i\! \cdot \! e_j$ to be rational numbers, which is consistent with 
$p_R^2$ being even. From now on we assume that $K$ has rank $d$.

The constraints on the $A_i$ and $E_{ij}$ are compatible with having a gauge group of maximal enhancing, 
which is the case under study. 
In fact, recall that to this end there must exist solutions to ${\bf p_R}=0$ and ${\bf p_L}^2=2$. The former implies
the conditions \eqref{pR0}, which can be achieved with quantized $A_i$ and rational $E_{ij}$.

The even lattice $K \subset \nld$ has signature $(d,0)$ by construction.
Applying Nikulin's Theorem 1.12.4 in \cite{Nikulin80}, we learn that $K$ admits a primitive embedding in $\nld$.
It follows that the orthogonal complement of $K$ in $\nld$ also admits a primitive embedding in $\nld$.
This orthogonal complement is just the sublattice of $\nld$ defined by ${\bf p_R}=0$ which we denote $M$, i.e.
\be
M=\{({\bf p_R};{\bf p_L}) \in \nld \ ||\  {\bf p_R}=0 \} \, .
\label{Mheterotic}
\ee
The name $M$ is appropriate because it is indeed the overlattice of criteria 3 with $M_{\text{root}}=L$.
The reason is that $M_{\text{root}}$ is the sublattice of $M$ generated by vectors with 
${\bf p_L}^2=2$ and it has rank $(d+16)$ by the assumption of maximal enhancing.

So far we have argued that $M$ of signature $(0,d+16)$ is the orthogonal complement in $\nld$ of $K$ of signature $(d,0)$,
and that $K$ as well as $M$ are primitively embedded in $\nld$. 
In fact, $\nld$ is an overlattice of $M \oplus K$. We can then apply Lemma 2.4 in \cite{SZ}
to conclude that there is an isomorphism $(A_M, q_M) \cong (A_K, -q_K)$. A proof of this lemma is
presented in appendix \ref{ap:extraT}. Finally, by Nikulin's Proposition 1.12.1 \cite{Nikulin80} there exists $T$ of
signature $(0,d)$ satisfying $(A_M, q_M) \cong (A_T, q_T)$. It is obtained by changing the sign of the Gram matrix
of $K$, i.e. 
\be
T=K\langle -1 \rangle \ .  
\ee
Summarizing, the two rationality conditions $N_i A_i \in \Upsilon_{16}$ and $E_{ij} \in \mathbb{Q}$,
guarantee the existence of the even $(0,d)$ lattice $T$, which in turn implies the existence
of the even $(0,d+16)$ lattice $M$ with $(A_M, q_M) \cong (A_T, q_T)$. Thus, the rationality conditions
are necessary to have maximal enhancing to a group of rank $d+16$. However, these conditions are
not sufficient to ensure that the sub-lattice $M_{\text{root}}$ has rank $d+16$. The additional constraint
in criterion 3 is precisely that the gauge lattice $L$ of rank $d+16$ coincides with $M_{\text{root}}$.

\subsubsection{Lattice data from moduli}
\label{ss:ldata}

 Once we know the data $(L,T)$ of the allowed groups $\gr$ we still have to determine specific moduli
$A_i$ and $E_{ij}$ that give rise to them. 
Conversely, given $A_i$ and $E_{ij}$, in principle $L$ is obtained from the solutions 
of ${\bf p_R}=0$, ${\bf p_L}^2=2$, which correspond to the roots of $\gr$. 
On the other hand, $T$ can be derived directly from the moduli as explained below.

The elements of $T$ are of the form \eqref{pL0}. Besides, the moduli must comply with the conditions
\eqref{pI0} and \eqref{pL0n}. To make more concrete statements, consider first the case in which
the $E_{ij}$ are integers so that \eqref{pL0} is satisfied by any $w^i$. Then, a class of allowed
values for the $w^i$ are multiples of the Wilson lines orders, namely $w^i=\ell_i N_i$ (no sum over $i$), with
$\ell_i \in \ZZ$. If we assume that this class exhausts all possibilities, $T$ will be generated by a basis
\be
u_1 = \sqrt 2 N_1 e_1, \ u_2 = \sqrt 2 N_2 e_2,  \ldots, \ u_d = \sqrt 2 N_d e_d \, ,
\label{naivef}
\ee
where we dropped an irrelevant sign. The Gram matrix of $T$ will then be given by
\be
Q_{ij} = u_i \cdot u_j = 2 N_i N_j g_{ij} = N_i N_j (E_{ij} + E_{ji} - A_i \cdot A_j) \, .
\label{naiveQ}
\ee
Since this is valid for $E_{ij}$ integers and $N_i A_i \in \Upsilon_{16}$, we see that the $Q_{ij}$ are integers
and the diagonal components are even, as required for an even lattice. 

In some cases there might be more admissible values of the winding numbers $w^i$. In general, the allowed values
are sets of integers $(M_1, M_2, \ldots, M_d)$ that satisfy
\begin{subequations}
\begin{align}
&{} M_1 A_1 + M_2 A_2 + \cdots + M_d A_d \in \Upsilon_{16}\, , \label{MA}\\
&{} M_1 E_{1i} + M_2 E_{2i} + \cdots + M_d E_{di} \in \ZZ \, , \ i=1,\ldots, d 
\, .
\label{ME}
\end{align}
\label{genM}
\end{subequations}
In this situation a way to proceed is to obtain $d$ solutions $(M_1^{(k)}, \ldots, M_d^{(k)})$, $k=1, \ldots, d$, 
linearly independent (with Euclidean metric), such that the vectors
\be
u_k=\sum_{\ell=1}^d \sqrt2 M_\ell^{(k)} e_\ell
\label{genf}
\ee
generate a lattice with the least volume. For instance, the vectors in \eqref{naivef} are recovered when $E_{ij} \in \ZZ$ and 
the only solutions of \eqref{MA} are $M_\ell^{(k)}=N_\ell \delta_{\ell k}$ (no sum over $\ell$).
In the general case we have to impose the condition of least volume. To be more precise, define the matrix $C$ with
elements $C_{k\ell} = M_\ell^{(k)}$, i.e. the rows of $C$ are the solutions of \eqref{genM}. The Gram matrix of $T$ then reads 
\be
Q_{k\ell} = u_k \cdot u_\ell = 2 (C \, g\,  C^t)_{k\ell} \, ,
\label{genQ}
\ee
where we used $g_{ij} = e_i \! \cdot \! e_j$.  
Therefore, $\det Q=2^d (\det C)^2 \det g$. Since the determinant of the torus metric is fixed by the choice of
moduli $A_i$ and $E_{ij}$, to obtain the least lattice volume it suffices to choose $C$ with least determinant.
Hadamard's inequality then instructs us to choose $d$ independent solutions
$(M_1^{(k)}, \ldots, M_d^{(k)})$ of $\eqref{genM}$ with the least norm.
To check that $Q_{k\ell}$ are integers and the diagonal elements are even, we write 
\mbox{$g_{ij}=\frac12(E_{ij} + E_{ji} - A_i \cdot A_j)$}, and take into account that the $M_i^{(k)}$
verify \eqref{genM}. Finally, $Q$ is unique up to the action of $\mathrm{GL}(d,\ZZ)$. For $d=2$ we can use
the procedure described in section 3, Chapter 15, of \cite{CS} to bring $Q$ to the standard reduced form
used in \cite{SZ}.

In the next sections we will discuss systematic methods to determine moduli associated to
groups of maximal enhancing when $d=1$ and $d=2$. We will then exemplify further how $T$ computed 
from the moduli matches the $T$ from the lattice embedding data. Meanwhile it is instructive to 
illustrate the main points in cases with generic $d$.

For a simple example, consider moduli $A_i=0$,  $g_{ij} = \frac12 \tilde g_{ij}$, where $\tilde g_{ij}$ is the 
Cartan matrix of an ADE group $\tilde {\mathrm G}_d$ of rank $d$, and $b_{ij}$ is given in \eqref{bginsparg}.
The $E_{ij}$ moduli are found to be 
\be
E_{ij} = \left\{\!\!\!\begin{array}{ll}
\tfrac12 \op \tilde g_{ij}, & i = j, \\
\tilde g_{ij}, & i < j, \\
0, & i > j
\end{array}
\right.
\, .
\label{Eginsparg}
\ee
Therefore, the $E_{ij}$ are either $1$, $-1$ or $0$.
In this setup the gauge group of the heterotic string on $T^d$ is $2 \mre_8 + \tilde{\mathrm G}_d$ in the HE or   
$\mrd_{16} + \tilde{\mathrm G}_d$ in the HO. This example is of the general type in which all Wilson lines are set to zero
and $p_L - p_R\in \widetilde \Gamma_d$, where $\widetilde \Gamma_d$ is the root lattice of 
$\tilde {\mathrm G}_d$ \cite{Goddard:1983at, Ginsparg:1986bx}.
From the lattice formalism we find that $T=\widetilde \Gamma_d$. From the moduli we obtain the same result
for $T$ because the basis is given in \eqref{naivef} with $e_i =\frac1{\sqrt2} \tilde e_i$ and $N_i=1$.

A second example in the HO on $T^d$ has moduli \cite{Goddard:1983at, Ginsparg:1986bx} 
\be
e_i^a = \frac1{\sqrt{2}} \delta_i^a \, , \quad b_{ij}=0 \, , \quad A_i^I=\delta_i^I \, \ {\rm with}\, \,  i\le d
\label{exd16md}
\ee
It can be shown that the resulting group is $\mrd_{d+16}$. All Wilson lines have order $N_i=2$.
Besides, $E_{ij}=\delta_{ij}$ so that the condition \eqref{ME} does not constrain the $M_i$.
For $d=1$ we can just take $M_1=N_1=2$ so that $u_1=2$ and $T=\mra_1\langle 2 \rangle$
as we found with the lattice formalism in section \ref{ss:d1}. For $d \ge 2$ there are solutions to \eqref{MA}
other than $M_i^{(j)} = 2 \delta_{ij}$. For instance, $A_1 \pm A_2 \in \Gamma_{16}$.
The $M_i^{(j)}$ can be chosen so that the $u_i$ are the roots of $\mrd_d$. Thus, $T=\mrd_d$. 

Another important question in the heterotic context is the meaning of the quadratic discriminant form $q_T$.
The answer is that the values that $p_R^2$ can take are precisely given by $q_T\!\!\mod 2$. This follows because $p_R$ 
generically lies in the dual lattice $T^*$. When $T$ has basis \eqref{naivef}, it is easy to see from \eqref{pR} that $p_R$ 
indeed takes values in a lattice generated by $u^{*i} = \frac{1}{\sqrt 2 N_i} \hat e^{*i}$, with Gram matrix
the inverse of $Q$ in \eqref{naiveQ}. When there are additional solutions to \eqref{genM}, 
so that the basis for $T$ is given by \eqref{genf}, $p_R$ lies in a lattice spanned by 
\be
u^{*i}=\frac1{\sqrt2} \sum_{k=1}^d C^{ki} \hat e^{*k} \, ,
\label{dualT}
\ee
where $C^{ki}=C^{-1}_{ki}$ and as before $C_{k\ell} = M_\ell^{(k)}$. 
Thus, $u^{*i} \cdot  u^{*j}=Q^{ij}=Q^{-1}_{ij}$, with $Q$ the Gram matrix in \eqref{genQ}.
The fact that $q_T$ gives the values of $p_R^2$ is useful to determine the spectrum of massive states. 
In particular, it could be relevant in the double field theory analysis of gauge enhancements \cite{Gerardo}.

\section{Compactifications on $S^1$}
\label{sec:circle}

In this section we consider in more detail compactifications of the heterotic string on the circle, where
the moduli are the radius $R$ and the 16-dimensional Wilson line $A^I$.  
The problems of finding all possible gauge groups $\gr\times \uo^{17-r}$ and the corresponding moduli $(R,A^I)$,
were  solved in \cite{Fraiman:2018ebo}
by means of the extended Dynkin diagram (EDD) associated to $\mathrm{II}_{1,17}$,
depicted in Figure \ref{dynkindiag2}. We will first review the procedure and the results. We will also explain 
how they can be put in a form that can be generalized to compactification on $T^d$ .
Afterwards we will discuss the connection with the lattice embedding formalism.
In Table \ref{tab:alld1} we collect the relevant lattice and moduli data for all the 44 groups of maximal rank that appear
in heterotic string compactifications on $S^1$.

For generic moduli the elements of $\nlc$ are given in \eqref{momenta}, with $e_1=R$, $\hat e_1^*=1/R$, i.e.
\be
p_R = \frac{1}{\sqrt2\, R} (n - Ew - \pi\cdot A), \quad p_L=\sqrt2 R w + p_R, \quad p^I= \pi^I + w A^I \, ,
\label{momc}
\ee
where $E=R^2 + \frac12 A^2$ is just the $E$-tensor of \eqref{genmetric} for $d=1$.  
Recall that $n$ and $w$ are the quantized momenta and winding numbers, while $\pi^I$ belongs to the
 lattice $\Gamma_8 \times \Gamma_8$ in the HE or $\Gamma_{16}$ in the HO.

As in \cite{Ginsparg:1986bx}, ${\bf p} =(p_R;p_L,p^I)$ can be expanded as
\be
{\bf p}=w\overline k + nk +  \pi\cdot l\ ,
\label{gbasis}
\ee
with basis
\be
k=\frac1{\sqrt2}\left(\frac1R; \frac1R,0\right),  \quad 
\overline k=\frac1{\sqrt2}\left( -R-\frac{A^2}{2R}; R-\frac{A^2}{2R},\sqrt2A^I\right), \quad 
l^I=\left( -\frac{A^I}{\sqrt2R}; -\frac{A^I}{\sqrt2R},u^I\right)\, .
\label{ginsp1}
\ee
Here $u^I$ is a Cartesian 16-dimensional basis vector. The inner product is taken with the Lorentzian metric $(-;+,\ldots,+)$.
Thus $k\cdot k=\overline k\cdot \overline k=0,$ $k\cdot\overline k=1$,  $l^I\cdot l^J=\delta^{IJ}$, $k\cdot l^I=\overline k\cdot l^I=0$. 
For many purposes it is simpler to work with the charge vector $|Z\rangle = |w, n; \pi^I\rangle$. 
The change of basis to ${\bf p}$ is easily read from \eqref{gbasis}. Besides, 
$\langle Z^\prime | Z\rangle = w^\prime n + n^\prime w + \pi^\prime \cdot \pi$.

\subsection{Moduli and gauge group from the EDD diagram}
\label{sec:EDD}

We refer to \cite{Goddard:1983at} for an introduction to root systems and associated EDDs of Lorentzian 
$\mathrm{II}_{1,8m+1}$ lattices. The special case of $\nlc$ is discussed in detail in \cite{Ginsparg:1986bx} and 
\cite{Cachazo:2000ey}, precisely in connection to circle compactifications of the heterotic string.
It was originally considered by Vinberg \cite{Vinberg72}.
The reflective part of its group of automorphisms, which is actually the duality group $\rO(1,17,\ZZ)$ \cite{Cachazo:2000ey}, 
can be encoded in the EDD as we review shortly.

\subsubsection{Embedding of $\Gamma_8 \times \Gamma_8$}
\label{sss:he}

\begin{figure}[htb]
\begin{center}
\begin{tikzpicture}[scale=.175]
   \draw[thick] (0 cm,0) circle (5 mm) node [shift={(0.0,-0.5)}] {\mfn{1}} node [shift={(0.0,-1)}] {\msc{\red{3}}};
    \draw[thick] (.5 cm, 0) -- (3.5 cm,0);    
       \draw[thick] (4 cm,0) circle (5 mm) node [shift={(0.0,-0.5)}] {\mfn{2}}node [shift={(0.0,-1)}] {\msc{\red{6}}};
     \draw[thick] (4.5 cm, 0) -- (7.5 cm,0); 
        \draw[thick] (8 cm,0) circle (5 mm) node [shift={(0.0,-0.5)}] {\mfn{3}} node [shift={(0.0,-1)}] {\msc{\red{5}}};
     \draw[thick] (8.5 cm, 0) -- (11.5 cm,0);     
    \draw[thick] (12 cm,0) circle (5 mm) node [shift={(0.0,-0.5)}] {\mfn{4}}node [shift={(0.0,-1)}] {\msc{\red{4}}};
     \draw[thick] (12.5 cm, 0) -- (15.5 cm,0);     
         \draw[thick] (16 cm,0) circle (5 mm) node [shift={(0.0,-0.5)}] {\mfn{5}}node [shift={(0.0,-1)}] {\msc{\red{3}}};
             \draw[thick] (16.5 cm, 0) -- (19.5 cm,0);     
         \draw[thick] (20cm,0) circle (5 mm) node [shift={(0.0,-0.5)}] {\mfn{6}}node [shift={(0.0,-1)}] {\msc{\red{2}}};
                             \draw[thick] (4 cm, 0.5cm) -- (4 cm, 3.5cm);     
         \draw[thick] (4cm,4cm) circle (5 mm) node [shift={(-0.5,0.0)}] {\mfn{7}}node [shift={(-1,0)}] {\msc{\red{4}}};
 \draw[thick] (4 cm, 4.5cm) -- (4 cm, 7.5cm);     
         \draw[thick] (4cm,8cm) circle (5 mm) node [shift={(-0.5,0.0)}] {\mfn{8}}node [shift={(-1,0)}] {\msc{\red{2}}};
                 \draw[thick] (20.5 cm, 0) -- (23.5 cm,0);     
         \draw[thick] (24cm,0) circle (5 mm) node [shift={(0.0,-0.5)}] {\mfn{0}}node [shift={(0.0,-1)}] {\msc{\red{1}}};
            \draw[thick] (24.5 cm, 0) -- (27.5 cm,0);     
         \draw[thick] (28cm,0) circle (5 mm) node [shift={(0.0,-0.5)}] {\footnotesize{\texttt{C}}};
   \draw[thick] (28.5 cm, 0) -- (31.5 cm,0);     
   \draw[thick] (32 cm,0) circle (5 mm) node [shift={(0.0,-0.5)}] {\mfn{0^\prime}} node [shift={(0.0,-1)}] {\msc{\red{1}}};
    \draw[thick] (32.5 cm, 0) -- (35.5 cm,0);    
       \draw[thick] (36 cm,0) circle (5 mm) node [shift={(0.0,-0.5)}] {\mfn{6^\prime}}node [shift={(0.0,-1)}] {\msc{\red{2}}};
     \draw[thick] (36.5 cm, 0) -- (39.5 cm,0); 
        \draw[thick] (40 cm,0) circle (5 mm) node [shift={(0.0,-0.5)}] {\mfn{5^\prime}}node [shift={(0.0,-1)}] {\msc{\red{3}}};
     \draw[thick] (40.5 cm, 0) -- (43.5 cm,0);     
    \draw[thick] (44 cm,0) circle (5 mm) node [shift={(0.0,-0.5)}] {\mfn{4^\prime}}node [shift={(0.0,-1)}] {\msc{\red{4}}};
     \draw[thick] (44.5 cm, 0) -- (47.5 cm,0);     
         \draw[thick] (48 cm,0) circle (5 mm) node [shift={(0.0,-0.5)}] {\mfn{3^\prime}}node [shift={(0.0,-1)}] {\msc{\red{5}}};
             \draw[thick] (48.5 cm, 0) -- (51.5 cm,0);     
         \draw[thick] (52cm,0) circle (5 mm) node [shift={(0.0,-0.5)}] {\mfn{2^\prime}}node [shift={(0.0,-1)}] {\msc{\red{6}}};
                 \draw[thick] (52.5 cm, 0) -- (55.5 cm,0);     
         \draw[thick] (56cm,0) circle (5 mm) node [shift={(0.0,-0.5)}] {\mfn{1^\prime}}node [shift={(0.0,-1)}] {\msc{\red{3}}};
                    \draw[thick] (52 cm, 0.5cm) -- (52 cm, 3.5cm);     
 \draw[thick] (52cm,4cm) circle (5 mm) node [shift={(0.5,0.0)}] {\mfn{7^\prime}}node [shift={(1,0)}] {\msc{\red{4}}};
 \draw[thick] (52 cm, 4.5cm) -- (52 cm, 7.5cm);     
         \draw[thick] (52cm,8cm) circle (5 mm) node [shift={(0.5,0.0)}] {\mfn{8^\prime}}node [shift={(1,0)}] {\msc{\red{2}}};
     \end{tikzpicture}
     \end{center}
     \caption{Extended Dynkin diagram for the $\mathrm{II}_{1,17}$ lattice, with labels showing the embedding of the
     extended Dynkin diagrams of $\mre_8 + \mre^{\prime}_8$. The Kac marks are shown in red.}\label{dynkindiag2}
\end{figure}
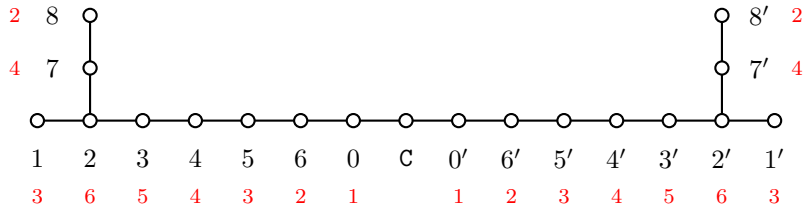

We begin by describing the embedding of the HE lattice $\Gamma_8 \times \Gamma_8$ in $\nlc$. 
The EDD is shown in Figure \ref{dynkindiag2}. It is composed by the extended Dynkin diagrams of $\mre_8$ and $\mre^\prime_8$
joined by a central node. The nodes can be specified in terms of the charge vectors
\begin{equation}
\begin{split}
\varphi_i&=| 0,0; \alpha_i, 0^8\rangle, \qquad \varphi_{i'}=|0,0; 0^8, \alpha'_i\rangle, \qquad  i=1,...,8\, , \\
\varphi_{0}&=|0,-1;\alpha_0, 0^8\rangle, \quad \
\varphi_{{\texttt{C}}}=|1,1; 0^{8}, 0^8\rangle,  \qquad
\varphi_{0'}=|0,-1; 0^8, \alpha'_0\rangle\, .
\end{split}
\label{basishe}
\end{equation}
where $\alpha_i$  and $\alpha'_i$ are the simple roots of $\mre_8$ and  $\mre^\prime_8$, given in Table \ref{tab:e8} (note that for convenience in regard to the EDD diagram, we take different conventions for simple roots of the two groups) . Our conventions
for the simple roots and fundamental weights $w_i$, $w^\prime_i$, of $\mre_8$ and $\mre^\prime_8$ are collected in 
Table \ref{tab:e8}. We have also written down the lowest root $\alpha_0=-\sum_{k=1}^8 \kappa_k \alpha_k$, and similarly for
$\alpha^\prime_0$. The $\kappa_i$ and $\kappa^\prime_i$ are the Kac marks,  shown in red in the Figure \ref{dynkindiag2}.
By definition $\kappa_0=\kappa^\prime_0=1$ and sometimes we will set $w_0=0$, $w^\prime_0=0$.

\begin{table}[h!]\begin{center}
\renewcommand{\arraystretch}{2}
{\footnotesize
\begin{tabular}{|c||c|c||c|c|}
\hline
$k$ & $\alpha_k$ & $w_k$ & $\alpha_k^\prime$ & $w_k^\prime$  \\[3pt]
\hline\hline
1 & $(1,\tm1,0,0,0,0,0,0)$ & $\tm(\tm\frac12,\frac12,\frac12,\frac12,\frac12,\frac12,\frac12,\tm\frac52)$ &
 $(0,0,0,0,0,0,1,\tm1)$ & $(\tm\frac52,\frac12,\frac12,\frac12,\frac12,\frac12,\frac12,\tm\frac12)$
 \\[3pt]\hline
2 &  $(0,1,\tm1,0,0,0,0,0)$ & $\tm(0,0,1,1,1,1,1,\tm5)$  &
$(0,0,0,0,0,1,\tm1,0)$ & $(\tm5,1,1,1,1,1,0,0)$
 \\[3pt]\hline
3 &  $(0,0,1,\tm1,0,0,0,0)$ & $\tm(0,0,0,1,1,1,1,\tm4)$  &
$(0,0,0,0,1,\tm1,0,0)$ & $(\tm4,1,1,1,1,0,0,0)$
 \\[3pt]\hline
4 &  $(0,0,0,1,\tm1,0,0,0)$ & $\tm(0,0,0,0,1,1,1,\tm3)$  &
$(0,0,0,1,\tm1,0,0,0)$ & $(\tm3,1,1,1,0,0,0,0)$
 \\[3pt]\hline
5 &   $(0,0,0,0,1,\tm1,0,0)$ & $\tm(0,0,0,0,0,1,1,\tm2)$  &
$(0,0,1,\tm1,0,0,0,0)$ & $(\tm2,1,1,0,0,0,0,0)$
 \\[3pt]\hline
6 &   $(0,0,0,0,0,1,\tm1,0)$ & $(0,0,0,0,0,0,\tm1,1)$ &
$(0,1,\tm1,0,0,0,0,0)$ & $(\tm1,1,0,0,0,0,0,0)$
 \\[3pt]\hline
 7 & $\tm(1,1,0,0,0,0,0,0)$ & $\tm(\frac12,\frac12,\frac12,\frac12,\frac12,\frac12,\frac12,\tm\frac72)$  &
$(0,0,0,0,0,0,1,1)$ & $(\tm\frac72,\frac12,\frac12,\frac12,\frac12,\frac12,\frac12,\frac12)$
 \\[3pt]\hline 
8 & $(\frac12,\frac12,\frac12,\frac12,\frac12,\frac12,\frac12,\frac12)$ & $(0,0,0,0,0,0,0,2)$ &
$\tm(\frac12,\frac12,\frac12,\frac12,\frac12,\frac12,\frac12,\frac12)$ &
$(\tm2,0,0,0,0,0,0,0)$  \\[3pt]\hline
0 & $(0,0,0,0,0,0,1,\tm1)$ & $(0,0,0,0,0,0,0,0)$ &
$(1,\tm1,0,0,0,0,0,0)$ &
$(0,0,0,0,0,0,0,0)$  \\[3pt]\hline
\end{tabular}
}
\caption{Simple roots and fundamental weights of $\mre_8$ and $\mre_8^\prime$.}
  \label{tab:e8}\end{center}\end{table}

In \cite{Cachazo:2000ey} (see also \cite{Vinberg72}), the generators of the duality group $\rO(1,17,\mathbb Z)$ were identified 
with Weyl reflections in the lattice. To be more concrete, let us consider the transformations of the charge vector $|Z\rangle$ about
the simple roots of $\nlc$ in \eqref{basishe}, denoted collectively $|\varphi\rangle$. Since $\langle\varphi|\varphi\rangle=2$,
the Weyl transformation is 
\be
|Z^\prime \rangle = |Z\rangle - \langle \varphi | Z \rangle |\varphi \rangle\, .
\label{weylref}
\ee
Once $|Z^\prime \rangle$ is found, the action on the moduli is deduced by imposing that $p_R=0$
transforms into $p^\prime_R=0$, 
i.e. $n- E w - \pi\cdot A =0$ goes into $n^\prime- E^\prime w^\prime - \pi^\prime \cdot A^\prime =0$. 
This is a shortcut to requiring invariance of the spectrum. 
For example, writing only the transformed quantities, from the nodes $1$, $0$, and $\texttt{C}$ we obtain 
\begin{subequations}
\begin{alignat}{2}
\varphi_1 &: \pi^{\prime 1} = \pi^2, \pi^{\prime 2} = \pi^1 \ \Rightarrow  \
A^{\prime 1} = A^2, A^{\prime 2} = A^1, \label{permu12} \\
\varphi_0 &: n^\prime= n - w + \pi^7 -\pi^8,  \pi^{\prime 7} = \pi^8 + w, \pi^{\prime 8} = \pi^7 - w \ \Rightarrow  \
A^{\prime 7} = A^8 - 1, A^{\prime 8} = A^7+1, \nn \\
{} & \hspace*{10cm} E^\prime = E + A^7 - A^8 + 1,  \label{shift0}\\
\varphi_{\texttt{C}} &: w^'= -n, n^\prime=-w \ \Rightarrow \ E^\prime=\frac1{E} , A^\prime=\frac{A}{E}.  
\label{tdual1}
\end{alignat}
\end{subequations}
Clearly,  \eqref{permu12} is a permutation of the first two components of the Wilson line. In general, the reflections about
nodes $\varphi_i$, or $\varphi^\prime_i$, $i=1,\ldots,8$, induce transformations of the Wilson line  $A^I$ which are
just elements of the Weyl group of $\mre_8$, or $\mre^\prime_8$.
In \eqref{shift0} we recognize a translation of $A^I$ by $\alpha_0 \times 0$, which belongs to $\Gamma_8 \times  \Gamma_8$,
combined with a permutation of $A^7$ and $A^8$.
Finally, \eqref{tdual1} is the generalization of the T-duality $R \to 1/R$ when $A\not=0$.

\begin{table}[h!]\begin{center}
\renewcommand{\arraystretch}{2}
{\footnotesize
 \begin{tabular}{|c|c|c|}
\hline 
Node  &  Fundamental region for $\Gamma_{8}\times \Gamma_8$\\
\hline\hline 
$1\le i\le 8$  & $ A\cdot (\alpha_i \times 0) \geq 0 $\\ 
$0$  & $ A\cdot (\alpha_0 \times 0) \geq -1$\\
$\texttt{C}$  & $E \geq 1$ \\
$0^\prime$  & $ A\cdot (0\times \alpha^\prime_0) \geq -1$\\
$1'\le i'\le 8'$  & $ A\cdot (0\times \alpha^\prime_i) \geq 0 $\\
\hline
\end{tabular}
}
\caption{Fundamental region for HE in $d=1$.} 
\label{tab:fundreghe}
\end{center}
\end{table}

The prescription to obtain a non-Abelian gauge group $G_r$ is to delete $19-r$ nodes of the EDD such that the remaining ones give the
Dynkin diagram of the desired semi-simple Lie Algebra.
The total gauge group is $\gr \times \uo^{17-r}$. 
The Wilson line and the radius are determined by saturating the inequalities in Table \ref{tab:fundreghe} corresponding to the $r$
undeleted nodes. In this manner one can obtain all the allowed groups and the corresponding moduli. For example, for maximal
enhancement, all but $2$ of the inequalities are saturated.
The allowed groups of maximal rank are precisely found by deleting one node in the $\mre_8$ side and 
one node in the $\mre^\prime_8$ side, while the central node ${\texttt{C}}$  corresponding  to $E=1$ cannot be erased. 
In section \ref{sec:shift} we will discuss a simplified way to implement this method, that we call saturation.

Conversely, if the Wilson line $A$ and the radius $R$ are supplied, the resulting group can
be determined by checking which boundary conditions are saturated and keeping only the associated nodes in the EDD.
To this end we might need to first bring the given $A$ and $R$ to the fundamental region by transformations including
shifts and Weyl reflections of $A$ in $\Gamma_8 \times \Gamma_8$,  and the T-duality \eqref{tdual1}.

From the EDD we can also determine  the automorphisms of the lattice corresponding to any enhanced gauge group. 
They are just generated by Weyl reflections \eqref{weylref} associated to the surviving nodes.
The fixed points of each reflection determine a 16-dimensional hyperplane in moduli space where the inequality associated to the 
given node is saturated. 
The intersection of $r$ of these hyperplanes gives the \mbox{$(17-r)$-dimensional} subspace of moduli space where the given rank $r$ gauge group is realized (maximal enhancements are realized at a point). This subspace is invariant under the subgroup of 
$\rO(1,17,\mathbb{Z})$ generated by the $r$ Weyl reflections associated to the surviving nodes.

Having explained how the EDD enables us to determine the allowed groups  $\gr \times \uo^{17-r}$ and the
corresponding moduli, we can draw some results.
For instance, it is easy to see that all ADE $\gr$ of $r \le 9$ are allowed, consistent with Theorem 1.12.4
in the Nikulin formalism \cite{Nikulin80}.
The diagram also shows that for $r=10$ all ADE $\gr$ can appear and that for $r=11$ only $11\mra_1$ is forbidden.
There are 44 allowed groups with maximal rank $r=17$. They were determined in \cite{Fraiman:2018ebo} and are
collected in Table \ref{tab:alld1}. On the other hand, there are 1093 forbidden groups with $r=17$, e.g. $2\mrd_8 + \mra_1$,
which clearly cannot be obtained from the EDD.  
The connection with the Nikulin formalism for the case of maximal rank will be further discussed in section \ref{sec:alld1}.

\subsubsection{Embedding of $\Gamma_{16}$}
\label{sss:ho}

The moduli in the HO theory can be obtained by adapting the EDD to embed $\Gamma_{16}$  explicitly.
To this end we need to write the charge vectors of the nodes in terms of the simple roots $\beta_k$, plus the
spinor weight $\tw_{16}$ of $\mathrm{SO}(32)$. The simple roots and the corresponding fundamental weights are
\be
\begin{split}
\beta_k&=(0^{k-1},1,-1,0^{15-k})\, , \quad \  \tw_k  =  (1^k, 0^{16-k}) \, , \quad k=1,...,14\\
\beta_{15}&=(0^{14},1,-1)\, ,  \hspace*{1.75cm} \tw_{15} =  (\msm{\frac12}^{15},-\msm{\frac12})\, ,\\
 \beta_{16}&=(0^{14},1,1)\, , \hspace*{2.1cm}  \tw_{16}=(\msm{\frac12}^{16}). 
\end{split}
\label{roots16}
\ee 
The lowest root of $\hog$ is 
\be
\beta_{17}=(-1,-1,0^{14})\, . 
\label{low16}
\ee
The Kac marks are $\kappa_k=1$ for $k=1,15,16,17$, and $\kappa_k=2$ for $k=2,...,14$.

\begin{figure}[htb]
\begin{center}
\begin{tikzpicture}[scale=.175]
   \draw[thick] (0 cm,0) circle (5 mm) node [shift={(0.0,-0.5)}] {\mfn{1}} node [shift={(0.0,-1)}] {\msc{\red{1}}};
    \draw[thick] (.5 cm, 0) -- (3.5 cm,0);    
       \draw[thick] (4 cm,0) circle (5 mm) node [shift={(0.0,-0.5)}] {\mfn{2}}node [shift={(0.0,-1)}] {\msc{\red{2}}};
     \draw[thick] (4.5 cm, 0) -- (7.5 cm,0); 
        \draw[thick] (8 cm,0) circle (5 mm) node [shift={(0.0,-0.5)}] {\mfn{3}} node [shift={(0.0,-1)}] {\msc{\red{2}}};
     \draw[thick] (8.5 cm, 0) -- (11.5 cm,0);     
    \draw[thick] (12 cm,0) circle (5 mm) node [shift={(0.0,-0.5)}] {\mfn{4}}node [shift={(0.0,-1)}] {\msc{\red{2}}};
     \draw[thick] (12.5 cm, 0) -- (15.5 cm,0);     
         \draw[thick] (16 cm,0) circle (5 mm) node [shift={(0.0,-0.5)}] {\mfn{5}}node [shift={(0.0,-1)}] {\msc{\red{2}}};
             \draw[thick] (16.5 cm, 0) -- (19.5 cm,0);     
         \draw[thick] (20cm,0) circle (5 mm) node [shift={(0.0,-0.5)}] {\mfn{6}}node [shift={(0.0,-1)}] {\msc{\red{2}}};
                             \draw[thick] (4 cm, 0.5cm) -- (4 cm, 3.5cm);     
         \draw[thick] (4cm,4cm) circle (5 mm) node [shift={(-0.5,0.0)}] {\mfn{17}}node [shift={(-1,0)}] {\msc{\red{1}}};
 \draw[thick] (4 cm, 4.5cm) -- (4 cm, 7.5cm);     
         \draw[thick] (4cm,8cm) circle (5 mm) node [shift={(-0.5,0.0)}] {\mfn{18}}node [shift={(-1,0)}] {\msc{\red{}}};
                 \draw[thick] (20.5 cm, 0) -- (23.5 cm,0);     
         \draw[thick] (24cm,0) circle (5 mm) node [shift={(0.0,-0.5)}] {\mfn{7}}node [shift={(0.0,-1)}] {\msc{\red{2}}};
            \draw[thick] (24.5 cm, 0) -- (27.5 cm,0);     
\draw[thick] (28cm,0) circle (5 mm) node [shift={(0.0,-0.5)}] {\mfn{8}}node [shift={(0.0,-1)}] {\msc{\red{2}}};
   \draw[thick] (28.5 cm, 0) -- (31.5 cm,0);     
   \draw[thick] (32 cm,0) circle (5 mm) node [shift={(0.0,-0.5)}] {\mfn{9}} node [shift={(0.0,-1)}] {\msc{\red{2}}};
    \draw[thick] (32.5 cm, 0) -- (35.5 cm,0);    
       \draw[thick] (36 cm,0) circle (5 mm) node [shift={(0.0,-0.5)}] {\mfn{10}}node [shift={(0.0,-1)}] {\msc{\red{2}}};
     \draw[thick] (36.5 cm, 0) -- (39.5 cm,0); 
        \draw[thick] (40 cm,0) circle (5 mm) node [shift={(0.0,-0.5)}] {\mfn{11}}node [shift={(0.0,-1)}] {\msc{\red{2}}};
     \draw[thick] (40.5 cm, 0) -- (43.5 cm,0);     
    \draw[thick] (44 cm,0) circle (5 mm) node [shift={(0.0,-0.5)}] {\mfn{12}}node [shift={(0.0,-1)}] {\msc{\red{2}}};
     \draw[thick] (44.5 cm, 0) -- (47.5 cm,0);     
         \draw[thick] (48 cm,0) circle (5 mm) node [shift={(0.0,-0.5)}] {\mfn{13}}node [shift={(0.0,-1)}] {\msc{\red{2}}};
             \draw[thick] (48.5 cm, 0) -- (51.5 cm,0);     
         \draw[thick] (52cm,0) circle (5 mm) node [shift={(0.0,-0.5)}] {\mfn{14}}node [shift={(0.0,-1)}] {\msc{\red{2}}};
                 \draw[thick] (52.5 cm, 0) -- (55.5 cm,0);     
         \draw[thick] (56cm,0) circle (5 mm) node [shift={(0.0,-0.5)}] {\mfn{15}}node [shift={(0.0,-1)}] {\msc{\red{1}}};
                    \draw[thick] (52 cm, 0.5cm) -- (52 cm, 3.5cm);     
 \draw[thick] (52cm,4cm) circle (5 mm) node [shift={(0.5,0.0)}] {\mfn{16}}node [shift={(1,0)}] {\msc{\red{1}}};
 \draw[thick] (52 cm, 4.5cm) -- (52 cm, 7.5cm);     
         \draw[thick] (52cm,8cm) circle (5 mm) node [shift={(0.5,0.0)}] {\mfn{19}}node [shift={(1,0)}] {\msc{\red{}}};
     \end{tikzpicture}
     \end{center}
     \caption{Extended Dynkin diagram for the $\mathrm{II}_{1,17}$ lattice, with labels showing the embedding of $\Gamma_{16}$.
     The Kac marks of the extended $\mathrm{SO}(32)$ diagram are shown in red.}\label{dynkinho}
\end{figure}
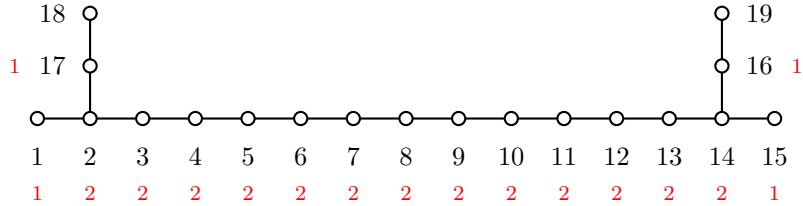

The EDD embedding $\Gamma_{16}$ is shown in Figure \ref{dynkinho}.
The charge vectors of the nodes read
\be
\label{extrar}
\begin{split}
\tilde{\varphi}_k=|0,0; \beta_k\rangle, \quad  k=1,...,16 \, , \qquad &{}
 \tilde{\varphi}_{17}=|0,-1; \beta_{17}\rangle=|0,-1; -1,-1,0^{14}\rangle\, , \\ 
\tilde{\varphi}_{18}=|1,1; 0^{16}\rangle,  \qquad &{} \tilde{\varphi}_{19}=|1,-1; -\tw_{16}\rangle \, .
\end{split}
\ee 
It is straightforward to carry out the analysis of the Weyl reflections \eqref{weylref} to identify the generators
of $\rO(1,17,\ZZ)$ and the boundaries of the fundamental region.
A choice of fundamental region for the moduli space of the HO theory is given in Table \ref{tab:fundregho}
(our conventions for the roots differ by a sign from those in \cite{Fraiman:2018ebo}).

\begin{table}[h!]\begin{center}
\renewcommand{\arraystretch}{2}
{\footnotesize
 \begin{tabular}{|c|c|c|}
\hline 
Node  &  Fundamental region for $\Gamma_{16}$\\
\hline\hline 
$1\le k\le 16$  & $  A\cdot \beta_k \geq 0 $\\
17  & $A \cdot \beta_{17} \geq -1$ \\
18  & $ E \geq 1$\\
19  & $E \geq A\cdot \tw_{16} -1$\\
\hline
\end{tabular}
}
\caption{Fundamental region for HO in $d=1$ } 
\label{tab:fundregho}
\end{center}
\end{table}

As in the HE theory, the procedure to determine the allowed groups $\gr \times \uo^{17-r}$, and the corresponding moduli, 
consists of deleting nodes such that those remaining give the Dynkin diagram of an ADE algebra. 
Obviously the groups will be the same as in the HE,
but the moduli will differ.
They are simply deduced by saturating the inequalities in Table \ref{tab:fundregho} that pertain to the undeleted
nodes. 

From the EDD we can also find the group due to some given moduli but, if necessary, $A$ and $R$ have to
first be brought to the fundamental region by dualities, namely shifts and Weyl reflections in $\Gamma_{16}$, e.g. 
$A^I \to - A^I$ or $A^I \to1-A^I$ in pairs, and the T-duality \eqref{tdual1}.
For instance, in this way $A = ({\frac23}^3, 0^{13}), R = \frac1{\sqrt3}$, can be transformed into
$A = (\frac12^2, 0^{14})=\frac12 \tw_2$, $R = \frac{\sqrt3}2$ . 
From the latter data we find that the nodes 2 and 19 must be deleted so that the gauge group is $\mra_1 + \mra_2 + \mrd_{14}$. 
Similarly, $A = (\frac12^3,0^{13})$, $R = \frac1{2\sqrt2}$, can be brought to $A = (1,0^{15})=\tw_1$, $R = \frac1{\sqrt2}$, which implies gauge group $\mrd_{17}$ because nodes 1 and 19 must be deleted.

\subsubsection{The shift algorithm}
\label{sec:shift}

As we have seen, Wilson lines of a given order are relevant to relate the moduli with lattice data 
obtained in the formalism of section \ref{sec:lattices}. 
Recall that the order of $A$ is defined as the smallest integer $N$ such that $NA \in \Upsilon_{16}$, with 
$\Upsilon_{16}$ equal to $\Gamma_8 \times \Gamma_8$ in the
HE and to $\Gamma_{16}$ in the HO.
There exists an algorithm, based on original work of Kac \cite{Kac}, to find Wilson lines (``shift vectors") of specific order. 
It was applied to heterotic compactifications originally in \cite{Dixon:1986yc}. The name shift vector comes from the orbifold 
terminology.  The algorithm also prescribes how to obtain the group left unbroken by the action of the shift. In fact, 
another motivation to review it is its relation to the method of saturating inequalities of undeleted nodes in a Dynkin diagram 
in order to find the moduli.

The shift algorithm can be applied to any ADE group starting with its extended Dynkin diagram. 
We will describe the $\mre_8$ case following \cite{Hollowood:1987hf}.  
The simple roots $\alpha_i$ and the fundamental weights $w_i$ are given in Table \ref{tab:e8}, while the extended
Dynkin diagram of $\mre_8$ is formed by the nodes $0,1, \ldots, 8$, in Figure \ref{dynkindiag2}.
Consider now a set of non-negative relative prime integers $(s_0, s_1, \ldots, s_8)$ and define
\be
N =\sum_{ i=0}^8\kappa_i s_i\, ,
\label{orderdelta} 
\ee
where $\kappa_i$ are the Kac marks. Then construct the shift vector
\be \delta =\frac1{N} \sum^8_{ i=1}s_iw_i\,  . 
\label{shiftdef}
\ee
Note that $N \, \delta \in \Gamma_{8}$ so that $\delta$ has order $N$. The subalgebra left invariant by this shift 
is obtained by deleting the nodes of the extended Dynkin diagram associated to non-zero $s_i$, and adding $\uo$'s to 
preserve the rank. The reason is that $\delta$ in \eqref{shiftdef} satisfies
\be 
\delta\cdot\alpha_0 = -1 +\frac{s_0}{N}\, , \qquad
\delta\cdot\alpha_j =\frac{s_j}N\, ,  \quad j = 1,\ldots,8 .
\label{dprop}
\ee
Notice also that in order to break to a group of rank 8, necessarily only one $s_k$, $k=0, \ldots,8$, is different from zero 
at a time. In this case, $\delta = w_k/\kappa_k$. In particular, since $w_0\equiv 0$, $k=0$ corresponds to $\delta=0$, consistent 
with deleting node $\alpha_0$ and leaving $\mre_8$ unbroken.
For the $\mre^\prime_8$ factor in the HE theory one constructs a shift $\delta^\prime$ in analogy to $\delta$ for $\mre_8$. 

From \eqref{dprop} one also obtains
\be
\delta\cdot\alpha_0\geq -1\, , \qquad \delta\cdot\alpha_j\geq  0\, .
\label{funddef}
\ee
These are the conditions for $\delta$ to be in a fundamental region \cite{Dixon:1986yc,MP84}. By translations in the
root lattice of $\mre_8$ and/or transformations in the Weyl group of $\mre_8$ one can obtain a shift that
gives the same breaking but is outside the fundamental region. For the shift $\delta^\prime$ of $\mre^\prime_8$
there are conditions analogous to \eqref{funddef}.

The shift algorithm can be extended to the HO theory taking care that $\Gamma_{16}$ is the root lattice of 
$\hog$ with the spinor weight $\tw_{16}$ added \cite{Dixon:1986yc}. The starting point is the extended Dynkin diagram 
of $\hog$ which is formed by the nodes $1$ to $17$ in Figure \ref{dynkinho} where the Dynkin marks are also shown. 
The simple roots $\beta_k$ and the fundamental weights $\tw_k$ are given in \eqref{roots16}, and the lowest root 
$\beta_{17}$ in \eqref{low16}.
We now introduce a set of non-negative relative prime integers $\tilde s_k$, $k=1,\ldots,17$, and define
the order $\widetilde N$ and the shift $\Delta$ as
\be
\widetilde N=\sum_{k=1}^{17} \tilde s_k \kappa_k \, , \qquad
\Delta =\frac1{\widetilde N} \sum_{k=1}^{16} \tilde s_k \tw_i \,  .
\label{aho}
\ee
It is necessary to further enforce the constraint
\be
\sum_{k\,  \text{odd}} \tilde s_k = \text{even} \, 
\ee
in order to guarantee that $\widetilde N \Delta \in \Gamma_{16}$. As before,
the subalgebra left invariant by the shift $\Delta$  is obtained by deleting the nodes of the extended Dynkin diagram 
associated to $\tilde s_k >0$, and adding $\uo$'s to preserve rank 16. 
The algorithm can produce pairs of shifts that are equivalent under a translation by $\tw_{16}$.

Let us now discuss the generalization of the shift algorithm to $\nlc$ in the HE theory. 
As in the saturation method, we begin by deleting some nodes in the EDD of Figure \ref{dynkindiag2} such that the
surviving ones form an allowed Dynkin diagram of a semi-simple Lie Algebra. As before
the emerging group is identified from this allowed Dynkin diagram, appending enough $\uo$
factors to add to rank 17. The Wilson line that produces the emerging group is simply given by
\be
A = \delta\times \delta'\, ,
\label{adelta}
\ee
with $\delta$ given in \eqref{shiftdef}, and similarly for $\delta^\prime$. 
The values of $s_i$ are now fixed to be zero or one according to whether the $i$-th node is undeleted or not, 
and likewise for the $s'_i$. Indeed, the inequalities that would have to be saturated to find $A$ are a subset of those connected to the 
nodes $i,i^\prime=0,1,\ldots,8$, in Table \ref{tab:fundreghe}, which precisely amount to the conditions \eqref{funddef}.
The value of the radius depends on whether the node $\texttt{C}$ is undeleted or not. If it is not, the constraint $E=1$ must be 
imposed. Since $\delta$ and $\delta'$  are in the fundamental region, it is not hard to show that  $A^2 \leq 2$. This
guarantees that $R^2=E-\frac12 A^2$ is positive.

It is useful to work out the case of maximal enhancing with the shift algorithm.
As mentioned before, maximal rank 17 requires deleting one node in the $\mre_8$ side and 
one node in the $\mre^\prime_8$ side, while keeping the central node ${\texttt{C}}$.  The moduli are then
\be
A=\frac{w_k}{\kappa_k}\times \frac{w^\prime_m}{\kappa^\prime_m}\, , \qquad 
E=1 \Rightarrow R^2=1- \tfrac12 A^2  \, . 
\label{wlmax}
\ee 
Here $k,m=0,1,\ldots, 8$, but the choice $k=m=8$ is excluded because it would lead to $A^2=2$ and $R=0$, which is
unphysical. Thus, altogether there are 44 different groups with maximal rank. The moduli in \eqref{wlmax} agree
with the results in Table 2 of \cite{Fraiman:2018ebo}, except for irrelevant overall minus signs in the Wilson line due to 
different conventions. The groups of maximal rank and the corresponding moduli are collected in Table \ref{tab:alld1}.
 
The algorithm can also be used to determine the moduli corresponding to groups of lower rank. 
For example, $\sug(16)\times \sug(2)\times \uo$ can be obtained dropping the nodes $1, 1^\prime, 7^\prime$.
From the algorithm we deduce
\be
A =\tfrac13w_1 \times\tfrac17(w'_1 + w'_7) = (\tfrac16,-\tfrac16^6,\tfrac56)\times (-\tfrac67,\tfrac17^6,\msm{0})\, .
\ee
Since node {\texttt{C}} is undeleted, $E=1$ and the radius is fixed to be $R =\sqrt{\frac8{63}}$.

We will not attempt to generalize the shift algorithm to $\nlc$ with HO embedding.
For one reason, for the HO the allowed groups and the corresponding moduli can be obtained by the saturation method
discussed in section \ref{sss:ho}. In particular, the moduli for the 44 groups of maximal enhancing are presented
in Table 1 in \cite{Fraiman:2018ebo}.
Moreover, we can use the map \eqref{duality1app} to obtain a point $(\RO,\AO)$  in the moduli space of the HO theory from a given one $(\REE,\AEE)$ in the HE theory, or vice versa. 
For all the 44 cases of maximal enhancement we have verified that $(\RO,\AO)$ obtained from the $(\REE, \AEE)$ in 
\eqref{wlmax} agree with the data found using the saturation method \cite{Fraiman:2018ebo}. 
These results are listed in Table \ref{tab:alld1}.

\subsection{All maximal rank groups for $d=1$}
\label{sec:alld1}

As mentioned previously, there are 44 different groups of maximal rank that are realized
in heterotic compactification on $S^1$. We collect them in Table \ref{tab:alld1} in appendix \ref{app:tablesmaximal}, where 
they are denoted by its root lattice $L$. The Table includes the moduli
$(R_{\small{\rm E}}, A_{\small{\rm E}})$ and $(R_{\small{\rm O}},A_{\small{\rm O}})$ in the HE and HO theories 
respectively. For both the moduli lie in the fundamental regions defined in Tables \ref{tab:fundreghe} and \ref{tab:fundregho}.
They can be obtained using the saturation method, or equivalently the shift
algorithm in the HE. The moduli for the HO can be derived from the map \eqref{duality1app} too.
In all cases $E_{\small{\rm E}}=E_{\small{\rm O}}=1$.

For each maximal group in Table \ref{tab:alld1} we also give its discriminant group $A_L=L^*/L$, its 
appropriate isotropic subgroup $H_L$,  and its complementary lattice $T$. 
For the lattice $T$, the notation $\mra_1\langle m \rangle$ is simplified
to $\langle m \rangle$. Besides, $d(T)=2 m$. It is easy to check that in all cases $d(L)=d(T)|H_L|^2$ holds.
For all groups we have verified the isomorphism $(A_M, q_M) \cong (A_T, q_T)$, which is less trivial when 
$H_L \not=\mathbb{1}$. Some examples were worked out in section \ref{ss:d1}.

It is a compelling exercise to deduce the lattice $T$ from the moduli as explained in section \ref{ss:nikhet}.
For $d=1$ there is only one Wilson line and the simple result \eqref{naivef} is valid. Thus, $T$ is generated by 
\be
u=\sqrt 2 N R \, ,
\label{fbasis1}
\ee
where $N$ is the order of $A$ and we used $e_1=R$. The Gram matrix is then $Q=2 N^2 R^2 = d(T)$.
On the lattice side, $T=\mra_1\langle m \rangle$ with $d(T)=2m$.
Therefore, it must be that
\be
2 N^2 R^2 = 2 N^2 (1 - \tfrac12 A^2) = 2m \, ,
\label{checkT1}
\ee
where we used $E=1$ in all cases of maximal enhancing. It is straightforward to confirm this relation
using the data for $m$ and $A$ in Table \ref{tab:alld1}.
In the HE case the Wilson line $\AEE$ is given in \eqref{wlmax} and the order is 
\be
\NEE= \frac{\kappa_k \op \kappa^\prime_m}{\text{gcd}(\kappa_k, \kappa^\prime_m)} \, .
\label{orderAE}
\ee
In the HO, $\AO$ and its order $\NO$ are of the form in \eqref{aho}.

Another interesting question is the relation of generic $p_R$ to the complementary lattice $T$. 
In section \ref{ss:nikhet} we argued that in general $p_R$ takes values in $T^*$. When $d=1$ the
proof is rather simple.
Since $E=1$, \eqref{momc} reduces to
\be
p_R =\frac{1}{\sqrt2 \op R} (n - w - \pi \cdot A) \, .
\label{pr1}
\ee
We now use that $A$ has order $N$ to set $\pi \cdot A = \tilde l/N$, $\tilde l \in \ZZ$. Inserting in $p_R$ above gives
$p_R=\frac{l}{\sqrt 2 N R}$, with $l$ integer. Hence, $p_R$ lies in a lattice generated by $u^*$, with
$u$ the generator of $T$ in \eqref{fbasis1}. We conclude that $p_R$ lies on $T^*$ and the allowed values of $p_R^2$
are  $q_T\!\!\!\mod 2$.

\section{Compactifications on $T^2$}
\label{sec:d2}

In heterotic compactification on $T^2$ there are 36 real moduli, namely $\{g_{11}, g_{12}, g_{22}, b_{12}\}$, plus two
16-dimensional Wilson lines $\{A_1^I, A_2^I\}$.  
The $\mathrm{II}_{2,18}$ lattice vectors $(p_R; p_L, p^I)$, which depend on these moduli, are given in \eqref{momenta}.
For the purpose of studying enhancement of symmetries it is actually more appropriate to use  as moduli the components $E_{ij}$, cf. \eqref{genmetric}, together with the $A_i^I$. Indeed,
as we have seen in section \ref{ss:nikhet}, enhancement requires the $E_{ij}$ to be rational numbers and the
$A_i$ to be quantized in the sense of eq.~\eqref{orderA}.
On the other hand, to discuss the moduli space and duality symmetries it is also convenient to work with complex parameters. 
In  section \ref{sec:complex}, we  introduce the complex moduli and their duality transformations, and 
review the action of $\rO(2,3;\mathbb Z)$, a subgroup of the duality group, on a particular slice of the moduli space.
Then we turn to the problem of determining all gauge groups \mbox{$\gr\times \uo^{18-r}$} that can appear, and the
corresponding moduli. 

The extension of the  systematic procedure discussed in the previous section to compactifications on $T^2$ would
require the construction of a generalized Dynkin diagram for $\mathrm{II}_{2,18}$. However, it has been argued that the even, 
self-dual lattices of signature $ (p,q)$ with both $p,q > 1$ (that is, with a signature with more that one negative sign), do not 
possess a system of simple roots and  cannot be described in terms of generators and relations similar to Kac-Moody or 
Borcherds algebras \cite{axel}. Nevertheless, although  the addition of a new Kac-Moody simple root introduces 
multiple links and loops in the structure of the quadruple extension of simple Lie algebras,  it was shown in \cite{forte} 
that the ``simple-links''  structure can be preserved if the extra root is a Borcherds (imaginary) simple root. 
In any case, a generalized Dynkin diagram for $\mathrm{II}_{2,18}$  is not known and 
it is not even clear whether it exists. Hence, we will proceed in a constructive way.

In section \ref{sec:lattices} we explained that all allowed groups \mbox{$\gr\times \uo^{d+16-r}$} in 
heterotic compactification on $T^d$ can be obtained by lattice embedding techniques. For $T^2$ the full results
are known from the work of Shimada and Zhang who classified all possible ADE types of singular fibers
in elliptic K3 surfaces \cite{SZ, ShimadaK3}. The classification translates into all possible heterotic 
gauge groups because the lattice embedding conditions are the same in
the K3 and heterotic contexts. This can also be seen as a further element in favor of the conjectured 
duality between heterotic on $T^2$ and F-theory on K3.

Knowing all allowed groups it remains to compute the corresponding moduli. We will focus in the HE since the
moduli in the HO can be derived from the map elaborated in section \ref{ss:hehomap}.
We will mostly consider the case of maximal enhancing, i.e. $r=18$. 
As argued in section \ref{ss:nikhet}, this can occur only if the $E_{ij}$ are rational numbers and the $A_i$ are quantized.
In section \ref{sec:shift} we explained a shift algorithm to find such Wilson lines. 
In particular, in the HE we can find all pairs of quantized Wilson lines that break $\mre_8 \times \mre'_8$ to a subgroup
of rank 16, hence with a Dynkin diagram having 16 nodes. We can then look for
values of the $E_{ij}$ that allow to add two additional nodes, thereby leading to a semisimple group of rank 18. 
This is analogous to the procedure of finding all maximal enhancements from the EDD in the circle compactification.

In section \ref{sec:shiftd2} we will explain the EDD inspired method in more detail. We will see that 
it fails to give several of the known groups of maximal rank. In section \ref{sec:explore} we will then develop 
more general procedures in order to obtain all such groups. The results are summarized in 
section \ref{ss:alld2}.

\subsection{Complex moduli}
\label{sec:complex}

Without Wilson lines we know that it is revealing to combine the parameters from the metric and
the antisymmetric field into complex structure and K\"ahler moduli, denoted $\tau$ and $\rho$ respectively.
In particular, the duality transformations and the fundamental moduli region can be described very efficiently
in terms of $\tau$ and $\rho$.  It is then reasonable to use these complex parameters in the presence 
of the $A^I_i$, which in turn can be combined into complex moduli $\xi^I$ as well. Altogether we have the 18 complex moduli
\be
\begin{split}
\tau&=\frac{g_{12}}{g_{11}}+i\frac{\sqrt g}{g_{11}}\, , \qquad 
\rho=b_{12}+i\sqrt g+\frac 12 A_1^I\zeta^I\, ,  \\
 \zeta^I&=A_1^I\tau-A_2^I \ ,
 \end{split}
\label{cplx}
\ee
where $g = \det g_{ij}$. 
The conditions $g_{ii} > 0$ and $g > 0$ imply the restrictions
\be
\tau_2 > 0, \quad \rho_2 > 0, \quad  \tau_2\rho_2 -\tfrac12 \zeta_2^2 > 0 \, ,
\label{cplxregion}
\ee
where the subscript $2$ refers to the imaginary parts.
The moduli $(\tau,\rho,\zeta^I)$ were considered in \cite{Kiritsis:1997hf}, see also 
\cite{LopesCardoso, DaflonBarrozo:1999vk}.
As expected, the K\"ahler modulus, which is more stringy, receives corrections depending on the Wilson lines
whereas $\tau$, purely geometrical, is not affected.

The $\mathrm{II}_{2,18}$ lattice vectors $(p_R; p_L, p^I)$ can also be written
in terms of the complex moduli. Now, we are mostly interested in the duality transformations of the moduli which 
can be derived from invariance of the spectrum. By virtue of \eqref{evencond} it then suffices to determine $p_R^2$.
We obtain 
\be
p_R^2 = \frac{1}{2\big(\rho_2 \tau_2 - \frac{1}{2}\zeta^2_2\big)}
\big|n_2 - \tau n_1 + \rho w^1 + (\rho\tau - \tfrac{1}{2}\zeta^2)w^2 + \pi \cdot \zeta\big|^2\, .
\label{pRmod}	
\ee
Imposing invariance of $p^2_R$ and $(p_L+p^I)^2-p_R^2=\pi\cdot \pi + 2 n_i w^i$ we deduce the duality transformations
\be
\begin{split}
\mathcal{Z}_1 &:\quad \tau' = \rho, \quad \rho' = \tau, \quad \zeta' = \zeta, \\[2mm]
\mathcal{Z}_2 &: \quad \tau' = - \bar  \tau, \quad \rho' = - \bar \rho, \quad \zeta' = \bar \zeta, \\[2mm]
\mathcal{A}_1 &: \quad \tau' = \tau+1, \quad \rho' = \rho, \quad \zeta' = \zeta, \\[2mm]
\mathcal{S}_1 &: \quad \tau' =-\frac{1}{\tau}, 
\quad \rho' =\rho -  \tfrac12 \frac{\zeta^2}{\tau}, \quad \zeta' = \frac{\zeta}{\tau},   \\[2mm]
\Gamma_1 &: \quad \tau' =\tau, 
\quad \rho' =\rho +  \zeta \cdot \Lambda + \tfrac12 \Lambda^2 \tau,
\quad \zeta' = \zeta + \Lambda \tau, \quad \Lambda \in \Upsilon_{16} ,
\end{split}
\label{dualt2}
\ee
where we have dropped the superscript $I$ in $\zeta$ to simplify the expressions. 
These transformations were also found in \cite{Kiritsis:1997hf}. 

Together with Weyl automorphisms in $\Upsilon_{16}$,  $\{\mathcal{Z}_1, \mathcal{Z}_2, \mathcal{A}_1,
\mathcal{S}_1, \Gamma_1\}$ generate the duality group $\rO(2,18,{\mathbb Z})$.
We recognize $\mathcal{A}_1$ and $\mathcal{S}_1$ as the generators of $\mathrm{SL}(2,\ZZ)$ changes of the $(e_1, e_2)$ basis,
whereas $\mathcal{Z}_2$ is the parity $e_1 \to -e_1$.
The transformation $\Gamma_1$ is the translation of $A_1^I$ by the lattice vector $\Lambda$. 
The shift $b_{12} \to b_{12} + 1$, implying $\rho \to \rho + 1$,
is just $\mathcal{Z}_1 \mathcal{A}_1 \mathcal{Z}_1$. 
The composition $\mathcal{S}_1 \mathcal{Z}_1 \mathcal{S}_1 \mathcal{Z}_1$
gives the full T-duality (i.e. in directions $e_1$ and $e_2$), generalizing $R \to 1/R$, with action
\be
\mathcal D : \quad \tau^\prime = -\frac{\rho}{\rho\tau - \frac12\zeta^2} \, , 
\qquad \rho^\prime = -\frac{\tau}{\rho\tau - \frac12\zeta^2}\, , 
\qquad \zeta^\prime = \frac{\zeta}{\rho\tau - \frac12\zeta^2} \, .
\label{daction}
\ee
The factorized duality in the direction $e_1$ of $T^2$ is $\mathcal{Z}_2 \mathcal{Z}_1$, while $\mathcal{Z}_1$ is
`mirror symmetry'. The $\Upsilon_{16}$ automorphisms include the transformation $\tau'=\tau$, $\rho'=\rho$,
$\zeta'=-\zeta$, which amounts to $A_i'=-A_i$.

The moduli $E_{ij}$ are related to $(\tau, \rho, \zeta)$ by
\be
E_{11} \tau - E_{21} = \rho, \qquad E_{12} \tau - E_{22} = \tau\rho - \tfrac12 \zeta^2 \equiv \xi \, .
\label{Ecplx}
\ee
The duality transformations of $E_{ij}$ and $A_i$ can be efficiently derived as explained in section \ref{ss:duality}.
For instance, the factorized duality in the direction $e_1$, i.e. $\mathcal{Z}_2 \mathcal{Z}_1$, is given in \eqref{fac1d2}.
Analogously, the factorized duality in the direction $e_2$, i.e.  $\mathcal{S}_1 \mathcal{Z}_1 \mathcal{S}_1  \mathcal{Z}_2$,
amounts to
\be\label{fac2d2}
E'=\frac1{E_{22}}\left(\begin{matrix}\det E &E_{12}\\
-E_{21}&1 \end{matrix}\right)\, , \quad A'_1= A_1-\frac{E_{12}}{E_{22}}A_2\, , \quad 
A'_2=-\frac{A_2}{E_{22}}\, .
\ee
The product of the two factorized dualities yields
\be\label{dactionEA}
E'=E^{-1}, \quad \left(\!\begin{matrix} A'_1 \\ A'_2 \end{matrix} \right)= 
- E^{-1} \left(\!\begin{matrix} A_1 \\ A_2 \end{matrix} \!\right) \, ,
\ee
which corresponds to the transformation in \eqref{daction}.

It is instructive to consider a particular slice of moduli space defined by restricting the Wilson lines to break
an $\sug(2)$ in $\mre_8$. This can be achieved taking $A_i=a_i w_6 \times 0$, so that 
\be
\zeta=\beta w_6 \times 0, \quad \beta=a_1 \tau - a_2\, . 
\label{betadef}
\ee
There are then three complex parameters $(\tau,\rho,\beta)$. The duality
group acting on them reduces to $\rO(2,3,\ZZ)$, whose generators are given in \eqref{dualt2}, with 
$\Lambda=w_6 \times 0$ in $\Gamma_1$.
It is known that $\rO(2,3,\ZZ)$ has a subgroup which can be identified with $\mathrm{Sp}(4,\ZZ)$, see e.g. \cite{Malmendier:2014uka}.
A minimal set of generators is provided by $\{\mathcal{Z}_1, \mathcal{A}_1, \mathcal{S}_1, \Gamma_1\}$. 
The standard Dehn twists  (shown e.g. in \cite{Font:2016odl}) can be expressed in terms of the elements of this set.
In fact, there is an isomorphism from the moduli space of $(\tau,\rho,\beta)$ to the genus-two Siegel upper half-plane
parametrized by $\Omega=\left(\begin{smallmatrix} \tau & \beta \\ \beta & \rho \end{smallmatrix}\right)$, 
see \cite {Font:2016odl} and references therein.
Thus, $(\tau,\rho,\beta)$ can be regarded as the moduli of a genus-two surface.
Several useful results about the moduli space of genus-two curves are known. 
In particular, the fundamental region and fixed points of finite subgroups have been determined \cite{Gottschling1, Gottschling2, Gottschling3}.
Some special duality transformations, needed for future purposes, are
\be
\Omega' = 
\begin{pmatrix}
\rho & \rho-\beta \\
\rho-\beta & \tau +\rho - 2\beta
\end{pmatrix},
\quad
\Omega' = -\frac{1}{\xi  + \rho}
\begin{pmatrix}
\rho & \xi-\beta \\
\xi-\beta & 2\rho +\tau -2\beta+ \xi+1  
\end{pmatrix}\, ,
\label{octa}
\ee
where $\xi=\rho\tau - \beta^2$.
At generic values of the moduli the gauge group is $\uo^3 \times \mre_7 \times \mre'_8$, but at the fixed points
the $\uo^3$ can enhance for instance to $\sug(2)\times \sug(3)$ or $\sug(4)$ \cite{afcern}. More details will be
given in section \ref{ss:alld2}. 
This slice of heterotic moduli space is specially interesting because an explicit map to the moduli of
elliptic K3 surfaces with $\mre_7$ and $\mre_8$ singularities was established recently \cite{Malmendier:2014uka},
see also \cite {Font:2016odl} and references therein.

\subsection{Generalizing the EDD algorithm to two Wilson lines}
\label{sec:shiftd2}

The EDD algorithm in circle compactifications uses the fact that  the T-duality group $\rO(1,17,\mathbb{Z})$ is completely generated by simple reflections. This ceases to be true for $d > 1$ and so it cannot be generalized with its full power. What we can do, instead, is to develop a more general method to find maximal groups and their associated moduli which works for all $d$, and reduces to the EDD algorithm in $d = 1$. 

The key idea is that the EDD algorithm in $d=1$ can be stated in an equivalent but qualitatively different way. Instead of breaking two nodes of the 19-node generalized diagram, we do a step by step procedure: we first break $\mre_8 \times \mre'_8$ to a maximal subgroup with a Wilson line given by the shift algorithm, and then enhance this subgroup by adding the node 
$\tC$ which corresponds to a massless state only when $E=1$.  
The completeness of this algorithm relies on the fact that there is a finite number of ways of breaking 
$\mre_8 \times \mre'_8$ because a fundamental region for a single Wilson line is known, and then the choice of $E$ 
which enhances the resulting group for the given Wilson line is unique. 

In higher dimensions we lack a complete description of the fundamental domain.
Wilson lines $A_i$, $i=1,\ldots,d$,  can be turned on and there are more possibilities to break 
$\mre_8 \times \mre'_8$ to a maximal subgroup, but in general all $A_i$ cannot be brought simultaneously into a fundamental
region  of $\mre_8 \times \mre'_8$. Besides, the options for the moduli $E_{ij}$ are less constrained.
Nonetheless, we will describe how a systematic choice of the $A_i$, and the $E_{ij}$, leads to a class of extended Dynkin
diagrams with $d+18$ nodes such that by deleting nodes in an appropriate way allows to read off the gauge group and the
corresponding moduli.

Before outlining the procedure, let us remark that the generalization of the EDD algorithm does not capture all the maximal enhancements. As we discuss in more detail in section \ref{sec:explore}, there exist maximal rank $d+16$ groups that cannot be obtained by enhancing a rank 16  subgroup of $\mre_8 \times \mre_8$, for example $\sug(7)^3$ in $d=2$.

\subsubsection{Reformulating and generalizing the algorithm}
\label{sss:gdd}

In section \ref{sec:shift} we explained how the shift algorithm can be used to find a Wilson line in the fundamental
region of $\mre_8 \times \mre'_8$ and which breaks to a maximal subgroup. Writing the Wilson line as 
$A=\delta \times \delta'$, we obtained
\be
\label{shiftmax}
\delta=\frac{w_k}{\kappa_k} \, , \qquad \delta^\prime=\frac{w'_m}{\kappa'_m} \, ,
\ee
for $k$ and $m$ taking fixed values in $0, \ldots, 8$, but with $k=m=8$ excluded.
This choice in the circle compactification then implies that in the basis \eqref{basishe} 
the nodes $\varphi_i$, $\varphi'_j$, $i,j=0,\ldots,8$, with $i\not=k$ and $j\not=m$, 
correspond to massless states which satisfy the conditions $n=E w + \pi\cdot A$ and
$\pi^2 + 2 n w=2$. These conditions are also satisfied by 
node $\varphi_\tC$ provided $E=1$, while it is not satisfied by the nodes $\varphi_k$ and $\varphi'_m$ which are deleted.
Notice that the node $\varphi_{\texttt{C}}$  gives the extension to a group of rank 17 and
that actually $\varphi_{\texttt{C}} \in \text{II}_{1,1}$. 

These observations motivate a similar procedure for the $T^2$ compactification. The nodes in the generalized diagram
now have charge vectors $\ket{w^1, w^2, n_1, n_2; \pi}$. As before it is convenient to introduce nodes associated to the simple
roots of $\mre_8 \times \mre'_8$, namely
\be
\varphi_i=\ket{0,0,0,0; \alpha_i, 0^8}\, , \quad \varphi'_i=\ket{0,0,0,0; 0^8,\alpha'_i}\, \quad i=1,\ldots,8 \ .
\label{simple2}
\ee
They will correspond to roots of the resulting gauge group whenever
they satisfy the massless conditions ${\bf p_R}=0$ and ${\bf p_L}^2=2$, leading in turn to \eqref{pR0} and \eqref{pL2}.
Explicitly,
\begin{subequations}\label{massless2}
\begin{align}
&{} n_1=E_{11} w^1 + E_{12} w^2 + \pi \cdot A_1,  \quad n_2 =E_{21} w^1 + E_{22} w^2 + \pi \cdot A_2 \, ,
\label{quant2}\\
&{} \hspace*{3cm} \pi^2 + 2 w^1 n_1 + 2 w^2 n_2 =2
\, . \label{norm2}
\end{align}
\end{subequations}
To proceed we need to specify the moduli. 

The Wilson lines are conveniently written as
\be \label{twoWilson}
A_1=\delta_1 \times \delta_1' \, , \qquad A_2= \delta_2 \times \delta'_2 \, .
\ee
We are interested in the case in which the two Wilson lines together break $\mre_8 \times \mre'_8$ to a subgroup 
of rank 16 for generic $E_{ij}$. To achieve this we first take $\delta_1$ and $\delta'_1$ exactly as in \eqref{shiftmax}. 
Thus, the subgroup left invariant by $A_1$, denoted  $\mrh_k \times \mrh_m'$, is found deleting the nodes 
corresponding to the roots $\alpha_k$ and $\alpha'_m$ in the extended Dynkin diagram of $\mre_8 \times \mre'_8$. 
For $A_2$ we basically use the shift algorithm applied to $\mrh_k \times \mrh'_m$. 
To this end we first append two affine roots $\hat\alpha_k$ and $\hat\alpha'_m$ of the 
subgroup $\mrh_k \times \mrh'_m$ and delete two additional nodes, say those corresponding to the
roots $\alpha_p$, $p\not=k$,  and $\alpha'_q$, $q\not=m$,  of $\mre_8 \times \mre'_8$, which are also 
roots of the subgroup. The new affine roots are given by the lowest roots of one of the factors
in $\mrh_k$ and  $\mrh_m'$ respectively.  The precise way will be explained shortly.

The combined effect of $A_1$ and $A_2$ is to leave a subgroup of $\mre_8 \times \mre'_8$ unbroken.
The simple roots that survive are $\alpha_i$, $i\not=k,p$, $\alpha'_j$, $j\not=m,q$, $i,j=0,\ldots,8$, plus
$\hat\alpha_k$ and $\hat\alpha'_m$. For $A_1$ we have $\delta_1=w_k/\kappa_k$, 
$\delta'_1=w'_m/\kappa'_m$, and by construction 
\be
\delta_1 \cdot \hat\alpha_k=0 \, \quad \delta'_1 \cdot \hat\alpha'_m=0 \, .
\ee
For $A_2$ the shift algorithm dictates that
\be
\label{shiftconstraints}
\delta_2 \cdot \alpha_i = 0, \ i\not=k,p, \ \ \delta_2 \cdot \hat\alpha_k=-1\, ; \quad
\delta'_2 \cdot \alpha'_j = 0, \ j\not=m,q, \ \ \delta'_2 \cdot \hat\alpha'_m =-1
\, ; \quad i,j=0,\ldots,8
\, .
\ee
Here we are assuming that $\delta_2 \not=0$ and $\delta'_2\not=0$. 
If $\delta_2=0$, then $\hat\alpha_k$ is not appended and $\alpha_p$ is not deleted.
Likewise, if $\delta'_2=0$, $\hat\alpha'_m$ is absent and $\alpha'_q$ remains.

The advantage of choosing $A_1$ and $A_2$ as just described is that we can now
construct extended nodes that satisfy the massless conditions in \eqref{massless2}.
Indeed, to the original affine roots of $\mre_8 \times \mre'_8$ we associate  
two extended nodes with momentum number in the direction 1 
\be
\varphi_{0}=|0,0,-1,0;\alpha_0, 0^8\rangle, \quad \
\varphi'_{0}=|0,0,-1,0; 0^8, \alpha'_0\rangle \, .
\label{zero2}
\ee
Actually, when $k=0$, so $\delta_1=0$, and/or $m=0$, so $\delta'_1=0$,
$\varphi_0$ and/or $\varphi'_0$ do not verify \eqref{massless2}, but in these
cases they are meant to be deleted.
The new affine roots of the subgroup $\mrh_k \times \mrh'_m$ lead instead 
to two different extended roots with momentum number in the direction 2
\be
\varphi_{-1}=\ket{0,0,0,-1; \hat\alpha_k, 0^8}, \quad \
\varphi'_{-1}=\ket{0,0,0,-1; 0^8,  \hat\alpha'_m} \ .
\label{minus2}
\ee
In section \ref{sss:nontrivialA2} we will explain in more detail how $\hat\alpha_k$ and $\hat\alpha'_m$ are
determined.

To continue with the analogy with the EDD of $\rO(1,17,\ZZ)$ we still have to 
add two nodes corresponding to $\nltt$. For this purpose we need to 
make a choice of tensor $E_{ij}$ such that these extra roots do correspond to massless states. 
In this section we allow for two possibilities only, which cover most of the enhancement groups. 
Other possibilities are explored in the next sections. The two choices are
\begin{equation}
E_1 = \begin{pmatrix}1 & 0 \\ 0 & 1\end{pmatrix}, ~~~~~ 
E_2 = \begin{pmatrix}1 & -1 \\ 0 & \phantom{-}1\end{pmatrix}. \label{E_standard}
\end{equation}
For $E = E_1$ the following charge vectors satisfy the massless conditions \eqref{massless2}
\begin{equation} \label{nodesc1c2}
\varphi_{{\texttt{C}_1}} = \ket{1,0,1,0;0^8,0^8},~~~~~ \varphi_{{\texttt{C}_2}} = \ket{0,1,0,1;0^8,0^8} \ . 
\end{equation}
Since they are orthogonal, they are not connected to one another in the Dynkin diagram. 
Notice that $\varphi_{{\texttt{C}_1}}$ corresponds to $\varphi_{\texttt{C}}$ in the $S^1$ compactification.
On the other hand, setting $E = E_2$ gives the vectors
\begin{equation}\label{nodesc1c3}
\varphi_{{\texttt{C}_1}} = \ket{1,0,1,0;0^8,0^8},~~~~~ \varphi_{{\texttt{C}_3}} = \ket{0,1,-1,1;0^8,0^8},
\end{equation} 
which enter our extended diagram as an $\mra_2$ subdiagram joined to ${\varphi_0}$ and ${\varphi_0}'$.
We finally have to delete two nodes\footnote{We can also construct models with partial enhancement by deleting more nodes, but this is not the main focus of the present work.}. 

Before giving some examples, let us point out once again that this algorithm does not give all the possible enhancements. 
As we explain in more detail later, further generalizations that do not involve extended diagrams are required to get all the 
possibilities, as explored in section \ref{sec:explore}.

\subsubsection{Extended diagrams with trivial second breaking}
\label{subsec:eddtriv}

We now give some examples, starting from the simplest. For the sake of clarity,
we will use a color coding for the nodes which partly or completely lie in the $\text{II}_{2,2}$ sublattice. 
We will paint with green the roots $\varphi_0$, $\varphi_0'$ and $\varphi_{{\texttt{C}_1}}$, 
and with blue the roots $\varphi_{-1}$, $\varphi_{-1}'$, $\varphi_{{\texttt{C}_2}}$ or $\varphi_{{\texttt{C}_3}}$. 
This will help us keep track of the extensions of the diagram and how they relate to the Wilson lines. 

The simplest example of an extended diagram in $d = 2$ compactifications is obtained by taking our second breaking to be trivial, namely taking $A_2 = 0$. For this choice, our task is easier because there is no need at all to apply the conditions
\eqref{shiftconstraints}. In practice we just have to supplement the EDD for $S^1$ with
the node $\varphi_{{\texttt{C}_2}}$ or $\varphi_{{\texttt{C}_3}}$. Concretely, 
taking $E=E_1$, we get the extended diagram shown in Figure \ref{edd2:1}, where to obtain the rank 18 maximal groups we have to delete two nodes. With this we can obtain all the groups of the form $\text{G}_{17}\times \mra_1$, where $\text{G}_{17}$ is one of the 44 maximally enhanced groups in $S^1$ compactifications. 
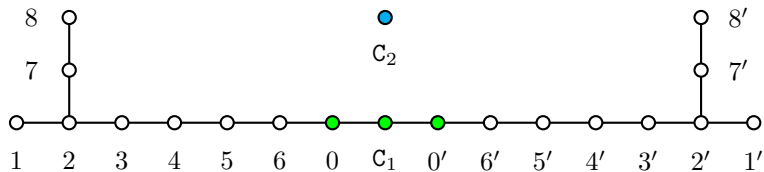
\begin{figure}[htb]
	\begin{center}
		\begin{tikzpicture}[scale=.175]
		\draw[thick] (0 cm,0) circle (5 mm) node [shift={(0.0,-0.5)}] {\mfn{1}} ;
		\draw[thick] (.5 cm, 0) -- (3.5 cm,0);    
		\draw[thick] (4 cm,0) circle (5 mm) node [shift={(0.0,-0.5)}] {\mfn{2}};
		\draw[thick] (4.5 cm, 0) -- (7.5 cm,0); 
		\draw[thick] (8 cm,0) circle (5 mm) node [shift={(0.0,-0.5)}] {\mfn{3}};
		\draw[thick] (8.5 cm, 0) -- (11.5 cm,0);     
		\draw[thick] (12 cm,0) circle (5 mm) node [shift={(0.0,-0.5)}] {\mfn{4}};
		\draw[thick] (12.5 cm, 0) -- (15.5 cm,0);     
		\draw[thick] (16 cm,0) circle (5 mm) node [shift={(0.0,-0.5)}] {\mfn{5}};
		\draw[thick] (16.5 cm, 0) -- (19.5 cm,0);     
		\draw[thick] (20cm,0) circle (5 mm) node [shift={(0.0,-0.5)}] {\mfn{6}};
		\draw[thick] (4 cm, 0.5cm) -- (4 cm, 3.5cm);     
		\draw[thick] (4cm,4cm) circle (5 mm) node [shift={(-0.5,0.0)}] {\mfn{7}};
		\draw[thick] (4 cm, 4.5cm) -- (4 cm, 7.5cm);     
		\draw[thick] (4cm,8cm) circle (5 mm) node [shift={(-0.5,0.0)}] {\mfn{8}};
		\draw[thick] (20.5 cm, 0) -- (23.5 cm,0);     
		\draw[thick, fill = green] (24cm,0) circle (5 mm) node [shift={(0.0,-0.5)}] {\mfn{0}};
		\draw[thick] (24.5 cm, 0) -- (27.5 cm,0);     
		\draw[thick, fill=green] (28cm,0) circle (5 mm) node [shift={(0.0,-0.5)}] {\footnotesize{$\texttt{C}_1$}};
		\draw[thick, fill=cyan] (28cm,8cm) circle (5 mm) node [shift={(0.0,-0.5)}] {\footnotesize{$\texttt{C}_2$}};
		\draw[thick] (28.5 cm, 0) -- (31.5 cm,0);     
		\draw[thick, fill = green] (32 cm,0) circle (5 mm) node [shift={(0.0,-0.5)}] {\mfn{0^\prime}} ;
		\draw[thick] (32.5 cm, 0) -- (35.5 cm,0);    
		\draw[thick] (36 cm,0) circle (5 mm) node [shift={(0.0,-0.5)}] {\mfn{6^\prime}};
		\draw[thick] (36.5 cm, 0) -- (39.5 cm,0); 
		\draw[thick] (40 cm,0) circle (5 mm) node [shift={(0.0,-0.5)}] {\mfn{5^\prime}};
		\draw[thick] (40.5 cm, 0) -- (43.5 cm,0);     
		\draw[thick] (44 cm,0) circle (5 mm) node [shift={(0.0,-0.5)}] {\mfn{4^\prime}};
		\draw[thick] (44.5 cm, 0) -- (47.5 cm,0);     
		\draw[thick] (48 cm,0) circle (5 mm) node [shift={(0.0,-0.5)}] {\mfn{3^\prime}};
		\draw[thick] (48.5 cm, 0) -- (51.5 cm,0);     
		\draw[thick] (52cm,0) circle (5 mm) node [shift={(0.0,-0.5)}] {\mfn{2^\prime}};
		\draw[thick] (52.5 cm, 0) -- (55.5 cm,0);     
		\draw[thick] (56cm,0) circle (5 mm) node [shift={(0.0,-0.5)}] {\mfn{1^\prime}};
		\draw[thick] (52 cm, 0.5cm) -- (52 cm, 3.5cm);     
		\draw[thick] (52cm,4cm) circle (5 mm) node [shift={(0.5,0.0)}] {\mfn{7^\prime}};
		\draw[thick] (52 cm, 4.5cm) -- (52 cm, 7.5cm);     
		\draw[thick] (52cm,8cm) circle (5 mm) node [shift={(0.5,0.0)}] {\mfn{8^\prime}};
		\end{tikzpicture}
	\end{center}
	\caption{Simplest extended diagram for $T^2$ compactifications, reproducing the 44 maximal enhancements of $S^1$ compactifications times $\mra_1$. All models have $A_2 = 0$ and $E = E_1$.}\label{edd2:1}
\end{figure}

If we take instead $E = E_2$, we get the diagram shown in Figure \ref{edd2:2}. With this simple construction we are now able to get 
non-trivial enhancements by deleting two nodes such that the resulting diagram is ADE. 
For example, by deleting nodes $5$ and $5'$ we get the group $\mre_6^3$, with moduli 
$A_1 = \tfrac{1}{3}w_5 \times \tfrac{1}{3}w_5'$, $A_2 = 0$, $E = E_2$.

\begin{figure}[htb]
	\begin{center}
		\begin{tikzpicture}[scale=.175]
		\draw[thick] (0 cm,0) circle (5 mm) node [shift={(0.0,-0.5)}] {\mfn{1}} ;
		\draw[thick] (.5 cm, 0) -- (3.5 cm,0);    
		\draw[thick] (4 cm,0) circle (5 mm) node [shift={(0.0,-0.5)}] {\mfn{2}};
		\draw[thick] (4.5 cm, 0) -- (7.5 cm,0); 
		\draw[thick] (8 cm,0) circle (5 mm) node [shift={(0.0,-0.5)}] {\mfn{3}};
		\draw[thick] (8.5 cm, 0) -- (11.5 cm,0);     
		\draw[thick] (12 cm,0) circle (5 mm) node [shift={(0.0,-0.5)}] {\mfn{4}};
		\draw[thick] (12.5 cm, 0) -- (15.5 cm,0);     
		\draw[thick] (16 cm,0) circle (5 mm) node [shift={(0.0,-0.5)}] {\mfn{5}};
		\draw[thick] (16.5 cm, 0) -- (19.5 cm,0);     
		\draw[thick] (20cm,0) circle (5 mm) node [shift={(0.0,-0.5)}] {\mfn{6}};
		\draw[thick] (4 cm, 0.5cm) -- (4 cm, 3.5cm);     
		\draw[thick] (4cm,4cm) circle (5 mm) node [shift={(-0.5,0.0)}] {\mfn{7}};
		\draw[thick] (4 cm, 4.5cm) -- (4 cm, 7.5cm);     
		\draw[thick] (4cm,8cm) circle (5 mm) node [shift={(-0.5,0.0)}] {\mfn{8}};
		\draw[thick] (20.5 cm, 0) -- (23.5 cm,0);     
		\draw[thick, fill = green] (24cm,0) circle (5 mm) node [shift={(0.0,-0.5)}] {\mfn{0}};
		\draw[thick] (24.5 cm, 0) -- (27.5 cm,0);     
		\draw[thick, fill=green] (28cm,0) circle (5 mm) node [shift={(0.0,-0.5)}] {\footnotesize{$\texttt{C}_1$}};
		\draw[thick, fill=cyan] (28cm,4cm) circle (5 mm) node [shift={(0.0,0.5)}] {\footnotesize{$\texttt{C}_3$}};
		\draw[thick] (28 cm, 0.5 cm) -- (28 cm, 3.5 cm);
		\draw[thick] (28.5 cm, 0) -- (31.5 cm,0);     
		\draw[thick, fill = green] (32 cm,0) circle (5 mm) node [shift={(0.0,-0.5)}] {\mfn{0^\prime}} ;
		\draw[thick] (32.5 cm, 0) -- (35.5 cm,0);    
		\draw[thick] (36 cm,0) circle (5 mm) node [shift={(0.0,-0.5)}] {\mfn{6^\prime}};
		\draw[thick] (36.5 cm, 0) -- (39.5 cm,0); 
		\draw[thick] (40 cm,0) circle (5 mm) node [shift={(0.0,-0.5)}] {\mfn{5^\prime}};
		\draw[thick] (40.5 cm, 0) -- (43.5 cm,0);     
		\draw[thick] (44 cm,0) circle (5 mm) node [shift={(0.0,-0.5)}] {\mfn{4^\prime}};
		\draw[thick] (44.5 cm, 0) -- (47.5 cm,0);     
		\draw[thick] (48 cm,0) circle (5 mm) node [shift={(0.0,-0.5)}] {\mfn{3^\prime}};
		\draw[thick] (48.5 cm, 0) -- (51.5 cm,0);     
		\draw[thick] (52cm,0) circle (5 mm) node [shift={(0.0,-0.5)}] {\mfn{2^\prime}};
		\draw[thick] (52.5 cm, 0) -- (55.5 cm,0);     
		\draw[thick] (56cm,0) circle (5 mm) node [shift={(0.0,-0.5)}] {\mfn{1^\prime}};
		\draw[thick] (52 cm, 0.5cm) -- (52 cm, 3.5cm);     
		\draw[thick] (52cm,4cm) circle (5 mm) node [shift={(0.5,0.0)}] {\mfn{7^\prime}};
		\draw[thick] (52 cm, 4.5cm) -- (52 cm, 7.5cm);     
		\draw[thick] (52cm,8cm) circle (5 mm) node [shift={(0.5,0.0)}] {\mfn{8^\prime}};
		\end{tikzpicture}
	\end{center}
	\caption{Extended diagram with $A_2 = 0$ and $E = E_2$. This is the simplest example of an extended diagram with 
non-trivial new results.}\label{edd2:2}
\end{figure}
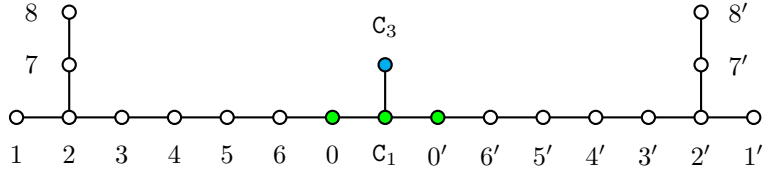

At this point it is useful to introduce an operation on the diagrams which consists of interchanging the last eight components of the two Wilson lines, namely 
\be \label{twisting}
A_1 \to \delta_1 \times \delta'_2, ~~~~~ A_2  \to \delta_2 \times \delta'_1.
\ee
This amounts to exchanging $n'_1 \leftrightarrow n'_2$ in $\varphi_0'$ and $\varphi_{-1}'$. If we follow the rule that nodes of the same color couple together, then this operation  simply exchanges the colors of the affine roots relating to $\mre'_8$. Because of the way the diagrams transform, we call this operation ``twisting".

Applying this operation to the diagram in Figure \ref{edd2:1} we get the one shown in Figure \ref{edd2:3}, which gives an explicit 
realization of the embedding $\text{II}_{1,9}+ \text{II}_{1,9} \subset \text{II}_{2,18}$. Since the automorphisms of 
$\text{II}_{1,9}$ form a Coxeter group (as in the $\text{II}_{1,17}$ case), this diagram yields all ADE lattices which are 
products of rank $9$ positive definite lattices 
admitting an embedding in $\text{II}_{1,9}$.
In this diagram $E=E_1$ and effectively $A_1= \delta_1 \times 0$ and $A_2=0 \times \delta'_2$, with shifts depending on the 
deleted nodes. For instance, $\delta_1=\frac13 w_1$, $\delta'_2=\frac13 w'_1$, gives the group $\sug(10)^2$.

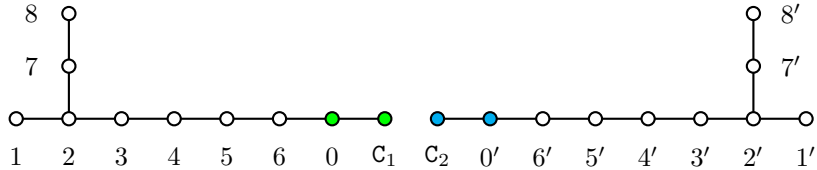
\begin{figure}[htb]
	\begin{center}
		\begin{tikzpicture}[scale=.175]
		\draw[thick] (0 cm,0) circle (5 mm) node [shift={(0.0,-0.5)}] {\mfn{1}} ;
		\draw[thick] (.5 cm, 0) -- (3.5 cm,0);    
		\draw[thick] (4 cm,0) circle (5 mm) node [shift={(0.0,-0.5)}] {\mfn{2}};
		\draw[thick] (4.5 cm, 0) -- (7.5 cm,0); 
		\draw[thick] (8 cm,0) circle (5 mm) node [shift={(0.0,-0.5)}] {\mfn{3}};
		\draw[thick] (8.5 cm, 0) -- (11.5 cm,0);     
		\draw[thick] (12 cm,0) circle (5 mm) node [shift={(0.0,-0.5)}] {\mfn{4}};
		\draw[thick] (12.5 cm, 0) -- (15.5 cm,0);     
		\draw[thick] (16 cm,0) circle (5 mm) node [shift={(0.0,-0.5)}] {\mfn{5}};
		\draw[thick] (16.5 cm, 0) -- (19.5 cm,0);     
		\draw[thick] (20cm,0) circle (5 mm) node [shift={(0.0,-0.5)}] {\mfn{6}};
		\draw[thick] (4 cm, 0.5cm) -- (4 cm, 3.5cm);     
		\draw[thick] (4cm,4cm) circle (5 mm) node [shift={(-0.5,0.0)}] {\mfn{7}};
		\draw[thick] (4 cm, 4.5cm) -- (4 cm, 7.5cm);     
		\draw[thick] (4cm,8cm) circle (5 mm) node [shift={(-0.5,0.0)}] {\mfn{8}};
		\draw[thick] (20.5 cm, 0) -- (23.5 cm,0);     
		\draw[thick, fill = green] (24cm,0) circle (5 mm) node [shift={(0.0,-0.5)}] {\mfn{0}};
		\draw[thick] (24.5 cm, 0) -- (27.5 cm,0);     
		\draw[thick, fill=green] (28cm,0) circle (5 mm) node [shift={(0.0,-0.5)}] {\footnotesize{$\texttt{C}_1$}};
		\draw[thick, fill=cyan] (32cm,0cm) circle (5 mm) node [shift={(0.0,-0.5)}] {\footnotesize{$\texttt{C}_2$}};
		\draw[thick] (32.5 cm, 0) -- (35.5 cm,0);     
		\begin{scope}[shift = {(4cm,0)}]
		\draw[thick, fill = cyan] (32 cm,0) circle (5 mm) node [shift={(0.0,-0.5)}] {\mfn{0^\prime}} ;
		\draw[thick] (32.5 cm, 0) -- (35.5 cm,0);    
		\draw[thick] (36 cm,0) circle (5 mm) node [shift={(0.0,-0.5)}] {\mfn{6^\prime}};
		\draw[thick] (36.5 cm, 0) -- (39.5 cm,0); 
		\draw[thick] (40 cm,0) circle (5 mm) node [shift={(0.0,-0.5)}] {\mfn{5^\prime}};
		\draw[thick] (40.5 cm, 0) -- (43.5 cm,0);     
		\draw[thick] (44 cm,0) circle (5 mm) node [shift={(0.0,-0.5)}] {\mfn{4^\prime}};
		\draw[thick] (44.5 cm, 0) -- (47.5 cm,0);     
		\draw[thick] (48 cm,0) circle (5 mm) node [shift={(0.0,-0.5)}] {\mfn{3^\prime}};
		\draw[thick] (48.5 cm, 0) -- (51.5 cm,0);     
		\draw[thick] (52cm,0) circle (5 mm) node [shift={(0.0,-0.5)}] {\mfn{2^\prime}};
		\draw[thick] (52.5 cm, 0) -- (55.5 cm,0);     
		\draw[thick] (56cm,0) circle (5 mm) node [shift={(0.0,-0.5)}] {\mfn{1^\prime}};
		\draw[thick] (52 cm, 0.5cm) -- (52 cm, 3.5cm);     
		\draw[thick] (52cm,4cm) circle (5 mm) node [shift={(0.5,0.0)}] {\mfn{7^\prime}};
		\draw[thick] (52 cm, 4.5cm) -- (52 cm, 7.5cm);     
		\draw[thick] (52cm,8cm) circle (5 mm) node [shift={(0.5,0.0)}] {\mfn{8^\prime}};
		\end{scope}
		\end{tikzpicture}
	\end{center}
	\caption{Extended diagram with $A_1= \delta_1 \times 0$, $A_2=0 \times \delta'_2$, $E=E_1$, which corresponds to the lattice $\text{II}_{1,9} + \text{II}_{1,9}$. Interestingly, these roots form a basis for $\text{II}_{2,18}$.}\label{edd2:3}
\end{figure}

If  we twist the diagram in Figure \ref{edd2:2}, we get the one shown in Figure \ref{edd2:4}.
Now we can get groups such as $\sug(19)$ with $A_1=\frac13 w_1 \times 0$ and 
$A_2=0 \times \frac13 w'_1$, as well as $\sog(36)$  with $A_1=\frac13 w_1 \times 0$ and 
$A_2=0 \times \frac12 w'_8$. In both cases $E=E_2$.

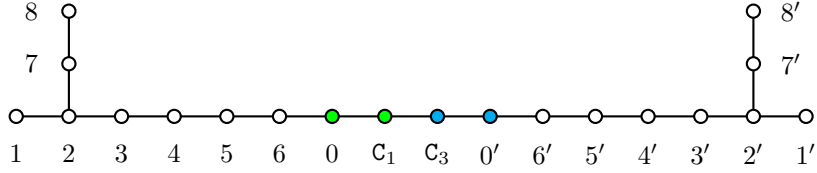
\begin{figure}[htb]
	\begin{center}
		\begin{tikzpicture}[scale=.175]
		\draw[thick] (0 cm,0) circle (5 mm) node [shift={(0.0,-0.5)}] {\mfn{1}} ;
		\draw[thick] (.5 cm, 0) -- (3.5 cm,0);    
		\draw[thick] (4 cm,0) circle (5 mm) node [shift={(0.0,-0.5)}] {\mfn{2}};
		\draw[thick] (4.5 cm, 0) -- (7.5 cm,0); 
		\draw[thick] (8 cm,0) circle (5 mm) node [shift={(0.0,-0.5)}] {\mfn{3}};
		\draw[thick] (8.5 cm, 0) -- (11.5 cm,0);     
		\draw[thick] (12 cm,0) circle (5 mm) node [shift={(0.0,-0.5)}] {\mfn{4}};
		\draw[thick] (12.5 cm, 0) -- (15.5 cm,0);     
		\draw[thick] (16 cm,0) circle (5 mm) node [shift={(0.0,-0.5)}] {\mfn{5}};
		\draw[thick] (16.5 cm, 0) -- (19.5 cm,0);     
		\draw[thick] (20cm,0) circle (5 mm) node [shift={(0.0,-0.5)}] {\mfn{6}};
		\draw[thick] (4 cm, 0.5cm) -- (4 cm, 3.5cm);     
		\draw[thick] (4cm,4cm) circle (5 mm) node [shift={(-0.5,0.0)}] {\mfn{7}};
		\draw[thick] (4 cm, 4.5cm) -- (4 cm, 7.5cm);     
		\draw[thick] (4cm,8cm) circle (5 mm) node [shift={(-0.5,0.0)}] {\mfn{8}};
		\draw[thick] (20.5 cm, 0) -- (23.5 cm,0);     
		\draw[thick, fill = green] (24cm,0) circle (5 mm) node [shift={(0.0,-0.5)}] {\mfn{0}};
		\draw[thick] (24.5 cm, 0) -- (31.5 cm,0);     
		\draw[thick, fill=green] (28cm,0) circle (5 mm) node [shift={(0.0,-0.5)}] {\footnotesize{$\texttt{C}_1$}};
		\draw[thick, fill=cyan] (32cm,0cm) circle (5 mm) node [shift={(0.0,-0.5)}] {\footnotesize{$\texttt{C}_3$}};
		\draw[thick] (32.5 cm, 0) -- (35.5 cm,0);     
		\begin{scope}[shift = {(4cm,0)}]
		\draw[thick, fill = cyan] (32 cm,0) circle (5 mm) node [shift={(0.0,-0.5)}] {\mfn{0^\prime}} ;
		\draw[thick] (32.5 cm, 0) -- (35.5 cm,0);    
		\draw[thick] (36 cm,0) circle (5 mm) node [shift={(0.0,-0.5)}] {\mfn{6^\prime}};
		\draw[thick] (36.5 cm, 0) -- (39.5 cm,0); 
		\draw[thick] (40 cm,0) circle (5 mm) node [shift={(0.0,-0.5)}] {\mfn{5^\prime}};
		\draw[thick] (40.5 cm, 0) -- (43.5 cm,0);     
		\draw[thick] (44 cm,0) circle (5 mm) node [shift={(0.0,-0.5)}] {\mfn{4^\prime}};
		\draw[thick] (44.5 cm, 0) -- (47.5 cm,0);     
		\draw[thick] (48 cm,0) circle (5 mm) node [shift={(0.0,-0.5)}] {\mfn{3^\prime}};
		\draw[thick] (48.5 cm, 0) -- (51.5 cm,0);     
		\draw[thick] (52cm,0) circle (5 mm) node [shift={(0.0,-0.5)}] {\mfn{2^\prime}};
		\draw[thick] (52.5 cm, 0) -- (55.5 cm,0);     
		\draw[thick] (56cm,0) circle (5 mm) node [shift={(0.0,-0.5)}] {\mfn{1^\prime}};
		\draw[thick] (52 cm, 0.5cm) -- (52 cm, 3.5cm);     
		\draw[thick] (52cm,4cm) circle (5 mm) node [shift={(0.5,0.0)}] {\mfn{7^\prime}};
		\draw[thick] (52 cm, 4.5cm) -- (52 cm, 7.5cm);     
		\draw[thick] (52cm,8cm) circle (5 mm) node [shift={(0.5,0.0)}] {\mfn{8^\prime}};
		\end{scope}
		\end{tikzpicture}
	\end{center}
	\caption{Extended diagram with $A_1= \delta_1 \times 0$, $A_2=0 \times \delta'_2$, $E=E_1$. This also gives a basis for $\text{II}_{2,18}$.}\label{edd2:4}
\end{figure}

Summarizing, the generalized diagrams with 20 nodes can be used to obtain maximal enhancings which are read off
from residual ADE diagrams found by deleting two nodes. The $E_{ij}$ moduli are either of type $E_1$ or type $E_2$ in
 \eqref{E_standard}, while the $A_i$ are determined from the deleted nodes.
However, it should be noted that in some cases the gauge group determined from the predicted moduli
might not be represented by the residual diagram. 
The problem is that it is not enough to find a set of 18 nodes,
specified by charge vectors $\ket{\varphi_\mu}$,  $\mu=1,\ldots,18$, such that these nodes form a proper ADE Dynkin 
diagram with links given by $\langle \varphi_\mu | \varphi_\nu\rangle$, defined in \eqref{innerprod}. For these nodes
to correspond to roots $(0;{\bf p_L})$, ${\bf p_L}^2=2$, belonging to $\nlt$, there must exist moduli such that the
charge vectors satisfy \eqref{massless2}. If these moduli exist, we then have to check if they allow 
other roots such that the ones in the set $\ket{\varphi_\mu}$ are indeed simple and can appear in the Dynkin diagram.
For this reason, the diagrams presented here and below have been confirmed to work as intended. 
We will see that the same problem arises in the algorithms of section \ref{sec:explore}, but there is a systematic
prescription to determine the correct gauge group.

\subsubsection{Extended diagrams with nontrivial second breaking}
\label{sss:nontrivialA2}

Now we construct some extended diagrams for models with $A_2 \neq 0$. To keep things clear and unambiguous, we impose the restriction that the affine nodes $\varphi_0$ and $\varphi_{-1}$ cannot belong to the same connected component of the diagram, and similarly for $\varphi_0'$ and $\varphi_{-1}'$.

As a first example we take $A_1 = \tfrac{1}{2}w_6 \times \tfrac{1}{2}w_6'$. In the notation of 
section \ref{sss:gdd}, $k=m=6$ and the unbroken subgroup is the product of $\mrh_6=\mre_7 \times \mra_1$ and
$\mrh'_6=\mre'_7 \times \mra'_1$.
Our algorithm dictates that we add two affine nodes, and the restriction above says that these cannot extend  
$\mra_1$ nor $\mra'_1$, since these include the nodes ${\varphi_0}$ and $\varphi_{0'}$. Hence we should add the affine roots 
for $\mre_7$ and $\mre'_7$ and color them blue. With $E = E_1$ we then get the extended diagram shown in Figure \ref{edd2:5}. 

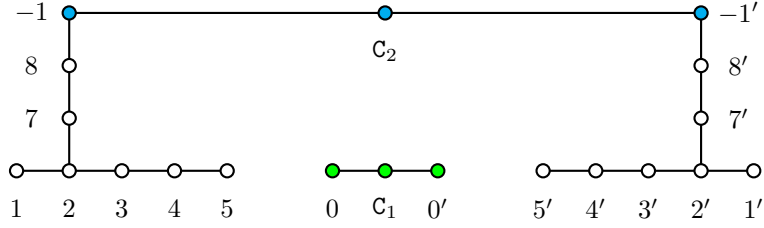
\begin{figure}[htb]
	\begin{center}
		\begin{tikzpicture}[scale=.175]
		\draw[thick] (0 cm,0) circle (5 mm) node [shift={(0.0,-0.5)}] {\mfn{1}} ;
		\draw[thick] (.5 cm, 0) -- (3.5 cm,0);    
		\draw[thick] (4 cm,0) circle (5 mm) node [shift={(0.0,-0.5)}] {\mfn{2}};
		\draw[thick] (4.5 cm, 0) -- (7.5 cm,0); 
		\draw[thick] (8 cm,0) circle (5 mm) node [shift={(0.0,-0.5)}] {\mfn{3}};
		\draw[thick] (8.5 cm, 0) -- (11.5 cm,0);     
		\draw[thick] (12 cm,0) circle (5 mm) node [shift={(0.0,-0.5)}] {\mfn{4}};
		\draw[thick] (12.5 cm, 0) -- (15.5 cm,0);     
		\draw[thick] (16 cm,0) circle (5 mm) node [shift={(0.0,-0.5)}] {\mfn{5}};
		
		\draw[thick] (4 cm, 0.5cm) -- (4 cm, 3.5cm);     
		\draw[thick] (4cm,4cm) circle (5 mm) node [shift={(-0.5,0.0)}] {\mfn{7}};
		\draw[thick] (4 cm, 4.5cm) -- (4 cm, 7.5cm);     
		\draw[thick] (4cm,8cm) circle (5 mm) node [shift={(-0.5,0.0)}] {\mfn{8}};
		\draw[thick] (4 cm, 8.5cm) -- (4 cm, 11.5cm);     
		\draw[thick, fill = cyan] (4cm,12cm) circle (5 mm) node [shift={(-0.5,0.0)}] {\mfn{-1}};
		\draw[thick, fill = green] (24cm,0) circle (5 mm) node [shift={(0.0,-0.5)}] {\mfn{0}};
		\draw[thick] (24.5 cm, 0) -- (27.5 cm,0);     
		\draw[thick, fill=green] (28cm,0) circle (5 mm) node [shift={(0.0,-0.5)}] {\footnotesize{$\texttt{C}_1$}};
		\draw[thick, fill=cyan] (28cm,12cm) circle (5 mm) node [shift={(0.0,-0.5)}] {\footnotesize{$\texttt{C}_2$}};
		\draw[thick] (4.5 cm, 12cm) -- (27.5 cm,12cm);
		\draw[thick] (28.5 cm, 12cm) -- (51.5 cm,12cm);
		\draw[thick] (28.5 cm, 0) -- (32.5 cm,0);
		\draw[thick, fill = green] (32 cm,0) circle (5 mm) node [shift={(0.0,-0.5)}] {\mfn{0^\prime}} ;
		\draw[thick] (40 cm,0) circle (5 mm) node [shift={(0.0,-0.5)}] {\mfn{5^\prime}};
		\draw[thick] (40.5 cm, 0) -- (43.5 cm,0);     
		\draw[thick] (44 cm,0) circle (5 mm) node [shift={(0.0,-0.5)}] {\mfn{4^\prime}};
		\draw[thick] (44.5 cm, 0) -- (47.5 cm,0);     
		\draw[thick] (48 cm,0) circle (5 mm) node [shift={(0.0,-0.5)}] {\mfn{3^\prime}};
		\draw[thick] (48.5 cm, 0) -- (51.5 cm,0);     
		\draw[thick] (52cm,0) circle (5 mm) node [shift={(0.0,-0.5)}] {\mfn{2^\prime}};
		\draw[thick] (52.5 cm, 0) -- (55.5 cm,0);     
		\draw[thick] (56cm,0) circle (5 mm) node [shift={(0.0,-0.5)}] {\mfn{1^\prime}};
		\draw[thick] (52 cm, 0.5cm) -- (52 cm, 3.5cm);     
		\draw[thick] (52cm,4cm) circle (5 mm) node [shift={(0.5,0.0)}] {\mfn{7^\prime}};
		\draw[thick] (52 cm, 4.5cm) -- (52 cm, 7.5cm);     
		\draw[thick] (52cm,8cm) circle (5 mm) node [shift={(0.5,0.0)}] {\mfn{8^\prime}};
		\draw[thick] (52 cm, 8.5cm) -- (52 cm, 11.5cm);     
		\draw[thick, fill = cyan] (52cm,12cm) circle (5 mm) node [shift={(0.5,0.0)}] {\mfn{-1^\prime}};
		\end{tikzpicture}
	\end{center}
	\caption{Extended diagram for double breakings corresponding to the choice $A_1 = \frac{1}{2}w_6 \times \frac{1}{2}w_6'$, $E=E_1$.}\label{edd2:5}
\end{figure}

In this example the new affine roots that build $\varphi_{-1}$ and $\varphi_{-1}'$ in \eqref{minus2} are the lowest roots of $\mre_7$ and $\mre'_7$  given by
\be
\begin{split}
\hat\alpha_6 &= -(2 \alpha_1 + 4 \alpha_2 + 3 \alpha_3 + 2 \alpha_4 + \alpha_5 + 3 \alpha_7 + 2 \alpha_8)=
w_6 - w_8=(0^6,-1,-1) \, , \\
\hat\alpha'_6 &= -(2 \alpha'_1 + 4 \alpha'_2 + 3 \alpha'_3 + 2 \alpha'_4 + \alpha'_5 + 3 \alpha'_7 + 2 \alpha'_8)=
w'_6 - w'_8=(1,1,0^6) \, .
\label{hata6}
\end{split}
\ee
The coefficients in the root expansion of $\hat\alpha_6$ are the Kac labels for $\mre_7$, and likewise for $\hat\alpha'_6$.
In other examples the new affine roots are found in an
analogous way. For example if $k=5$, $\mrh_5=E_6 \times A_2$, and $\hat\alpha_5$ is the lowest root of $\mre_6$, i.e.
$\hat\alpha_5 = -(2 \alpha_1 + 3 \alpha_2 + 2\alpha_3 + \alpha_4 + 2 \alpha_7 + \alpha_8)$.

Deleting any one of $\varphi_{-1}$ or $\varphi'_{-1}$ in Figure \ref{edd2:5}
gives us a group that could have been obtained with a simpler diagram, setting the first and/or last eight components of $A_2$ to zero. Similarly, the affine roots $\varphi_0$ and $\varphi_0'$ cannot be deleted, since this would lead to a non-ADE diagram.
But there are many other possibilities. For illustration we will
derive the moduli for the gauge group $\mra_1 \times \mra_3 \times\mrd_{14}$, found by deleting nodes $1$ and $4'$. 
According to \eqref{shiftconstraints}, for $A_2$ we require 
\begin{equation}
\begin{aligned}
\delta_2 \cdot \alpha_i &= 0, \ i = 0,2,3,4,5,7,8, ~~~ \delta_2 \cdot \hat\alpha_6 = -1,\\
\delta_2' \cdot \alpha_j' &= 0,  \ j = 0,1,2,3,5,7,8, ~~~ \delta_2' \cdot \hat\alpha'_6 = -1.
\end{aligned}
\end{equation}
These constraints are solved by
\begin{equation}
A_2 = \left(\tfrac12 w_1 - \tfrac34 w_6\right) \times \left(\tfrac12 w'_4 - w'_6\right) =
\left(\tfrac{1}{4}, -\tfrac{1}{4}^5, \tfrac{1}{2}^2\right) \times \left(-\tfrac{1}{2}^2,\tfrac{1}{2}^2,0^4  \right).
\end{equation}
Since we already know the values for $E$ and $A_1$, we are done. 

If we take $E = E_2$, we get an extended diagram in which the node $\texttt{C}_2$ in Figure \ref{edd2:5} is replaced by 
$\texttt{C}_3$, and is connected to $\texttt{C}_1$. Here one cannot delete any pair of nodes as we would not get an ADE group. 
This means that for $A_1= \frac{1}{2}w_6 \times \frac{1}{2}w_6'$ and $E=E_2$ there is  
no second Wilson line with $\delta_2 \neq 0$ and  $\delta_2' \neq 0$ that gives maximal enhancement.
What we can do is apply the twisting operation \eqref{twisting}, interchanging the colors of $\varphi_0'$ and $\varphi_{-1}'$. 
The resulting diagram is shown in Figure \ref{edd2:6}. To get for example the group $\mra_1 \times \mra_9 \times \mrd_8$ we delete the nodes $1$ and $4'$. The Wilson lines are then obtained from those in the previous example by exchanging the last eight components. 
\begin{figure}[htb]
	\begin{center}
		\begin{tikzpicture}[scale=.175]
		\draw[thick] (0 cm,0) circle (5 mm) node [shift={(0.0,-0.5)}] {\mfn{1}} ;
		\draw[thick] (.5 cm, 0) -- (3.5 cm,0);    
		\draw[thick] (4 cm,0) circle (5 mm) node [shift={(0.0,-0.5)}] {\mfn{2}};
		\draw[thick] (4.5 cm, 0) -- (7.5 cm,0); 
		\draw[thick] (8 cm,0) circle (5 mm) node [shift={(0.0,-0.5)}] {\mfn{7}};
		\draw[thick] (8.5 cm, 0) -- (11.5 cm,0);     
		\draw[thick] (12 cm,0) circle (5 mm) node [shift={(0.0,-0.5)}] {\mfn{8}};
		\draw[thick] (12.5 cm, 0) -- (15.5 cm,0);     
		\draw[thick, fill = cyan] (16 cm,0) circle (5 mm) node [shift={(0.0,-0.5)}] {\mfn{-1}};
		
		\draw[thick] (4 cm, 0.5cm) -- (4 cm, 3.5cm);     
		\draw[thick] (4cm,4cm) circle (5 mm) node [shift={(-0.5,0.0)}] {\mfn{3}};
		\draw[thick] (4 cm, 4.5cm) -- (4 cm, 7.5cm);     
		\draw[thick] (4cm,8cm) circle (5 mm) node [shift={(-0.5,0.0)}] {\mfn{4}};
		\draw[thick] (4 cm, 8.5cm) -- (4 cm, 11.5cm);     
		\draw[thick, ] (4cm,12cm) circle (5 mm) node [shift={(-0.5,0.0)}] {\mfn{5}};
		\draw[thick, fill = cyan] (24cm,0) circle (5 mm) node [shift={(0.0,-0.5)}] {\mfn{0^\prime}};
		\draw[thick] (23.5 cm, 0) -- (20.5 cm,0);     
		\draw[thick, fill=green] (36cm,0) circle (5 mm) node [shift={(0.0,-0.5)}] {\footnotesize{$\texttt{C}_1$}};
		\draw[thick, fill=cyan] (20cm,0cm) circle (5 mm) node [shift={(0.0,-0.5)}] {\footnotesize{$\texttt{C}_2$}};
		\draw[thick] (32.5 cm, 0) -- (35.5 cm,0);
		\draw[thick] (36.5 cm, 0) -- (39.5 cm,0);
		\draw[thick] (16.5 cm, 0) -- (19.5 cm,0);
		\draw[thick, fill = green] (32 cm,0) circle (5 mm) node [shift={(0.0,-0.5)}] {\mfn{0}} ;
		\draw[thick, fill = green] (40 cm,0) circle (5 mm) node [shift={(0.0,-0.5)}] {\mfn{-1^\prime}};
		\draw[thick] (40.5 cm, 0) -- (43.5 cm,0);     
		\draw[thick] (44 cm,0) circle (5 mm) node [shift={(0.0,-0.5)}] {\mfn{8^\prime}};
		\draw[thick] (44.5 cm, 0) -- (47.5 cm,0);     
		\draw[thick] (48 cm,0) circle (5 mm) node [shift={(0.0,-0.5)}] {\mfn{7^\prime}};
		\draw[thick] (48.5 cm, 0) -- (51.5 cm,0);     
		\draw[thick] (52cm,0) circle (5 mm) node [shift={(0.0,-0.5)}] {\mfn{2^\prime}};
		\draw[thick] (52.5 cm, 0) -- (55.5 cm,0);     
		\draw[thick] (56cm,0) circle (5 mm) node [shift={(0.0,-0.5)}] {\mfn{1^\prime}};
		\draw[thick] (52 cm, 0.5cm) -- (52 cm, 3.5cm);     
		\draw[thick] (52cm,4cm) circle (5 mm) node [shift={(0.5,0.0)}] {\mfn{3^\prime}};
		\draw[thick] (52 cm, 4.5cm) -- (52 cm, 7.5cm);     
		\draw[thick] (52cm,8cm) circle (5 mm) node [shift={(0.5,0.0)}] {\mfn{4^\prime}};
		\draw[thick] (52 cm, 8.5cm) -- (52 cm, 11.5cm);     
		\draw[thick] (52cm,12cm) circle (5 mm) node [shift={(0.5,0.0)}] {\mfn{5^\prime}};
		\end{tikzpicture}
	\end{center}
	\caption{Extended diagram for Wilson lines $A_1=\frac12 w_6 \times \delta_1'$, $A_2=\delta_2 \times \frac12 w'_6$ and $E=E_2$. Curiously it corresponds to the product of what are referred to as over over-extended diagrams for $\mre_7$, in this case written as $\mre_7^{++}+\mre_7^{++}$. }\label{edd2:6}
\end{figure}
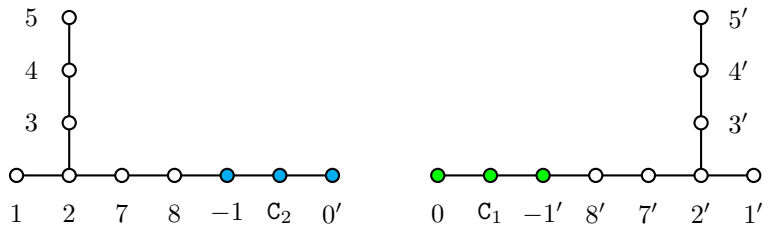

\subsubsection{Exceptional extended diagrams}
\label{sss:exceptional}

The construction of extended diagrams considered so far can be thought of as gluing two subdiagrams of nine nodes via the nodes 
$\{\varphi_{{\texttt{C}_1}}, \varphi_{{\texttt{C}_2}}\}$ or $\{\varphi_{{\texttt{C}_1}}, \varphi_{{\texttt{C}_3}}\}$.
The two subdiagrams are in turn assembled via the two-step shift algorithm applied to $\mre_8$ and $\mre_8'$. 
There are, however, three extra subdiagrams which do not exactly conform to this procedure, but arise naturally when one considers how the affine roots $\hat \alpha_i$, with $i = 4,5,6$, described in section \ref{sss:gdd}, are linked to the simple roots of $\mre_8$. Similar considerations for the other affine roots do not lead to analogous conclusions, in part due to the fact that they extend $\mra_n$ diagrams.

In Figure \ref{fig:afroots} we have drawn the extended Dynkin diagram of $\mre_8$, with its usual lowest root $\alpha_0$, together with the three affine roots mentioned above. The black (red) links represent inner products with value -1 (+1). The inner products between 
the $\hat \alpha_i$ are not shown, as they are not of interest. The color coding is exactly as before, meaning that the charge vectors
of the nodes corresponding to $\hat \alpha_i$ have $n_2 = -1$ and $n_1 = w^1 = w^2 = 0$. We see that 
deleting the $i$-th node, and adding the affine root $\hat \alpha_i$, gives us three of the subdiagrams which are predicted by the method of \ref{sss:gdd}. 
However, as suggested by the right side of the figure, if we flip the sign of the $\hat \alpha_i$ we are now able to construct three more subdiagrams. These are shown in Figure \ref{edd:e1}, with the blue extending nodes defined in each case as
\begin{equation}
\varphi_{-1} =
\begin{cases}
 	\ket{0,0,0,-1;-\hat\alpha_6,0^8} = \ket{0,0,0,-1;0^6,1,1,0^8} & \text{(a)}\\
	\ket{0,0,0,-1;-\hat \alpha_5,0^8} = \ket{0,0,0,-1; -\tfrac{1}{2},\tfrac{1}{2}^4,-\tfrac{1}{2}^3,0^8} & \text{(b)}\\
	\ket{0,0,0,-1; -\hat \alpha_4,0^8} = \ket{0,0,0,-1; -\tfrac{1}{2}^4,\tfrac{1}{2}^4,0^8} & \text{(c)}
\end{cases}
\end{equation}

\begin{figure}
	\centering
	\begin{tikzpicture}[scale=.175]   
	\draw[thick] (0 cm,0) circle (5 mm) node [shift={(-0.5,0.0)}] {\mfn{1}} ;
	\draw[thick] (.5 cm, 0) -- (3.5 cm,0); 
	\draw[thick] (4 cm,0) circle (5 mm) node [shift={(0.0,-0.5)}] {\mfn{2}};
	\draw[thick] (4.5 cm, 0) -- (7.5 cm,0); 
	\draw[thick] (8 cm,0) circle (5 mm) node [shift={(0.0,-0.5)}] {\mfn{3}};
	\draw[thick] (8.5 cm, 0) -- (11.5 cm,0);     
	\draw[thick] (12 cm,0) circle (5 mm) node [shift={(0.0,-0.5)}] {\mfn{4}};
	\draw[thick] (12.5 cm, 0) -- (15.5 cm,0);     
	\draw[thick] (16 cm,0) circle (5 mm) node [shift={(0.0,0.5)}] {\mfn{5}};
	\draw[thick] (16.5 cm, 0) -- (19.5 cm,0);     
	\draw[thick] (20 cm,0) circle (5 mm) node [shift={(0.0,-0.5)}] {\mfn{6}};
	\draw[thick] (20.5 cm, 0) -- (24.5 cm,0);   
	
	\draw[thick] (4 cm, 0.5cm) -- (4 cm, 3.5cm);     
	\draw[thick] (4cm,4cm) circle (5 mm) node [shift={(-0.5,0.0)}] {\mfn{7}};  
	
	\draw[thick] (4 cm, 4.5cm) -- (4 cm, 7.5cm);     
	\draw[thick] (4cm,8cm) circle (5 mm) node [shift={(-0.5,0.0)}] {\mfn{8}};   
	
	\draw[thick] (8 cm, -6cm) -- (0 cm, -6cm); 
	\draw[thick,red] (8 cm, -6cm) -- (16 cm, -6cm);
	\draw[thick] (0 cm, -6cm) -- (0 cm, -0.5cm); 
	\draw[thick,red] (16 cm, -0.5cm) -- (16 cm, -6cm);     
	\draw[thick,fill = cyan] (8cm,-6cm) circle (5 mm) node [shift={(0.0,-0.5)}] {$\hat\alpha_5$};   
	
	\draw[thick] (20 cm, 8cm) -- (4.5 cm, 8cm); 
	\draw[thick,red] (20 cm, 0.5cm) -- (20 cm, 7.5cm);     
	\draw[thick,fill = cyan] (20cm,8cm) circle (5 mm) node [shift={(0.5,0.0)}] {$\hat\alpha_6$};  
	
	\draw[thick] (12 cm, 4cm) -- (4.5 cm, 4cm); 
	\draw[thick,red] (12 cm, 0.5cm) -- (12 cm, 4cm);     
	\draw[thick,fill = cyan] (12cm,4cm) circle (5 mm) node [shift={(0.0,0.35)}] {$\hat\alpha_4$};  
	\draw[thick, fill = green] (24cm,0) circle (5 mm) node [shift={(0.0,-0.5)}] {\mfn{0}};5
	\draw(32cm,0)node{$\rightarrow$};
	\begin{scope}[shift={(40cm,0)}]
	\draw[thick] (0 cm,0) circle (5 mm) node [shift={(-0.5,0.0)}] {\mfn{1}} ;
	\draw[thick] (.5 cm, 0) -- (3.5 cm,0); 
	\draw[thick] (4 cm,0) circle (5 mm) node [shift={(0.0,-0.5)}] {\mfn{2}};
	\draw[thick] (4.5 cm, 0) -- (7.5 cm,0); 
	\draw[thick] (8 cm,0) circle (5 mm) node [shift={(0.0,-0.5)}] {\mfn{3}};
	\draw[thick] (8.5 cm, 0) -- (11.5 cm,0);     
	\draw[thick] (12 cm,0) circle (5 mm) node [shift={(0.0,-0.5)}] {\mfn{4}};
	\draw[thick] (12.5 cm, 0) -- (15.5 cm,0);     
	\draw[thick] (16 cm,0) circle (5 mm) node [shift={(0.0,0.5)}] {\mfn{5}};
	\draw[thick] (16.5 cm, 0) -- (19.5 cm,0);     
	\draw[thick] (20 cm,0) circle (5 mm) node [shift={(0.0,-0.5)}] {\mfn{6}};
	\draw[thick] (20.5 cm, 0) -- (24.5 cm,0);   
	
	\draw[thick] (4 cm, 0.5cm) -- (4 cm, 3.5cm);     
	\draw[thick] (4cm,4cm) circle (5 mm) node [shift={(-0.5,0.0)}] {\mfn{7}};  
	
	\draw[thick] (4 cm, 4.5cm) -- (4 cm, 7.5cm);     
	\draw[thick] (4cm,8cm) circle (5 mm) node [shift={(-0.5,0.0)}] {\mfn{8}};   
	
	\draw[thick,red] (8 cm, -6cm) -- (0 cm, -6cm); 
	\draw[thick] (8 cm, -6cm) -- (16 cm, -6cm);
	\draw[thick,red] (0 cm, -6cm) -- (0 cm, -0.5cm); 
	\draw[thick] (16 cm, -0.5cm) -- (16 cm, -6cm);     
	\draw[thick,fill = cyan] (8cm,-6cm) circle (5 mm) node [shift={(0.0,-0.5)}] {$-\hat\alpha_5$};   
	
	\draw[thick,red] (20 cm, 8cm) -- (4.5 cm, 8cm); 
	\draw[thick] (20 cm, 0.5cm) -- (20 cm, 7.5cm);     
	\draw[thick,fill = cyan] (20cm,8cm) circle (5 mm) node [shift={(0.5,0.0)}] {$-\hat\alpha_6$};  
	
	\draw[thick,red] (12 cm, 4cm) -- (4.5 cm, 4cm); 
	\draw[thick] (12 cm, 0.5cm) -- (12 cm, 4cm);     
	\draw[thick,fill = cyan] (12cm,4cm) circle (5 mm) node [shift={(0.0,0.35)}] {$-\hat\alpha_4$};  
	\draw[thick, fill = green] (24cm,0) circle (5 mm) node [shift={(0.0,-0.5)}] {\mfn{0}};5
	\end{scope}
	\end{tikzpicture}
	\caption{Links between the affine roots $\hat\alpha_4$, $\hat\alpha_5$, $\hat\alpha_6$ and the roots of the affine $\mre_8$ diagram.  
	Black (red) links correspond to inner products with value -1 (+1). The diagram to the right is obtained by flipping the signs of the 
	$\hat\alpha_i$ .} \label{fig:afroots}
\end{figure}
\begin{figure}[htb]
	\centering
	\begin{tabular}{ccc}
		\begin{tikzpicture}[scale=.175]
		\draw (12cm,8cm) node{(a)};
		\draw[thick] (0 cm,0) circle (5 mm) node [shift={(0.0,-0.5)}] {\mfn{1}} ;
		\draw[thick] (.5 cm, 0) -- (3.5 cm,0);    
		\draw[thick] (4 cm,0) circle (5 mm) node [shift={(0.0,-0.5)}] {\mfn{2}};
		\draw[thick] (4.5 cm, 0) -- (7.5 cm,0); 
		\draw[thick] (8 cm,0) circle (5 mm) node [shift={(0.0,-0.5)}] {\mfn{3}};
		\draw[thick] (8.5 cm, 0) -- (11.5 cm,0);     
		\draw[thick] (12 cm,0) circle (5 mm) node [shift={(0.0,-0.5)}] {\mfn{4}};
		\draw[thick] (12.5 cm, 0) -- (15.5 cm,0);     
		\draw[thick] (16 cm,0) circle (5 mm) node [shift={(0.0,-0.5)}] {\mfn{5}};
		\draw[thick] (16.5 cm, 0) -- (19.5 cm,0);     
		\draw[thick] (20 cm,0) circle (5 mm) node [shift={(0.0,-0.5)}] {\mfn{6}};
		\draw[thick] (20.5 cm, 0) -- (24.5 cm,0);   
		
		\draw[thick] (4 cm, 0.5cm) -- (4 cm, 3.5cm);     
		\draw[thick] (4cm,4cm) circle (5 mm) node [shift={(-0.5,0.0)}] {\mfn{7}};    
		
		\draw[thick] (20 cm, 0.5cm) -- (20 cm, 3.5cm);     
		\draw[thick,fill = cyan] (20cm,4cm) circle (5 mm) node [shift={(-0.5,0.0)}] {\mfn{-1}};   
		\draw[thick, fill = green] (24cm,0) circle (5 mm) node [shift={(0.0,-0.5)}] {\mfn{0}};
		\end{tikzpicture} &
		\begin{tikzpicture}[scale=.175]
		\draw (12cm,8cm) node{(b)};   
		\draw[thick] (4 cm,0) circle (5 mm) node [shift={(0.0,-0.5)}] {\mfn{2}};
		\draw[thick] (4.5 cm, 0) -- (7.5 cm,0); 
		\draw[thick] (8 cm,0) circle (5 mm) node [shift={(0.0,-0.5)}] {\mfn{3}};
		\draw[thick] (8.5 cm, 0) -- (11.5 cm,0);     
		\draw[thick] (12 cm,0) circle (5 mm) node [shift={(0.0,-0.5)}] {\mfn{4}};
		\draw[thick] (12.5 cm, 0) -- (15.5 cm,0);     
		\draw[thick] (16 cm,0) circle (5 mm) node [shift={(0.0,-0.5)}] {\mfn{5}};
		\draw[thick] (16.5 cm, 0) -- (19.5 cm,0);     
		\draw[thick] (20 cm,0) circle (5 mm) node [shift={(0.0,-0.5)}] {\mfn{6}};
		\draw[thick] (20.5 cm, 0) -- (24.5 cm,0);   
		
		\draw[thick] (4 cm, 0.5cm) -- (4 cm, 3.5cm);     
		\draw[thick] (4cm,4cm) circle (5 mm) node [shift={(-0.5,0.0)}] {\mfn{7}};  
		
		\draw[thick] (4 cm, 4.5cm) -- (4 cm, 7.5cm);     
		\draw[thick] (4cm,8cm) circle (5 mm) node [shift={(-0.5,0.0)}] {\mfn{8}};   
		
		\draw[thick] (16 cm, 0.5cm) -- (16 cm, 3.5cm);     
		\draw[thick,fill = cyan] (16cm,4cm) circle (5 mm) node [shift={(-0.5,0.0)}] {\mfn{-1}};   
		\draw[thick, fill = green] (24cm,0) circle (5 mm) node [shift={(0.0,-0.5)}] {\mfn{0}};5
		\end{tikzpicture} &
		\begin{tikzpicture}[scale=.175]
		\draw (12cm,8cm) node{(c)};
		\draw[thick] (0 cm,0) circle (5 mm) node [shift={(0.0,-0.5)}] {\mfn{1}} ;
		\draw[thick] (.5 cm, 0) -- (3.5 cm,0);    
		\draw[thick] (4 cm,0) circle (5 mm) node [shift={(0.0,-0.5)}] {\mfn{2}};
		\draw[thick] (4.5 cm, 0) -- (7.5 cm,0); 
		\draw[thick] (8 cm,0) circle (5 mm) node [shift={(0.0,-0.5)}] {\mfn{3}};
		\draw[thick] (8.5 cm, 0) -- (11.5 cm,0);     
		\draw[thick] (12 cm,0) circle (5 mm) node [shift={(0.0,-0.5)}] {\mfn{4}};
		\draw[thick] (12.5 cm, 0) -- (15.5 cm,0);     
		\draw[thick] (16 cm,0) circle (5 mm) node [shift={(0.0,-0.5)}] {\mfn{5}};
		\draw[thick] (16.5 cm, 0) -- (19.5 cm,0);     
		\draw[thick] (20 cm,0) circle (5 mm) node [shift={(0.0,-0.5)}] {\mfn{6}};
		\draw[thick] (20.5 cm, 0) -- (24.5 cm,0);

		\draw[thick] (4cm,8cm) circle (5 mm) node [shift={(-0.5,0.0)}] {\mfn{8}};  
		
		\draw[thick] (12 cm, 0.5cm) -- (12 cm, 3.5cm);     
		\draw[thick,fill = cyan] (12cm,4cm) circle (5 mm) node [shift={(-0.5,0.0)}] {\mfn{-1}};   
		\draw[thick, fill = green] (24cm,0) circle (5 mm) node [shift={(0.0,-0.5)}] {\mfn{0}};
		\end{tikzpicture}
		\\
	\end{tabular}
	\caption{Three extra subdiagrams which do not come from a two-step shift-vector construction. They can be inferred from the right side of figure \ref{fig:afroots}}\label{edd:e1}
\end{figure}

These new subdiagrams are qualitatively different from those obtained in the previous section in two ways. On one hand, they do not respect the restriction that a connected part cannot have two extending nodes. On the other hand, they are not associated to fixed values of $\delta_1$ or 
$\delta_1'$, as they do not come from a two-step shift algorithm. To illustrate this, consider the diagram (a) in Figure \ref{edd:e1} and break the fourth node, leaving out a $2\mrd_4$ diagram. Solving \eqref{massless2} for all the remaining nodes yields
\begin{equation}
A_1  = \tfrac12(w_4-w_8)\times \delta_1', ~~~~~ A_2 = \tfrac12(w_4 - 2w_8)\times \delta_2',
\end{equation}
with $\delta_1'$, $\delta_2'$ and $E_{ij}$ arbitrary, since the nodes are of the form $\ket{0,0,n_1,n_2;\pi^1,\ldots,\pi^8,0^8}$.
Instead, if we break the third node, corresponding to an $\mra_3 + \mrd_5$ diagram, we obtain
\begin{equation}
	A_1 = \tfrac14(2 w_3 - 3 w_8)\times \delta_1', ~~~~~ A_2 = \tfrac14(2 w_3 - 5w_8)\times \delta_2' \, .
\end{equation}
The Wilson line $A_1$ clearly differs from that of the previous breaking.

Apart from the two considerations mentioned above, the construction of EDD's
with the new subdiagrams is exactly as before. For example, we can take two copies of the subdiagram (a) of Figure \ref{edd:e1} and add the two nodes $\varphi_{{\texttt{C}_1}}, \varphi_{{\texttt{C}_2}}$ to get the EDD shown in Figure \ref{edd:e2}. Some enhancements obtained from this diagram are $2\mrd_6 + 2\mra_3$ and $\mrd_5 + \mrd_6 + \mra_7$.
\begin{figure}
\centering 
	\begin{tikzpicture}[scale=.175]
	\draw[thick] (24 cm, 4cm) -- (32 cm, 4cm); 
	\draw[thick] (24 cm, -4cm) -- (32 cm, -4cm); 
	\draw[thick] (0 cm, 4cm) -- (4 cm, 0cm); 
	\draw[thick] (0 cm, -4cm) -- (4 cm,0);
	\draw[thick, fill=white] (0 cm,-4cm) circle (5 mm) node [shift={(-0.5,0.0)}] {\mfn{1}} ;    
	\draw[thick, fill=white] (4 cm,0) circle (5 mm) node [shift={(0.0,-0.5)}] {\mfn{2}};
	\draw[thick] (4.5 cm, 0) -- (7.5 cm,0); 
	\draw[thick] (8 cm,0) circle (5 mm) node [shift={(0.0,-0.5)}] {\mfn{3}};
	\draw[thick] (8.5 cm, 0) -- (11.5 cm,0);     
	\draw[thick] (12 cm,0) circle (5 mm) node [shift={(0.0,-0.5)}] {\mfn{4}};
	\draw[thick] (12.5 cm, 0) -- (15.5 cm,0);     
	\draw[thick] (16 cm,0) circle (5 mm) node [shift={(0.0,-0.5)}] {\mfn{5}};
	\draw[thick] (16.5 cm, 0) -- (19.5 cm,0); 
	
	\draw[thick] (20 cm, 0) -- (24 cm, 4cm); 
	\draw[thick] (20 cm, 0) -- (24 cm,-4cm); 
	
	\draw[thick,fill=white] (20 cm,0) circle (5 mm) node [shift={(0.0,-0.5)}] {\mfn{6}};  
	
	\draw[thick, fill=white] (0cm,4cm) circle (5 mm) node [shift={(-0.5,0.0)}] {\mfn{7}};    
	
	\draw[thick,fill = cyan] (24cm,4cm) circle (5 mm) node [shift={(0.0,0.5)}] {\mfn{-1}};   
	\draw[thick, fill = green] (24cm,-4cm) circle (5 mm) node [shift={(0.0,-0.5)}] {\mfn{0}};
	
	\draw[thick, fill = green] (28cm,-4cm) circle (5 mm) node [shift={(0.0,-0.5)}] {\footnotesize{$\texttt{C}_1$}};
	
	\draw[thick, fill = cyan] (28cm,4cm) circle (5 mm) node [shift={(0.0,0.5)}] {\footnotesize{$\texttt{C}_2$}};
	
	\begin{scope}[shift={(32cm,0)}]
	\draw[thick] (0 cm, 4cm) -- (4 cm, 0cm);
	\draw[thick] (0 cm, -4cm) -- (4 cm,0);     
	\draw[thick, fill=green] (0 cm,-4cm) circle (5 mm) node [shift={(0.0,-0.5)}] {\mfn{0'}} ;  
	\draw[thick,fill=white] (4 cm,0) circle (5 mm) node [shift={(0.0,-0.5)}] {\mfn{6'}};
	\draw[thick] (4.5 cm, 0) -- (7.5 cm,0); 
	\draw[thick] (8 cm,0) circle (5 mm) node [shift={(0.0,-0.5)}] {\mfn{5'}};
	\draw[thick] (8.5 cm, 0) -- (11.5 cm,0);     
	\draw[thick] (12 cm,0) circle (5 mm) node [shift={(0.0,-0.5)}] {\mfn{4'}};
	\draw[thick] (12.5 cm, 0) -- (15.5 cm,0);     
	\draw[thick] (16 cm,0) circle (5 mm) node [shift={(0.0,-0.5)}] {\mfn{3'}};
	\draw[thick] (16.5 cm, 0) -- (19.5 cm,0);   
	
	\draw[thick] (20 cm, 0cm) -- (24 cm, 4cm); 
	\draw[thick] (20 cm, 0cm) -- (24 cm, -4cm); 
	  
	\draw[thick,fill=white] (20 cm,0) circle (5 mm) node [shift={(0.0,-0.5)}] {\mfn{2'}};   
	  
	\draw[thick, fill=cyan] (0cm,4cm) circle (5 mm) node [shift={(0.0,0.5)}] {\mfn{-1'}};    
	    
	\draw[thick,fill=white] (24cm,4cm) circle (5 mm) node [shift={(0.5,0.0)}] {\mfn{7'}};   
	\draw[thick,fill=white] (24cm,-4) circle (5 mm) node [shift={(0.5,0.0)}] {\mfn{1'}};
	\end{scope}
	\end{tikzpicture} 
	\caption{Extended Dynkin diagram constructed with the exceptional extension shown in Figure \ref{edd:e1} (a) for both $\mre$ and $\mre_{8}'$.}
	\label{edd:e2}
\end{figure}
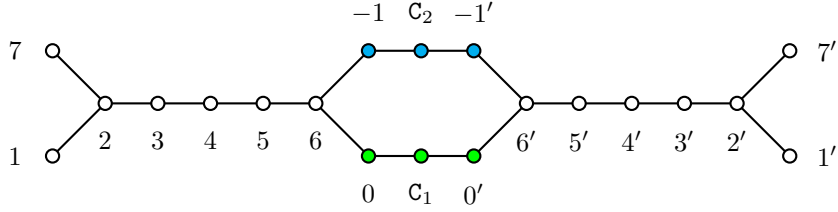

Exhausting the method of extended diagrams allows us to find 300 out of the 325 known maximal rank groups 
obtained in  \cite{SZ}.
Remarkably, without the three subdiagrams in Figure \ref{edd:e1}, this number is reduced to 150.
The incompleteness of the method is due in part to the complexity of the moduli space and the T-duality group $\rO(2,18,\mathbb{Z})$, which makes it hard to establish ways of obtaining global data. This is in contrast with the situation for $d = 1$, where a fundamental region can be easily constructed (see Tables \ref{tab:fundreghe} and \ref{tab:fundregho}).

The maximal rank groups which are missing from our results so far are
\begin{equation}
\begin{array}{c}
	3\mra_6, ~\mra_3 + \mra_6 + \mra_9,~6\mra_{3},  
~2\mra_{1}+4\mra_{4}, 
~2\mra_{2}+2\mra_{3}+2\mra_{4},
~3\mra_{1}+3\mra_{5},  \\
~\mra_{1}+2\mra_{2}+\mra_{3}+2\mra_{5}, 
~\mra_{1}+\mra_{2}+2\mra_{3}+\mra_{4}+\mra_{5}, ~2\mra_{3}+2\mra_{6}, \\
~2\mra_{1}+\mra_{2}+2\mra_{4}+\mra_{6},
~\mra_{2}+2\mra_{3}+\mra_{4}+\mra_{6}, 
~\mra_{1}+\mra_{3}+2\mra_{7}, 
~\mra_{2}+3\mra_{3}+\mra_{7},\\
 ~2\mra_{1}+\mra_{4}+\mra_{5}+\mra_{7}, ~\mra_{2}+\mra_{3}+\mra_{6}+\mra_{7}, 
~\mra_{2}+2\mra_{3}+\mra_{10}, ~3\mrd_{6}, ~2\mra_{2}+2\mrd_{7},\\ ~\mra_{2}+3\mra_{3}+\mrd_{7}, ~\mra_{1}+\mra_{2}+2\mra_{4}+\mrd_{7}, ~\mra_{2}+\mra_{3}+\mra_{6}+\mrd_{7},\\ ~2\mra_{2}+2\mra_{3}+\mrd_{8}, ~2\mrd_{5}+\mrd_{8}, ~\mrd_{5}+\mrd_{7}+\mre_{6}, ~2\mrd_{5}+\mre_{8}.
\end{array}
\end{equation}
As we will see, these can be obtained
with the more powerful algorithms developed in section \ref{sec:explore}. 
Actually, among the 300 groups found with the EDD method, there are 3 that can only be obtained with one of the possible $T$ lattices.
The algorithms presented shortly also determine the moduli corresponding to the other $T$ lattices. The full set of maximally enhanced models, taking into consideration inequivalent models with the same gauge group, are collected in Table \ref{tab:alld2} and further discussed in section \ref{ss:alld2}.

\subsection{Exploring the moduli space } 
\label{sec:explore}

In section \ref{sec:circle} we have seen that to find maximal enhancements in the circle compactification, 
it is enough to give the value of the Wilson line, since we can always take $E=1$.
Moreover, all the maximal enhancements can be obtained with a Wilson line that leaves unbroken a subgroup 
of $\mre_8 \times \mre_8$ of rank 16 and which can be obtained systematically using the shift algorithm described in \ref{sec:shift}.
These results actually rest on the existence of the EDD for $\rO(1,17,\ZZ)$.

In the case of $T^2$ compactifications things are not so simple. 
The techniques of section \ref{sec:shiftd2} do lead to
many maximal enhancement points starting from a collection of extended Dynkin diagrams,  
but this construction requires taking the particular values of $E_{ij}$ defined in 
\eqref{E_standard}. 
With this limitation it is impossible to get some maximal enhancements, such as $\sug(4)^6$, known to exist from the
lattice embedding results in \cite{SZ}. 
In section \ref{ss:swl} we will develop an algorithm that determines if there are maximal enhancements for other 
values of $E_{ij}$, but as in section \ref{sec:shiftd2} still starting from a pair of Wilson lines that leave unbroken a subgroup 
of $\mre_8 \times \mre_8$ of rank 16 for generic $E_{ij}$.
However, as argued shortly, such Wilson lines are not enough to reach all the known maximal enhancements: 
we miss the groups with algebras $3\mra_6$ and $\mra_3+\mra_6+\mra_9$. 
In section \ref{ss:neighbor} we will solve this issue by implementing an alternative algorithm  
which does not fix the Wilson lines.

We will apply the algorithms in the $\mre_8 \times \mre_8$ heterotic theory. 
The moduli for the $\hosp$ theory will then be determined using the map described in section \ref{ss:hehomap}.

\subsubsection{Fixed Wilson lines algorithm}
\label{ss:swl}

This algorithm assumes a pair of Wilson lines fixed by the shift algorithm in such a way that $\mre_8 \times \mre'_8$ is broken to
a maximal subgroup, say $\rG_{16}$. This is the same assumption of section \ref{sec:shiftd2} where we explained that $A_1$ and $A_2$
take the form \eqref{twoWilson} or \eqref{twisting}. For $A_1$, $\delta_1=\dfrac{w_k}{\kappa_k}$, 
$\delta'_1=\dfrac{w_m}{\kappa_m}$,
$k,m=0,\ldots,8$, but $k=m=8$ excluded. For $A_2$, $\delta_2$ and $\delta'_2$ are determined according to
\eqref{shiftconstraints}. 
Setting $E=E_1$, i.e. $E_{ij}=\delta_{ij}$, or $E=E_2$, i.e. $E_{11}=E_{22}=-E_{12}=1$, $E_{21}=0$,
as in \eqref{E_standard}, we can read off the maximal enhancement from the extended 
diagrams constructed in section \ref{sec:shiftd2}. Relaxing the choice of $E_{ij}$ would give the same diagrams but without 
the nodes $\texttt{C}_a$. 
For each choice of Wilson lines the resulting gauge group would generically be $\rG_{16} \times \uo^2$. 
We now want to explore the available four-dimensional region of the moduli space searching for values of $E_{ij}$ 
that give new maximal enhancements to a group of rank 18.

The great advantage of starting with Wilson lines fixed by the shift algorithm is that the 16 simple roots of $\rG_{16}$
are determined systematically. Moreover, we know the associated charge vectors $\ket{w^1,w^2,n_1,n_2;\pi}$
of the 16 nodes, cf. eqs.~\eqref{simple2}, \eqref{zero2} and \eqref{minus2}.
These charge vectors satisfy the massless conditions \eqref{massless2} regardless of the values of $E_{ij}$. 
Therefore, they will still correspond to roots of the enhanced gauge group if we take special values for $E_{ij}$. 
At points of maximal enhancement we must have these $16$ roots plus $2$ additional simple roots.
The algorithm first finds a subset of the possible pairs of extra roots and then computes the values of $E_{ij}$ 
by demanding that they satisfy the quantization conditions in \eqref{quant2}.
It is also necessary to check that the moduli correspond to a physical torus, i.e. that the resulting
torus metric satisfies $g_{ii} > 0$ and $\det g >0$.  
The gauge group is determined from the $18$ simple roots.  
In agreement with the lattice analysis of section \ref{ss:nikhet}, we will see that maximal enhancement can only be obtained
when the $E_{ij}$ take rational values.

The fact that the $E_{ij}$ can now take generic rational values means that we will get new
maximally enhanced groups that could not have appeared with the method of the previous section.
However, as already mentioned, the algorithm still misses known groups with maximal enhancement
as we now argue. For simplicity, we will mostly denote the groups by their algebras.
With only one Wilson line $A_1$ the first $\mre_8$ can only be broken to
\be
\mre_8, \ \ \mra_8, \ \  \mra_1+\mra_2+\mra_5, \ \  2\mra_4, \ \ \mra_3 + \mrd_5, \ \ \mra_2 + \mre_6, \ \
\mra_1+\mre_7, \ \ \mra_1+\mra_7, \ \ \mrd_8 \, .
\label{breakings1}
\ee
These subgroups are just obtained with $\delta_1=\dfrac{w_k}{\kappa_k}$, $k=0,\ldots,8$. Combining with $A_2$ gives
more possibilities. For example, $2\mra_1+\mrd_6$ can occur
breaking first to $\mra_1+\mre_7$ with $\delta_1=\frac12{w_6}$, then extending 
$\mre_7$ with $\hat\alpha_6$ and deleting the node 4, so that $\delta_2=\frac12 w_4-w_6$.
The additional distinct groups that can originate from two Wilson lines are{\footnote{\label{foot:4WL}With 3 and 4 Wilson lines one can obtain
$\mrd_4+4\mra_1$ and $8\mra_1$, respectively. Altogether there are 15 subalgebras of rank 8 that can be embedded in $\mre_8$.
The embeddings are unique up to Weyl automorphisms \cite{Nishiyama}.}
\be
2\mra_1+\mrd_6, \ \ 2\mra_1+2\mra_3, \ \  2\mrd_4, \ \ 4\mra_2 \, .
\label{breakings2}
\ee
Thus, necessarily $\rG_{16}=\rG_8 + \rG'_8$, where each factor can only be one of the above $13$ groups of rank $8$. 
Now, the possible maximal groups $\rG_{18}$ that can appear for specific values of $E_{ij}$ should have a 
Dynkin diagram (DD) that consists of the nodes of the $\mrg_{16}$ diagram plus two additional ones. 
If we want $\mrg_{18}=3\mra_6$, then we should be able to remove two nodes from its DD and get one 
of the algebras $\mrg_8 + \mrg_8'$. It is easy to see that there is no way of removing only two nodes without leaving
behind at least an $\mra_6$. Since none of the possible $\rG_8$ has an $\mra_6$ factor, we conclude that $3\mra_6$
cannot be found starting with Wilson lines fixed by the shift algorithm.
Although a bit longer, a similar reasoning shows that $\mra_3+\mra_6+\mra_9$ cannot be obtained either.
Except for these two groups, with the algorithm we can reproduce all the other known maximal enhancements found
in the K3 context \cite{SZ}. 

We will explain how the algorithm works with an example leading to $6\mra_3$, which
cannot appear with the $E_{ij}$ of \eqref{E_standard}. 
To begin, we delete the nodes $6$ and $6'$ and then $2$ and $2'$. The shift algorithm 
fixes the Wilson lines to be
\be
A_1 = \tfrac12 w_6 \times  \tfrac12  w_6',\quad
A_2 = \left(\tfrac14 w_2 - \tfrac34 w_6\right) \times \left(\tfrac14 w_2' - \tfrac34 w_6'\right)
\label{wlexample1}
\ee
The $16$ unbroken simple roots provide the nodes
\be
\begin{small}
\begin{array}{c}
\varphi_j = |0, 0, 0, 0; \alpha_j, 0^8 \rangle \, ,\quad 
\varphi'_{j}= |0, 0, 0, 0; 0^8, \alpha'_j \rangle \, ,\quad
j=1,\,3,\,4,\,5,\,7,\,8, \\
\varphi_{0}=|0, 0, -1, 0; \alpha_0, 0^8 \rangle \, ,\quad
\varphi'_{0}=|0, 0, -1, 0; 0^8, \alpha'_0 \rangle\, ,\quad \\
\varphi_{-1}=|0, 0, 0, -1; w_6 - w_8, 0^8 \rangle \, ,\quad
\varphi'_{-1}=|0, 0, 0, -1; 0^8, w'_6 - w'_8 \rangle
\end{array}
\end{small}
\label{nodesex1}
\ee
The two Wilson lines break $\mre_8 \times \mre'_8$ to the rank $16$ subgroup $4 \mra_1 + 4 \mra_3$ with DD shown in Figure \ref{dynkin6a3_1}. It can be obtained from the extended diagram in Figure \ref{edd2:5} by removing the nodes $2$ and $2'$, 
as well as the nodes $\tC_a$ associated to $\nltt$.

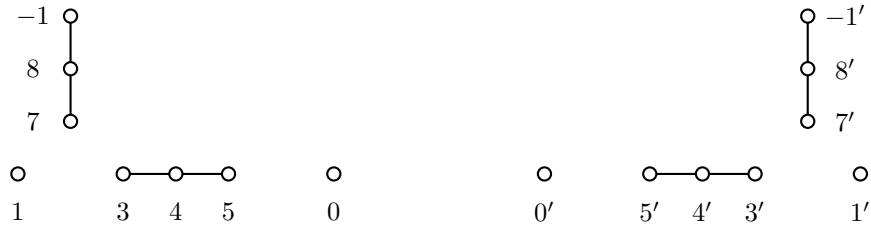
\begin{figure}[htb]
\centering
\begin{tikzpicture}[scale=.175]
   \filldraw[thick,fill=white] (0 cm,0) circle (5 mm) node [shift={(0.0,-0.5)}] {\mfn{1}} ;    
        \filldraw[thick,fill=white] (8 cm,0) circle (5 mm) node [shift={(0.0,-0.5)}] {\mfn{3}} ;     
  \filldraw[thick] (8.5 cm, 0) -- (11.5 cm,0);    
    \filldraw[thick,fill=white] (12 cm,0) circle (5 mm) node [shift={(0.0,-0.5)}] {\mfn{4}};    
  \filldraw[thick,fill=white] (12.5 cm, 0) -- (15.5 cm,0);     
         \filldraw[thick,fill=white] (16 cm,0) circle (5 mm) node [shift={(0.0,-0.5)}] {\mfn{5}};    
         \filldraw[thick,fill=white] (4cm,4cm) circle (5 mm) node [shift={(-0.5,0.0)}] {\mfn{7}};     
 \filldraw[thick,fill=white] (4 cm, 4.5cm) -- (4 cm, 7.5cm);    
         \filldraw[thick,fill=white] (4cm,8cm) circle (5 mm) node [shift={(-0.5,0.0)}] {\mfn{8}};
         \filldraw[thick,fill=white] (4cm, 8.5cm) -- (4cm, 11.5cm);    
         \filldraw[thick,fill=white] (4cm,12cm) circle (5 mm) node [shift={(-0.5,0.0)}] {\mfn{-1}};     
         \filldraw[thick,fill=white] (24cm,0) circle (5 mm) node [shift={(0.0,-0.5)}] {\mfn{0}};   
        \filldraw[thick,fill=white] (40 cm,0) circle (5 mm) node [shift={(0.0,-0.5)}] {\mfn{0'}};    
         \filldraw[thick,fill=white] (48 cm,0) circle (5 mm) node [shift={(0.0,-0.5)}] {\mfn{5'}}; 
  \filldraw[thick,fill=white] (48.5 cm, 0) -- (51.5 cm,0);       
         \filldraw[thick,fill=white] (52cm,0) circle (5 mm) node [shift={(0.0,-0.5)}] {\mfn{4'}};     
  \filldraw[thick,fill=white] (52.5 cm, 0) -- (55.5 cm,0);   
         \filldraw[thick,fill=white] (56cm,0) circle (5 mm) node [shift={(0.0,-0.5)}] {\mfn{3'}};     
   \filldraw[thick,fill=white] (64cm,0) circle (5 mm) node [shift={(0.0,-0.5)}] {\mfn{1'}};     
 \filldraw[thick,fill=white] (60cm,4cm) circle (5 mm) node [shift={(0.5,0.0)}] {\mfn{7'}};     
 \filldraw[thick,fill=white] (60 cm, 4.5cm) -- (60 cm, 7.5cm);    
         \filldraw[thick,fill=white] (60cm,8cm) circle (5 mm) node [shift={(0.5,0.0)}] {\mfn{8'}};
          \filldraw[thick,fill=white] (60cm, 8.5cm) -- (60cm, 11.5cm);    
         \filldraw[thick,fill=white] (60cm,12cm) circle (5 mm) node [shift={(0.5,0.0)}] {\mfn{{-1'}}};
     \end{tikzpicture}
          \caption{Dynkin diagram corresponding to the $16$ simple roots that survive the breaking by the Wilson lines \eqref{wlexample1}.}\label{dynkin6a3_1}
\end{figure}

For maximal enhancement we have to add two additional nodes. To illustrate the procedure we first add 
a single node denoted $\tN_1$. The charge vector $\varphi_{\tN_1}$
must have norm $2$ and the inner product with the 16 nodes in \eqref{nodesex1} must be $0$ or $-1$.
We then generate a list of all possible single nodes satisfying these conditions. The second node to be added is also picked
from this list.

Without demanding the corresponding DD to be ADE, we would have $2^{16}$ ways to connect the new node with the 
$16$ original ones. Since the nodes of ADE diagrams never have more than $3$ links,  the possibilities for the new node 
are reduced to $\sum_{i=0}^3 {16 \choose i} =697$.
Each of these $697$ ways of connecting gives a set of $16$ equations which we use to determine $16$ of the $20$ components of 
the new simple root. We solve the system of equations for $\pi^I$ and ${\pi'}^I$, $I=1,\ldots,8$, 
leaving the four $w^1$, $w^2$, $n_1$ and $n_2$ undetermined.  Afterwards, we compile a list of possible choices for $w^i$ and $n_i$. 
In principle, we could assign to these quantum numbers arbitrarily large values. Since we want to consider many (but finite number of) possibilities, 
we  truncate the possible choices by demanding $|w^i|\leq \lambda_1$ and $|n_i|\leq \lambda_2$, where $\lambda_1$ 
and $\lambda_2$ are two positive integers which we take as input parameters. 
In this example it is necessary to take at least $\lambda_1 = \lambda_2 = 2$, otherwise the algorithm would just not find the 
enhancement to $6\mra_3$. 
Considering the whole set of Wilson lines fixed by the shift algorithm, these bounds give all the maximal enhancements of the $T^2$ compactification except for $3\mra_6$ and $\mra_3 + \mra_6 + \mra_9$.

Some possibilities for the new node are depicted in Figure \ref{dynkin6a3_2}.
The links in cyan or red would give $5\mra_3 + 2\mra_1$, whereas those in blue or magenta 
would give $\mra_7+2\mra_3+4\mra_1$. The gray connections would lead to a DD
which is not ADE and will be discarded later. The orange line implies $\mra_4+3\mra_3+4\mra_1$. 
We could also disconnect the node from everything, obtaining $4\mra_3 + 5\mra_1$. 

\begin{figure}[htb]
\centering
\begin{tikzpicture}[scale=.175]
    \filldraw[cyan,thick] (40cm, 0cm) -- (32cm, 8cm);  
      \filldraw[cyan,thick] (24cm, 0cm) -- (32cm, 8cm);  
  \filldraw[orange,thick] (8cm, 0cm) -- (32cm, 8cm);  
  \filldraw[red,thick] (0cm, 0cm) -- (32cm, 8cm);  
  \filldraw[red,thick] (64cm, 0cm) -- (32cm, 8cm);   
   \filldraw[magenta,thick] (16cm, 0cm) -- (32cm, 8cm);  
  \filldraw[magenta,thick] (48cm, 0cm) -- (32cm, 8cm);
   \filldraw[blue,thick] (60cm, 12cm) -- (32cm, 8cm);  
  \filldraw[blue,thick] (4cm, 12cm) -- (32cm, 8cm);
    \filldraw[gray,thick] (60cm, 8cm) -- (32cm, 8cm);  
  \filldraw[gray,thick] (4cm, 8cm) -- (32cm, 8cm);    
   \filldraw[thick,fill=white] (0 cm,0) circle (5 mm) node [shift={(0.0,-0.5)}] {\mfn{1}} ;    
        \filldraw[thick,fill=white] (8 cm,0) circle (5 mm) node [shift={(0.0,-0.5)}] {\mfn{3}} ;     
  \filldraw[thick,fill=white] (8.5 cm, 0) -- (11.5 cm,0);    
    \filldraw[thick,fill=white] (12 cm,0) circle (5 mm) node [shift={(0.0,-0.5)}] {\mfn{4}};    
  \filldraw[thick,fill=white] (12.5 cm, 0) -- (15.5 cm,0);     
         \filldraw[thick,fill=white] (16 cm,0) circle (5 mm) node [shift={(0.0,-0.5)}] {\mfn{5}};    
         \filldraw[thick,fill=white] (4cm,4cm) circle (5 mm) node [shift={(-0.5,0.0)}] {\mfn{7}};     
 \filldraw[thick,fill=white] (4 cm, 4.5cm) -- (4 cm, 7.5cm);    
         \filldraw[thick,fill=white] (4cm,8cm) circle (5 mm) node [shift={(-0.5,0.0)}] {\mfn{8}};
         \filldraw[thick,fill=white] (4cm, 8.5cm) -- (4cm, 11.5cm);    
         \filldraw[thick,fill=white] (4cm,12cm) circle (5 mm) node [shift={(-0.5,0.0)}] {\mfn{-1}};     
         \filldraw[thick,fill=white] (24cm,0) circle (5 mm) node [shift={(0.0,-0.5)}] {\mfn{0}};   
        \filldraw[thick,fill=white] (40 cm,0) circle (5 mm) node [shift={(0.0,-0.5)}] {\mfn{0'}};    
         \filldraw[thick,fill=white] (48 cm,0) circle (5 mm) node [shift={(0.0,-0.5)}] {\mfn{5'}}; 
  \filldraw[thick,fill=white] (48.5 cm, 0) -- (51.5 cm,0);       
         \filldraw[thick,fill=white] (52cm,0) circle (5 mm) node [shift={(0.0,-0.5)}] {\mfn{4'}};     
  \filldraw[thick,fill=white] (52.5 cm, 0) -- (55.5 cm,0);   
         \filldraw[thick,fill=white] (56cm,0) circle (5 mm) node [shift={(0.0,-0.5)}] {\mfn{3'}};     
   \filldraw[thick,fill=white] (64cm,0) circle (5 mm) node [shift={(0.0,-0.5)}] {\mfn{1'}};     
 \filldraw[thick,fill=white] (60cm,4cm) circle (5 mm) node [shift={(0.5,0.0)}] {\mfn{7'}};     
 \filldraw[thick,fill=white] (60 cm, 4.5cm) -- (60 cm, 7.5cm);    
         \filldraw[thick,fill=white] (60cm,8cm) circle (5 mm) node [shift={(0.5,0.0)}] {\mfn{8'}};
          \filldraw[thick,fill=white] (60cm, 8.5cm) -- (60cm, 11.5cm);    
         \filldraw[thick,fill=white] (60cm,12cm) circle (5 mm) node [shift={(0.5,0.0)}] {\mfn{{-1'}}};
         \filldraw[thick,fill=white] (32cm,8cm) circle (5 mm) node
[shift={(0.0,0.5)}] {\mfn{\tN_1}};
     \end{tikzpicture}
          \caption{Dynkin diagram showing in different colors some (arbitrary) possible connections for the new node $\tN_1$.}\label{dynkin6a3_2}
\end{figure}
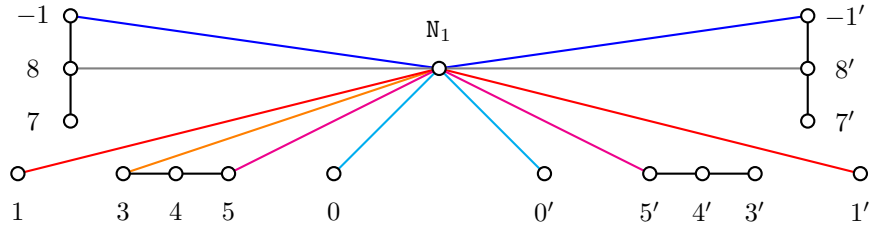

The next step is to determine the charge vector $\varphi_{\tN_1}$  for each of the connections. 
For example, for the cyan connections, putting $\lambda_1=\lambda_2=1$ we identify the candidates
\be
|1, 0, 1, 0; 0^8, 0^8 \rangle \, ,\quad  
|1, 0, 1, \pm 1; 0^8, 0^8 \rangle \, ,\quad
|-1, 0, 1, 0; w_6,w_6' \rangle\, ,\quad
|-1, 0, 1, \pm 1; w_6,w_6' \rangle\, .
\ee
For the red links, no states appear if the bounds $\lambda_1=\lambda_2=1$ are kept. It is thus necessary to 
consider higher winding and momentum numbers. Choosing $\lambda_1=\lambda_2=2$ we find
\begin{eqnarray}
&\begin{array}{ll}
|1, -2, 1, 2; \tilde w_1, \tilde w_1' \rangle \, , & |1, 2, 1, -2; \tilde w_3, \tilde w_3'\rangle \, , \\
|1, -2, -1, 1;\tilde w_1, \tilde w_1'  \rangle \, , & |1, 2, -1, -1; \tilde w_3, \tilde w_3'  \rangle \, , \\ 
|-1, -2, -1, 2; \tilde w_2, \tilde w_2'  \rangle\, , &  |-1, 2, -1, -2; \tilde w_4, \tilde w_4' \rangle \, , \\
|-1, -2, 1, 1; \tilde w_2, \tilde w_2' \rangle\, ,  &  |-1, 2, 1, -1; \tilde w_4, \tilde w_4' \rangle \, ,  
\end{array}
  \\
&\tilde w_1\equiv - w_1 + w_2 - 2 w_6 \ ,  \  \tilde w_2\equiv -w_1 + w_2 - w_6 \ , \  \tilde w_3\equiv -w_1 + w_6  \ , \  \tilde w_4 \equiv -w_1 + 2w_6 \ . \nonumber
\end{eqnarray}
The quantization conditions \eqref{quant2} will be imposed later, thereby determining the $E_{ij}$.

At this stage we have assembled a list of all the possible simple roots that can be added such that the resulting 
DD is admissible, although not necessarily ADE. This means that the Cartan matrix is symmetric, with diagonal elements equal to two,
and off-diagonal elements equal to $0$ or $-1$. In our example, considering all the possible connections, 
and with $\lambda_1=\lambda_2=2$, there are $1082$ possible simple roots. 
From this list we can extract all possible pairs of simple roots that can be adjoined to the original $16$.
The two roots must be compatible, i.e. their inner product must be $0$ or $-1$. We then collect
all the allowed pairs. In the case at hand there are $191501$ such pairs.
For example, some of the possible partners $\varphi_{\tN_2}$ for the simple root
$\varphi_{\tN_1}=|1, 0, 1, 0; 0^8, 0^8 \rangle$ (correlated with the cyan connections) are
\be
\begin{small}
\begin{array}{ll}
 \text{(1)\ }|-1, -2, 1, 1; -w_1 + w_2 - w_6, - w_1' + w_2' - w_6'\rangle, 
& \text{(2)\ } |-1, 0, 1, 0; w_6, w_6' \rangle,  \\
\text{(3)\ }|-1, 1, 1, 0; -w_5 + 2w_6,-w_5' + 2w_6' \rangle,
& \text{(4)\ } |0, 1, 0, 1; 0^8, 0^8 \rangle\,  .
\end{array}\label{4roots}
\end{small}
\ee
The corresponding Dynkin diagrams are shown in Figure \ref{dynkin6a3_3}.
The green connections for the node $\tN_2$ should be discarded because they give an affine $\mra_3$ subdiagram which is not ADE. 
If we choose the pink connections we would have $6\mra_3$, and $\mra_7+3\mra_3+2\mra_1$ if we choose the 
yellow or the brown. Next, for each of the possible pairs, distinguished by two sets of charged vectors $\ket{w^1,w^2,n_1,n_2;\pi}$, 
we substitute in \eqref{quant2} to compute the four components $E_{ij}$.
In all cases we find $E_{11}=1$ and $E_{21}=0$.
For the pink links, $E_{12}=-\frac12$, $E_{22}=1$; for the yellow $E_{12}=-1$, $E_{22}=\frac32$; and for the
brown, $E_{12}=0$, $E_{22}=1$. For the green connections $E_{12}$ and $E_{22}$ remain undetermined, reflecting the fact
that the associated DD is not ADE.
We still have to check that $g_{ij}=\frac12(E_{ij} + E_{ji} - A_i \cdot A_j)$ verifies $g_{ii}>0$, and $\det g>0$. 
In the end we have a list of all consistent pairs 
of simple roots that can be added, with the corresponding moduli. In this example there are $192$ elements on the list.

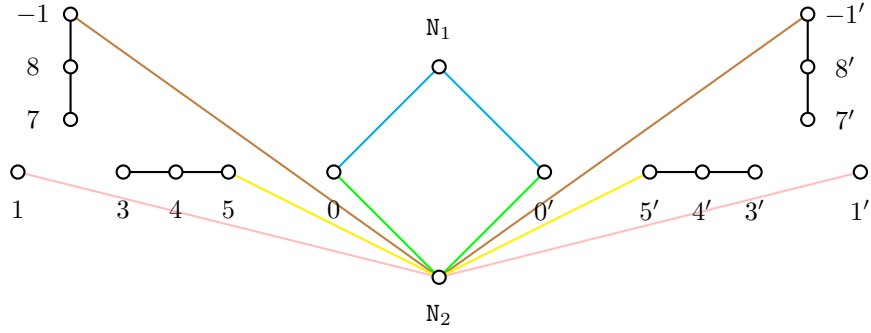
\begin{figure}[htb]
\centering
\begin{tikzpicture}[scale=.175]
    \filldraw[brown,thick] (15*4cm, 12cm) -- (32cm, -8cm);  
      \filldraw[brown,thick] (4cm, 12cm) -- (32cm, -8cm);  
    \filldraw[cyan,thick] (40cm, 0cm) -- (32cm, 8cm);  
      \filldraw[cyan,thick] (24cm, 0cm) -- (32cm, 8cm);  
         \filldraw[green,thick] (40cm, 0cm) -- (32cm, -8cm);  
      \filldraw[green,thick] (24cm, 0cm) -- (32cm, -8cm);  
  \filldraw[pink,thick] (0cm, 0cm) -- (32cm, -8cm);  
  \filldraw[pink,thick] (64cm, 0cm) -- (32cm, -8cm);   
   \filldraw[yellow,thick] (16cm, 0cm) -- (32cm, -8cm);  
  \filldraw[yellow,thick] (48cm, 0cm) -- (32cm, -8cm);           \filldraw[thick,fill=white] (32cm,8cm) circle (5 mm) node
[shift={(0.0,0.5)}] {\mfn{\tN_1}};
           \filldraw[thick,fill=white] (32cm,-8cm) circle (5 mm) node
[shift={(0.0,-0.5)}] {\mfn{\tN_2}};
   \filldraw[thick,fill=white] (0 cm,0) circle (5 mm) node [shift={(0.0,-0.5)}] {\mfn{1}} ;    
        \filldraw[thick,fill=white] (8 cm,0) circle (5 mm) node [shift={(0.0,-0.5)}] {\mfn{3}} ;     
  \filldraw[thick,fill=white] (8.5 cm, 0) -- (11.5 cm,0);    
    \filldraw[thick,fill=white] (12 cm,0) circle (5 mm) node [shift={(0.0,-0.5)}] {\mfn{4}};    
  \filldraw[thick,fill=white] (12.5 cm, 0) -- (15.5 cm,0);     
         \filldraw[thick,fill=white] (16 cm,0) circle (5 mm) node [shift={(0.0,-0.5)}] {\mfn{5}};    
         \filldraw[thick,fill=white] (4cm,4cm) circle (5 mm) node [shift={(-0.5,0.0)}] {\mfn{7}};     
 \filldraw[thick,fill=white] (4 cm, 4.5cm) -- (4 cm, 7.5cm);    
         \filldraw[thick,fill=white] (4cm,8cm) circle (5 mm) node [shift={(-0.5,0.0)}] {\mfn{8}};
         \filldraw[thick,fill=white] (4cm, 8.5cm) -- (4cm, 11.5cm);    
         \filldraw[thick,fill=white] (4cm,12cm) circle (5 mm) node [shift={(-0.5,0.0)}] {\mfn{-1}};     
         \filldraw[thick,fill=white] (24cm,0) circle (5 mm) node [shift={(0.0,-0.5)}] {\mfn{0}};   
        \filldraw[thick,fill=white] (40 cm,0) circle (5 mm) node [shift={(0.0,-0.5)}] {\mfn{0'}};    
         \filldraw[thick,fill=white] (48 cm,0) circle (5 mm) node [shift={(0.0,-0.5)}] {\mfn{5'}}; 
  \filldraw[thick,fill=white] (48.5 cm, 0) -- (51.5 cm,0);       
         \filldraw[thick,fill=white] (52cm,0) circle (5 mm) node [shift={(0.0,-0.5)}] {\mfn{4'}};     
  \filldraw[thick,fill=white] (52.5 cm, 0) -- (55.5 cm,0);   
         \filldraw[thick,fill=white] (56cm,0) circle (5 mm) node [shift={(0.0,-0.5)}] {\mfn{3'}};     
   \filldraw[thick,fill=white] (64cm,0) circle (5 mm) node [shift={(0.0,-0.5)}] {\mfn{1'}};     
 \filldraw[thick,fill=white] (60cm,4cm) circle (5 mm) node [shift={(0.5,0.0)}] {\mfn{7'}};     
 \filldraw[thick,fill=white] (60 cm, 4.5cm) -- (60 cm, 7.5cm);    
         \filldraw[thick,fill=white] (60cm,8cm) circle (5 mm) node [shift={(0.5,0.0)}] {\mfn{8'}};
          \filldraw[thick,fill=white] (60cm, 8.5cm) -- (60cm, 11.5cm);    
         \filldraw[thick,fill=white] (60cm,12cm) circle (5 mm) node [shift={(0.5,0.0)}] {\mfn{{-1'}}};
     \end{tikzpicture}
     \caption{Dynkin diagram showing the chosen connection for the node $\tN_1$ (in cyan, corresponding to the root 
     $|1,0,1,0;0^8,0^8 \rangle$) and the possible connections for the node $\tN_2$ 
     (in pink, green, yellow and brown; corresponding respectively to the roots (1)-(4) in \eqref{4roots}).}
     \label{dynkin6a3_3}
\end{figure}

We finally deduce the gauge group from the $18$ simple roots. We developed a routine that takes a base of 
simple roots and detects if its Dynkin diagram is of ADE type and, in that case, it identifies the group. 
We also compute the Gram matrix $Q$ corresponding to the moduli, as explained in section \ref{ss:ldata}. 
We apply this algorithm to all the elements in our list.
In our example, this process yields $53$ maximal enhancement points, but there are only $3$ inequivalent enhancements 
because $50$ of these points are T-dual to the $3$ presented in Table \ref{table_6a3}.
The corresponding diagrams are displayed in Figure \ref{dynkin6a3_4}. 

\begin{table}[h!]\begin{center}
\renewcommand{\arraystretch}{1.5}
\begin{tabular}{|cc|c|c|c|c|}
\hline
$\tN_1$ & $\tN_2$ & $L$ &  $E$ & $g$ & $Q$\\
\hline
\red{red} & \green{green} & $6\mra_3$ & $\left(\begin{smallmatrix}1 & -\frac12 \\ 0 & 1 \end{smallmatrix}\right)$  & $\left(\begin{smallmatrix} \frac12 & -\frac14 \\ -\frac14 & \frac14 \end{smallmatrix}\right)$
& $\left(\begin{smallmatrix} 4 & 0 \\ 0 & 4 \end{smallmatrix}\right)$
 \\
  \blue{blue} & \green{green} & $2 \mra_1 + 3 \mra_3 + \mra_7$ & $\left(\begin{smallmatrix}1 & 0 \\ 0 & 1 \end{smallmatrix}\right)$& $\left(\begin{smallmatrix} \frac12 & 0 \\ 0 & \frac14 \end{smallmatrix}\right)$
  & $\left(\begin{smallmatrix} 4 & 0 \\ 0 & 8 \end{smallmatrix}\right)$ \\
 \blue{blue} & \yellow{yellow}& $4 \mra_1 + 2 \mra_7$ & $\left(\begin{smallmatrix}1 & 0 \\ -\frac12 & 1 \end{smallmatrix}\right)$ & $\left(\begin{smallmatrix} \frac12 & -\frac14 \\ -\frac14 & \frac14 \end{smallmatrix}\right)$& $\left(\begin{smallmatrix} 4 & 0 \\ 0 & 4 \end{smallmatrix}\right)$\\
 \hline
 \end{tabular}
\caption{Three maximal enhancement points for the Wilson lines given in \eqref{wlexample1} and different values of $E$. 
The torus metric $g$ and the Gram matrix $Q$ of the complementary lattice $T$ are also given.}
  \label{table_6a3}\end{center}\end{table}

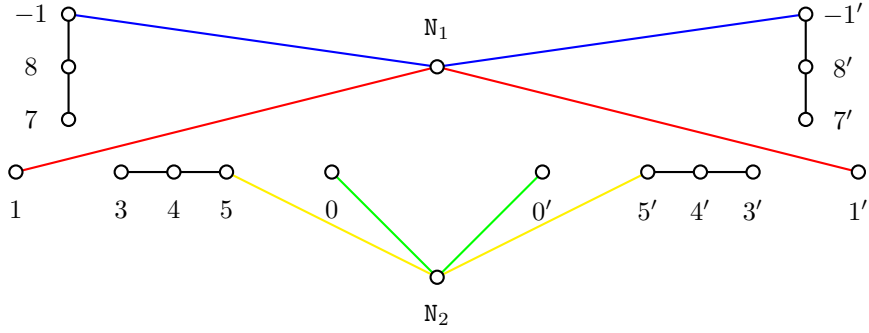
\begin{figure}[htb]
\centering
\begin{tikzpicture}[scale=.175]
    \filldraw[blue,thick] (15*4cm, 12cm) -- (32cm, 8cm);  
      \filldraw[blue,thick] (4cm, 12cm) -- (32cm, 8cm);  
    \filldraw[green,thick] (40cm, 0cm) -- (32cm, -8cm);  
      \filldraw[green,thick] (24cm, 0cm) -- (32cm, -8cm);  
  \filldraw[red,thick] (0cm, 0cm) -- (32cm, 8cm);  
  \filldraw[red,thick] (64cm, 0cm) -- (32cm, 8cm);   
   \filldraw[yellow,thick] (16cm, 0cm) -- (32cm, -8cm);  
  \filldraw[yellow,thick] (48cm, 0cm) -- (32cm, -8cm);           \filldraw[thick,fill=white] (32cm,8cm) circle (5 mm) node
[shift={(0.0,0.5)}] {\mfn{\tN_1}};
           \filldraw[thick,fill=white] (32cm,-8cm) circle (5 mm) node
[shift={(0.0,-0.5)}] {\mfn{\tN_2}};
   \filldraw[thick,fill=white] (0 cm,0) circle (5 mm) node [shift={(0.0,-0.5)}] {\mfn{1}} ;    
        \filldraw[thick,fill=white] (8 cm,0) circle (5 mm) node [shift={(0.0,-0.5)}] {\mfn{3}} ;     
  \filldraw[thick,fill=white] (8.5 cm, 0) -- (11.5 cm,0);    
    \filldraw[thick,fill=white] (12 cm,0) circle (5 mm) node [shift={(0.0,-0.5)}] {\mfn{4}};    
  \filldraw[thick,fill=white] (12.5 cm, 0) -- (15.5 cm,0);     
         \filldraw[thick,fill=white] (16 cm,0) circle (5 mm) node [shift={(0.0,-0.5)}] {\mfn{5}};    
         \filldraw[thick,fill=white] (4cm,4cm) circle (5 mm) node [shift={(-0.5,0.0)}] {\mfn{7}};     
 \filldraw[thick,fill=white] (4 cm, 4.5cm) -- (4 cm, 7.5cm);    
         \filldraw[thick,fill=white] (4cm,8cm) circle (5 mm) node [shift={(-0.5,0.0)}] {\mfn{8}};
         \filldraw[thick,fill=white] (4cm, 8.5cm) -- (4cm, 11.5cm);    
         \filldraw[thick,fill=white] (4cm,12cm) circle (5 mm) node [shift={(-0.5,0.0)}] {\mfn{-1}};     
         \filldraw[thick,fill=white] (24cm,0) circle (5 mm) node [shift={(0.0,-0.5)}] {\mfn{0}};   
        \filldraw[thick,fill=white] (40 cm,0) circle (5 mm) node [shift={(0.0,-0.5)}] {\mfn{0'}};    
         \filldraw[thick,fill=white] (48 cm,0) circle (5 mm) node [shift={(0.0,-0.5)}] {\mfn{5'}}; 
  \filldraw[thick,fill=white] (48.5 cm, 0) -- (51.5 cm,0);       
         \filldraw[thick,fill=white] (52cm,0) circle (5 mm) node [shift={(0.0,-0.5)}] {\mfn{4'}};     
  \filldraw[thick,fill=white] (52.5 cm, 0) -- (55.5 cm,0);   
         \filldraw[thick,fill=white] (56cm,0) circle (5 mm) node [shift={(0.0,-0.5)}] {\mfn{3'}};     
   \filldraw[thick,fill=white] (64cm,0) circle (5 mm) node [shift={(0.0,-0.5)}] {\mfn{1'}};     
 \filldraw[thick,fill=white] (60cm,4cm) circle (5 mm) node [shift={(0.5,0.0)}] {\mfn{7'}};     
 \filldraw[thick,fill=white] (60 cm, 4.5cm) -- (60 cm, 7.5cm);    
         \filldraw[thick,fill=white] (60cm,8cm) circle (5 mm) node [shift={(0.5,0.0)}] {\mfn{8'}};
          \filldraw[thick,fill=white] (60cm, 8.5cm) -- (60cm, 11.5cm);    
         \filldraw[thick,fill=white] (60cm,12cm) circle (5 mm) node [shift={(0.5,0.0)}] {\mfn{{-1'}}};
     \end{tikzpicture}
     \centering
          \caption{Dynkin diagrams for the maximal enhancements in Table \ref{table_6a3}}\label{dynkin6a3_4}
\end{figure} 

In general, there will be various pairs $(\tN_1,\tN_2)$ that return the same moduli.
In the simplest case, all corresponding sets of $18$ simple roots will have the same Dynkin diagram and, 
in consequence, the same gauge group. In this situation we simply discard all except one of the pairs. 
However, in some cases there might be pairs that, combined with the $16$ original roots, actually give a subgroup of the 
real group which is obtained with different $(\tN_1,\tN_2)$ but same moduli. This is the same problem
noticed at the end of section \ref{subsec:eddtriv}. The solution in this situation is to keep only one of the pairs 
belonging to the group of highest dimension.

\subsubsection{Neighborhood algorithm}
\label{ss:neighbor}

The previous algorithm starts with fixed Wilson lines that determine 16 initial
simple roots. It is then plausible to search for consistent ways of adding 
two nodes to the original Dynkin diagram, deducing in the process the remaining $E_{ij}$ moduli. 
If we do not want to make any assumptions on the $A_i$, nor the $E_{ij}$, 
for a procedure based on adding nodes to be feasible, it would be necessary to know beforehand most of the simple roots. 
The new Neighborhood algorithm goes in this direction. 

The main idea is to find new maximal enhancements that are {\it close} to those already found, but whose Wilson lines are not 
necessarily given by the shift algorithm. More precisely, we start at a point of maximal enhancement where the group $\rG_{18}$,
and its $18$ simple roots, are known. Then we move along surfaces in moduli space where the symmetry is broken to
$\rG_{17} \times \uo$. On each of these $18$ surfaces $\rG_{17}$ will have $17$ of the $18$ original simple roots. 
For each surface we collect the candidate extra simple root that would give back an ADE group of rank $18$. 
For each candidate we compute the moduli, $A_i$ and $E_{ij}$, by imposing that the $18$ simple roots correspond to states
that satisfy the massless conditions \eqref{massless2}. We then check that the torus metric $g_{ij}$ is well defined and
finally read the gauge group from the simple roots.
We end with a list of points of maximal enhancement that are on the neighborhood of the original point, i.e. they are connected 
through a $17$-dimensional enhancement surface. The algorithm can be repeated to explore regions of the moduli space that are far away
from the starting point.

We  illustrate the algorithm with an example defined by the starting point
$A_1=A_2=0$, $E_{ij}=\delta_{ij}$, where the gauge group is  $2\mra_1 + 2\mre_8$. The 
charge vectors of the $18$ simple roots are
\be
\begin{small}
\begin{array}{ll}
\varphi_j= \,\, |0, 0, 0, 0; \alpha_j, 0^8 \rangle \, ,\qquad &
\varphi'_{j}= \,\, |0, 0, 0, 0; 0^8, \alpha'_j \rangle \, ,\quad
j=1,\,\ldots, 8, \\
\varphi_{\tC_1}= \,\, |1, 0, 1, 0; 0^8, 0^8 \rangle \, ,\qquad &
\varphi_{\tC_2}= \,\, |0, 1, 0, 1; 0^8, 0^8 \rangle\, .  \end{array}
\end{small}
\label{nodesex}
\ee
They form the DD of Figure \ref{edd2:1}, with the nodes $0$ and $0'$ deleted.
Now we want to move along directions that preserve $17$ of the $18$ simple roots by deleting one node.
Since the DD is symmetric under the interchange of the node $[j]$ with the node $[j']$,
it suffices to remove one of the nodes $[j]$. 
We are then effectively breaking $\mre_8 + 2\mra_1$ by erasing one node.
The nodes $\tC_1$ and $\tC_2$ are also interchangeable. We choose to always keep $\tC_2$.
There are thus only $9$ inequivalent breakings, obtained by deleting either $\tC_1$ or one of the
8 nodes of $\mre_8$. 
Altogether, the $17$ surviving simple roots are the $18$ original ones in \eqref{nodesex}, except for the one corresponding to 
the removed node. 
Afterwards we add a new node which clearly cannot be connected to any
of the $8$ nodes $[j']$ associated to the second $\mre_8$, since the resulting diagram has to be of type ADE.
Hence, only algebras of the form $\rG_{10} + \mre_8$ can arise. For convenience we  ignore the second $\mre_8$ unless otherwise stated.

To further elaborate on the algorithm we analyze first the case in which the node $\tC_1$ is removed. The effect
is simply to break $\mre_8 + 2\mra_1$ to $\mre_8 + \mra_1$. 
We then add one node, called $\tN$, to its Dynkin diagram. The $2$ possibilities for the connections of the new node are 
displayed in Figure \ref{dynkina1+e8}. Generically, the charge vector corresponding to $\tN$ is
\be
\varphi_\tN = \ket{w^1, w^2, n_1, n_2; \pi^1,\ldots,\pi^8, 0^8} \, .
\label{phiN}
\ee
The last 8 components of $\pi$ are zero just because the new node is always disconnected from the second $\mre_8$. 
The way that $\tN$ is linked in each of the
possible Dynkin diagrams gives $9$ conditions for the $12$ unknowns $w^i$, $n_i$, plus 
the eight non-zero components of $\pi$. 
We use these conditions to determine all except $3$ of the unknowns. It is convenient, and always possible, to leave $w^1$ and $w^2$  undetermined. 

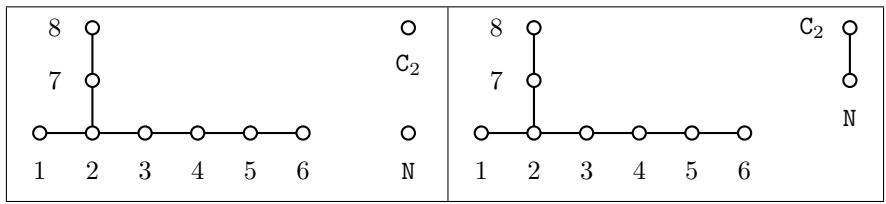
\begin{figure}[htb]
\centering
\begin{tabular}{|c|c|}
\hline
 \begin{tikzpicture}[scale=.175]
   \filldraw[thick,fill=white] (0 cm,0) circle (5 mm) node [shift={(0.0,-0.5)}] {\mfn{1}} ;
     \filldraw[thick,fill=white] (0.5 cm, 0) -- (3.5 cm,0);  
      \filldraw[thick,fill=white] (4 cm,0) circle (5 mm) node [shift={(0.0,-0.5)}] {\mfn{2}} ;  
        \filldraw[thick,fill=white] (4.5 cm, 0) -- (7.5 cm,0);    
          \filldraw[thick,fill=white] (4 cm, 0.5 cm) -- (4 cm, 3.5 cm);  
        \filldraw[thick,fill=white] (8 cm,0) circle (5 mm) node [shift={(0.0,-0.5)}] {\mfn{3}} ;     
  \filldraw[thick,fill=white] (8.5 cm, 0) -- (11.5 cm,0);    
    \filldraw[thick,fill=white] (12 cm,0) circle (5 mm) node [shift={(0.0,-0.5)}] {\mfn{4}};    
  \filldraw[thick,fill=white] (12.5 cm, 0) -- (15.5 cm,0);     
         \filldraw[thick,fill=white] (16 cm,0) circle (5 mm) node [shift={(0.0,-0.5)}] {\mfn{5}};
           \filldraw[thick,fill=white] (16.5 cm, 0) -- (19.5 cm,0);  
            \filldraw[thick,fill=white] (20 cm,0) circle (5 mm) node [shift={(0.0,-0.5)}] {\mfn{6}} ;    
         \filldraw[thick,fill=white] (4cm,4cm) circle (5 mm) node [shift={(-0.5,0.0)}] {\mfn{7}};     
 \filldraw[thick,fill=white] (4 cm, 4.5cm) -- (4 cm, 7.5cm);    
         \filldraw[thick,fill=white] (4cm,8cm) circle (5 mm) node [shift={(-0.5,0.0)}] {\mfn{8}}; 
        \filldraw[thick,fill=white] (28 cm,8) circle (5 mm) node [shift={(0.0,-0.5)}] {\mfn{\tC_2}};
                \filldraw[thick,fill=white] (28 cm,0) circle (5 mm) node [shift={(0.0,-0.5)}] {\mfn{\tN}};    
     \end{tikzpicture} & 
      \begin{tikzpicture}[scale=.175]
   \filldraw[thick,fill=white] (0 cm,0) circle (5 mm) node [shift={(0.0,-0.5)}] {\mfn{1}} ;
     \filldraw[thick,fill=white] (0.5 cm, 0) -- (3.5 cm,0);  
      \filldraw[thick,fill=white] (4 cm,0) circle (5 mm) node [shift={(0.0,-0.5)}] {\mfn{2}} ;  
        \filldraw[thick,fill=white] (4.5 cm, 0) -- (7.5 cm,0);    
          \filldraw[thick,fill=white] (4 cm, 0.5 cm) -- (4 cm, 3.5 cm);  
        \filldraw[thick,fill=white] (8 cm,0) circle (5 mm) node [shift={(0.0,-0.5)}] {\mfn{3}} ;     
  \filldraw[thick,fill=white] (8.5 cm, 0) -- (11.5 cm,0);    
    \filldraw[thick,fill=white] (12 cm,0) circle (5 mm) node [shift={(0.0,-0.5)}] {\mfn{4}};    
  \filldraw[thick,fill=white] (12.5 cm, 0) -- (15.5 cm,0);     
         \filldraw[thick,fill=white] (16 cm,0) circle (5 mm) node [shift={(0.0,-0.5)}] {\mfn{5}};
           \filldraw[thick,fill=white] (16.5 cm, 0) -- (19.5 cm,0);  
            \filldraw[thick,fill=white] (20 cm,0) circle (5 mm) node [shift={(0.0,-0.5)}] {\mfn{6}} ;    
         \filldraw[thick,fill=white] (4cm,4cm) circle (5 mm) node [shift={(-0.5,0.0)}] {\mfn{7}};     
 \filldraw[thick,fill=white] (4 cm, 4.5cm) -- (4 cm, 7.5cm);    
         \filldraw[thick,fill=white] (4cm,8cm) circle (5 mm) node [shift={(-0.5,0.0)}] {\mfn{8}}; 
        \filldraw[thick,fill=white] (28 cm,8) circle (5 mm) node [shift={(-0.5,0.0)}] {\mfn{\tC_2}};  
 \filldraw[thick,fill=white] (28 cm, 4.5cm) -- (28 cm, 7.5cm);    
         \filldraw[thick,fill=white] (28 cm,4) circle (5 mm) node [shift={(0.0,-0.5)}] {\mfn{\tN}};  
     \end{tikzpicture}\\ \hline
     \end{tabular}
         \caption{Dynkin diagrams corresponding to the possible ways of adding a node $\tN$ to the diagram of $\mre_8+\mra_1$.}\label{dynkina1+e8}
\end{figure}

The following steps are very similar to those in the algorithm described in \ref{ss:swl}. We just consider all possible values for the $3$ unknowns, with a fixed bound for the maximum of their absolute values. As in the previous section, 
for computational reasons, this truncation is necessary to avoid infinitely many possibilities. 
Concretely, we introduce two parameters $\lambda_1$ and $\lambda_2$, which define the truncation, and consider only states with
\be 
|w^i|\, \leq \lambda_1,\quad |n_i|\, \leq \lambda_2,\quad |\pi^I| \, \leq \lambda_2 \, .
\label{trunca}
\ee 
For this example it is enough to use $\lambda_1=1$ and $\lambda_2=2$. Afterwards, we filter all the candidates by imposing that 
$\varphi_\tN$ has norm squared $2$ and $\pi\in \Upsilon_{16}$. In some cases it might occur that, regardless of the values
of $\lambda_1$ and $\lambda_2$, there are actually no solutions with $w^i, n_i \in \ZZ$ and $\pi\in \Upsilon_{16}$.

The case of $\mre_8+2\mra_1$, on the left in Figure \ref{dynkina1+e8}, is rather trivial because we are just restoring the 
deleted node $\tC_1$. 
The algorithm will find charge vectors $\varphi_\tN$ which are not necessarily equal to $\varphi_{\tC_1}$,
but at the end of the day all of them should be equivalent to it. When we compute the moduli we obviously 
get $E_{ij}=\delta_{ij}$, $A_1=A_2=0$, or some T-dual point. 
We just restored the simple root that we removed, thus returning to the original point in the moduli space. In general, 
this possibility will occur in all the breakings.

In the less trivial case $\mre_8 + \mra_2$, on the right of Figure \ref{dynkina1+e8}, $\tN$ is  linked to $\tC_2$.
Imposing $\langle \varphi_{\tC_2} |\varphi_{\tN}\rangle = -1$, implies $n_2 = -1 - w^2$. 
Considering all the possible values for the $3$ unknowns $w^1$, $w^2$ and  $n_1$, with 
the bounds in \eqref{trunca}, and filtering by requiring $\langle \varphi_\tN |\varphi_{\tN}\rangle =2$, gives the list
\vspace*{-5mm}
\be
\small{
|1, 0, 1, -1; 0^8, 0^8 \rangle,\quad
|-1, -1, -1, 0; 0^8, 0^8 \rangle,\quad
|1, -1, 1, 0; 0^8, 0^8 \rangle, \quad
|-1, 0, -1, -1; 0^8, 0^8\rangle \, .
}
\label{extraex1}
\ee 
We next deduce the moduli by demanding that the charge vectors of the full set of $18$ simple roots satisfy the 
quantization conditions \eqref{quant2}.
This is a well posed problem because in general there are $36$ moduli to determine and the $18$ simple roots give two equations each. 
In this case we readily find $A_1=0$ and $A_2=0$.
From  $\varphi_{\texttt{C}_2}$ we obtain  $E_{12}=0$ and $E_{22}=1$, whereas from $\varphi_\tN$, 
$n_1=E_{11} w^1$ and $-2 w^2-1=E_{21} w_1$.
The $4$ elements in the list \eqref{extraex1} solve these equations with $E_{11}=1$, and
$E_{21}$ equal to $1$ or $-1$. It is easy to see that the corresponding $g_{ij}$ is well defined and that these points are T-dual
to each other.

The algorithm proceeds in the same fashion for all the $9$ possible breakings of $\mre_8 + 2\mra_1$. For a more fruitful example, let
us consider the breaking to $\mra_7+2\mra_1$, obtained by removing the node $\varphi_1$. Appending a new node $\tN$ 
leads to various possible enhancements. 
For instance, $\tN$ can connect only to $\varphi_8$ to form $\mra_8 + 2 \mra_1$. 
With $\lambda_1=1$ and $\lambda_2=2$ in the bounds \eqref{trunca}, we find that the charge vectors of $\varphi_N$ can be one of
 \be
 \small{
  |-1,0,1,0; -w_8,0^8 \rangle\,,\quad
   |0,-1,0,1; -w_8,0^8 \rangle\,,\quad 
     |1,0,-1,0; -w_8,0^8 \rangle\,,\quad
   |0,1,0,-1; -w_8,0^8 \rangle\, .
   \label{extraex2}
   }
 \ee
The moduli are determined as explained before. Taking into account all nodes except $\tN$, we arrive at 
\be
E_{ij}=\delta_{ij}, \quad A_1=\gamma_1 w_1 \times 0, \quad A_2=\gamma_2 w_1 \times 0 \, ,
\label{moduliex2}
\ee
where $(\gamma_1, \gamma_2)$ are some free parameters. The above moduli determine a slice of moduli space with 
group $\sug(8)\times \sug(2)^2 \times \mre_8 \times \uo$.
Finally imposing the quantization conditions \eqref{quant2} to the possible charge vectors for $\varphi_\tN$,
cf. \eqref{extraex2},  fixes $(\gamma_1,\gamma_2)= \pm(\underline{\frac25,0})$, where underlining means permutations.
With these values we reach the rank 18 group with algebra $\mra_8 + 2\mra_1 + \mre_8$.

There is a feature of the algorithm than can be explained considering again the enhancing to 
$\mra_8 + 2 \mra_1$, but now with $\mra_8$ formed 
by connecting $\varphi_\tN$ to $\varphi_6$.
The algorithm finds the charge vector $|0,0,0,0; -w_6,0^8 \rangle$ for $\varphi_\tN$. The moduli are again of
the form \eqref{moduliex2}, but now the quantization conditions from $\varphi_\tN$ imply $(\gamma_1,\gamma_2)=(0,0)$.
Thus, the predicted moduli are $A_1=A_2=0$, $E=\delta_{ij}$, and we know that this point has 
trivial enhancement to $2\mre_8+2\mra_1$. On the other hand, the Dynkin diagram that results
adding $\tN$ indicates enhancement to $\mra_8 + 2\mra_1 + \mre_8$. The problem here is that the 
$\varphi_\tN$, which has zero winding and momenta, corresponds to a root of $\mre_8$. In fact,
$-w_6=\alpha_0$ is the lowest root. 
Since the quantization conditions are linear equations, if we replace one of the original simple roots of $2\mre_8 + 2 \mra_1$ with 
any other root, the moduli that solve the system will be the same, but the other root is no longer simple.
This is the same issue discussed at the end of section \ref{subsec:eddtriv}.
Our prescription to  solve it is to classify all the enhancements, originating from the same starting point, by the resulting moduli.  
If there is more than one enhancement for the same moduli we just pick the one with higher dimensional group. In this case, 
we choose $2\mre_8 + 2 \mra_1$ over $\mre_8 + \mra_8 + 2 \mra_1$.

In Table \ref{vecindad} we collect the maximal enhancements in the neighborhood of the original point
$A_1=A_2=0$, $E=\delta_{ij}$, which has $\rG_{18}=2\mre_8 + 2\mra_1$. 
The node shown in the first column is removed from the set in \eqref{nodesex} at the start.
The effect is to break $\rG_{18}$ to $\rG_9 \times \mre_8 \times \uo$, with $\rG_9$ given
in the second column. Appending a new node then leads to  $\rG_{10} \times \mre_8$,
with the various possibilities for $\rG_{10}$ listed in the third column.  
To arrive at this list we have only kept the groups of higher dimension as explained before,
and we have used $\lambda_1=1$ and $\lambda_2=2$ in the bounds in \eqref{trunca}.  

\begin{table}[htb]\begin{center}
\renewcommand{\arraystretch}{1.25}
\setlength\tabcolsep{2pt}
\small{
\begin{tabular}{|M{1cm}|c|c|}\hline
\tiny{deleted node} & $\rG_9$ & $\rG_{10}$ \\ \hline
$\tC_1$ & $\mre_8 + \mra_1$ &  $\mre_8+2\mra_1$, \ $\mre_8+\mra_2$ \\ \hline
1 & 
$\mra_7 + 2 \mra_1$ & $\mra_9+\mra_1,\ \mra_8 + 2 \mra_1$, \ $\mrd_{10}$ \\ \hline
2 &
$\mra_4 + \mra_2 + 3\mra_1$ & $\mrd_7 + \mra_2 + \mra_1,\ \mrd_5 + \mra_4 + \mra_1,\ \mra_6 + \mra_2 + 2\mra_1,\ 
2 \mra_4 + 2 \mra_1$ \\ \hline
3 &
$\mra_4 + \mra_3 + 2\mra_1$ & $\mrd_6 + \mra_4, \mra_8 + 2 \mra_1, \mra_6 + \mra_3 + \mra_1, \mre_6 + \mra_3 + \mra_1, 
\mra_5 + \mra_4 + \mra_1$ 
\\ \hline
4 &
$\mrd_5 + \mra_2 + 2\mra_1$ & $2 \mrd_5,\ \mrd_7 + \mra_2 + \mra_1,\ \mre_7+\mra_2 + \mra_1,\ 
\mrd_5 + \mra_4 + \mra_1$  \\ \hline
5 &
$\mre_6 + 3\mra_1$ & $\mre_6 + \mrd_4,\ \mre_6+ \mra_3+\mra_1$ \\ \hline
6  & 
$\mre_7 + 2 \mra_1$ & $\mre_7 + \mra_3, \mre_7 + \mra_2 + \mra_1$
\\ \hline
7 & 
$\mra_6 + 3\mra_1$ & $\mrd_9 + \mra_1,\ \mra_8+2\mra_1,\ \mra_6 + \mra_3 + \mra_1, \ \mra_6 + \mra_2 + 2 \mra_1$ \\ \hline
8 & 
$\mrd_7 + 2 \mra_1$ & $\mrd_9+\mra_1,\ \mrd_7 + \mra_2 + \mra_1$ 
\\ \hline
\end{tabular}
}
\caption{Maximal enhancements $\rG_{10} +\mre_8$ in the neighborhood of $A_1=0$, $A_2=0$, $E_{ij}=\delta_{ij}$,
found setting $\lambda_1=1$ and $\lambda_2=2$ in the bounds of \eqref{trunca}.}
 \label{vecindad}\end{center}\end{table}

The Neighborhood algorithm can be iterated and can ramify from a different point of maximal rank. In particular, in this way
we can find the maximal enhancements $\mra_3+\mra_6+\mra_9$ and $3\mra_6$, which, as we have argued, cannot be deduced using the algorithm with fixed Wilson lines. To this end we will set the bounds \eqref{trunca} as before. We will see that this is enough
to obtain the missing groups, although a priori there was no guarantee for it.
We now start at a point with group $\rG_{18}=\mra_6 + \mra_3 + \mra_1 + \mre_8$, which in turn was found by the algorithm 
initiating from the point $E_{ij}=\delta_{ij}$, $A_1=0$, $A_2=0$, cf. Table \ref{vecindad}. 
Concretely, $\rG_{18}$ arises after deleting the node $\varphi_3$ in \eqref{nodesex} and then appending
the extra node $\tN$ with charge vector $\varphi_\tN=|0,-1,-1,1; w_3 - w_1,0^8 \rangle$.
The corresponding moduli are $A_1=-\frac18 w_3 \times 0$, $A_2=\frac14 w_3 \times 0$, $E_{ij}=\delta_{ij}$.
We can now readily apply the algorithm to $\rG_{18}$ whose Dynkin diagram is shown in Figure \ref{dynkin3a6}.a. 
All the enhancement points on the neighborhood of this point can be computed. However, to reach the desired maximal enhancements, 
the nodes $\tC_1$ and $\tC_2$ will be maintained during the whole process. Therefore, $E_{ij}$ will remain equal to the identity 
as we move through the neighborhood.
To proceed we remove the node $1'$, thereby breaking $\rG_{18}$ to $\rG_{17} \times \uo$, with
$\rG_{17}=\mra_1+\mra_3+\mra_6+\mra_7$, as shown in Figure \ref{dynkin3a6}.b.
The neighboring point is on the surface characterized by $A_1=-\frac18 w_3 \times \gamma_1 w'_1$, 
$A_2=\frac14 w_3 \times \gamma_2 w'_1$. The algorithm then searches for new nodes that can be consistently added.
It finds $\tN'$ with charge vector $|-1,-1,1,0; 0^8,-w_8' \rangle$, which leads to $\mra_3+\mra_6+\mra_9$, as seen
in Figure \ref{dynkin3a6}.c.  The point is $(\gamma_1,\gamma_2)=(-\frac25,-\frac{1}{5})$. 
Luckily, from this point we can attain $3\mra_6$ in a couple of steps. With the algorithm it is easy to see what is needed. 
As displayed in Figure \ref{dynkin3a6}.d, the node $8'$ is next removed to break the symmetry to 
$2\mra_6 + \mra_3 + \mra_2$, plus $\uo$. The surface is given by 
$A_1=-\frac18 w_3 \times (-\frac12 w_8'+ \mu_1(4w_1'-5 w_8'))$, 
$A_2=\frac14 w_3\times (-\frac14 w_8'+\mu_2(4w_1'-5 w_8'))$. 
The algorithm then discovers the extra node $\texttt{S}$, with charge vector
$|-1,0,1,-1; -w_6, w_8'-w_1' \rangle$, which has enhancement to $3 \mra_6$, as indicated in Figure \ref{dynkin3a6}.e.
The point is $(\mu_1,\mu_2)=(-\frac18,0)$.

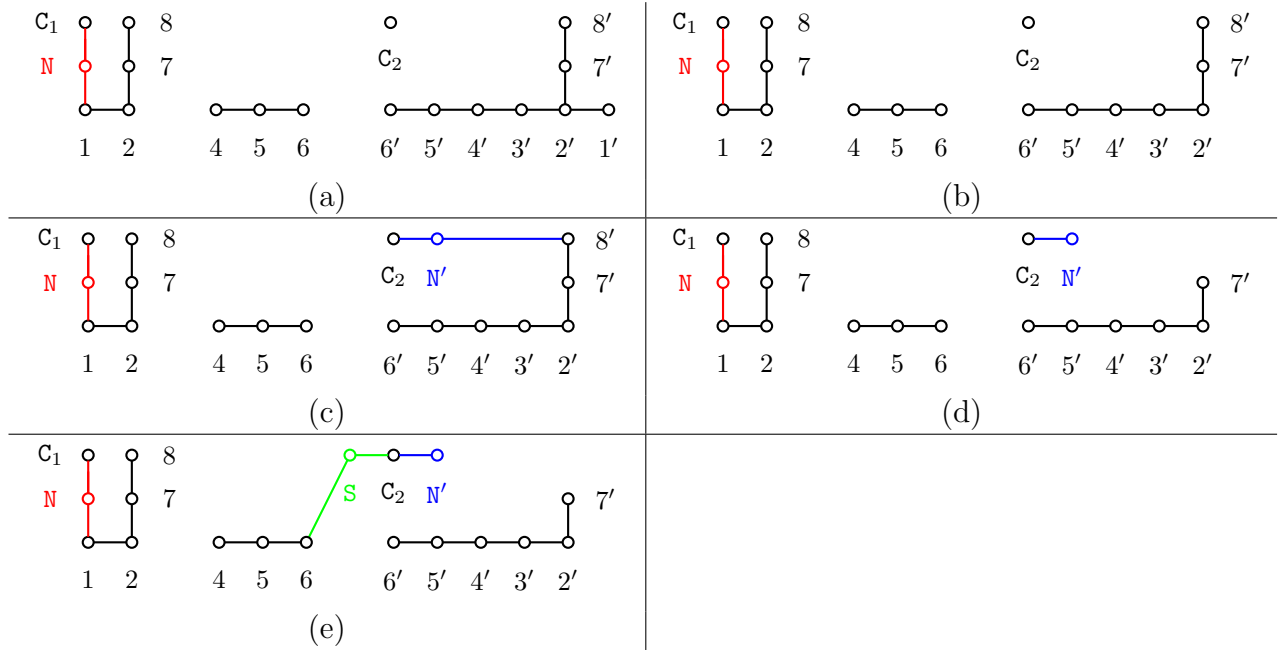
\begin{figure}[htb]
\centering
\begin{tabular}{c|c}
      \begin{tikzpicture}[scale=.145]
   \filldraw[thick,fill=white] (0 cm,0) circle (5 mm) node [shift={(0.0,-0.5)}] {\mfn{1}} ;
     \filldraw[thick,fill=white] (0.5 cm, 0) -- (3.5 cm,0);  
      \filldraw[thick,fill=white] (4 cm,0) circle (5 mm) node [shift={(0.0,-0.5)}] {\mfn{2}} ;  
          \filldraw[thick,fill=white] (4 cm, 0.5 cm) -- (4 cm, 3.5 cm);  
     \filldraw[thick,fill=white] (12 cm,0) circle (5 mm) node [shift={(0.0,-0.5)}] {\mfn{4}};    
  \filldraw[thick,fill=white] (12.5 cm, 0) -- (15.5 cm,0);     
         \filldraw[thick,fill=white] (16 cm,0) circle (5 mm) node [shift={(0.0,-0.5)}] {\mfn{5}};
           \filldraw[thick,fill=white] (16.5 cm, 0) -- (19.5 cm,0);  
            \filldraw[thick,fill=white] (20 cm,0) circle (5 mm) node [shift={(0.0,-0.5)}] {\mfn{6}} ;    
         \filldraw[thick,fill=white] (4cm,4cm) circle (5 mm) node [shift={(0.5,0.0)}] {\mfn{7}};     
 \filldraw[thick,fill=white] (4 cm, 4.5cm) -- (4 cm, 7.5cm);    
         \filldraw[thick,fill=white] (4cm,8cm) circle (5 mm) node [shift={(0.5,0.0)}] {\mfn{8}}; 
         \filldraw[thick,fill=white] (0cm,8) circle (5 mm) node [shift={(-0.5,0.0)}] {\mfn{\tC_1}};
                  \filldraw[thick,red,fill=white] (0 cm, 4.5cm) -- (0 cm, 7.5cm);  
                   \filldraw[thick,red,fill=white] (0 cm, 0.5cm) -- (0 cm, 6.5cm);     
                                    \filldraw[thick,red,fill=white] (0cm,4) circle (5 mm) node [shift={(-0.5,0.0)}] {\mfn{\tN}};
        \filldraw[thick,fill=white] (28 cm,8) circle (5 mm) node [shift={(0.0,-0.5)}] {\mfn{\tC_2}};    
 \filldraw[thick,fill=white] (28 cm,0) circle (5 mm) node [shift={(0.0,-0.5)}] {\mfn{6'}};        
 \filldraw[thick,fill=white] (28.5 cm, 0) -- (31.5 cm,0);  
         \filldraw[thick,fill=white] (32 cm,0) circle (5 mm) node [shift={(0.0,-0.5)}] {\mfn{5'}}; 
  \filldraw[thick,fill=white] (32.5 cm, 0) -- (35.5 cm,0);       
         \filldraw[thick,fill=white] (36cm,0) circle (5 mm) node [shift={(0.0,-0.5)}] {\mfn{4'}};     
  \filldraw[thick,fill=white] (36.5 cm, 0) -- (39.5 cm,0);   
     \filldraw[thick,fill=white] (40cm,0) circle (5 mm) node [shift={(0.0,-0.5)}] {\mfn{3'}}; 
     \filldraw[thick,fill=white] (40.5 cm, 0) -- (43.5 cm,0);   
  \filldraw[thick,fill=white] (44cm,0) circle (5 mm) node [shift={(0.0,-0.5)}] {\mfn{2'}};    
  \filldraw[thick,fill=white] (44.5 cm, 0) -- (47.5 cm,0); 
   \filldraw[thick,fill=white] (44 cm, 0.5 cm) -- (44 cm, 3.5 cm);    
   \filldraw[thick,fill=white] (48cm,0) circle (5 mm) node [shift={(0.0,-0.5)}] {\mfn{1'}};     
 \filldraw[thick,fill=white] (44cm,4cm) circle (5 mm) node [shift={(0.5,0.0)}] {\mfn{7'}};     
 \filldraw[thick,fill=white] (44 cm, 4.5cm) -- (44 cm, 7.5cm);    
         \filldraw[thick,fill=white] (44 cm,8cm) circle (5 mm) node [shift={(0.5,0.0)}] {\mfn{8'}};
     \end{tikzpicture} &
           \begin{tikzpicture}[scale=.145]
   \filldraw[thick,fill=white] (0 cm,0) circle (5 mm) node [shift={(0.0,-0.5)}] {\mfn{1}} ;
     \filldraw[thick,fill=white] (0.5 cm, 0) -- (3.5 cm,0);  
      \filldraw[thick,fill=white] (4 cm,0) circle (5 mm) node [shift={(0.0,-0.5)}] {\mfn{2}} ;  
          \filldraw[thick,fill=white] (4 cm, 0.5 cm) -- (4 cm, 3.5 cm);  
     \filldraw[thick,fill=white] (12 cm,0) circle (5 mm) node [shift={(0.0,-0.5)}] {\mfn{4}};    
  \filldraw[thick,fill=white] (12.5 cm, 0) -- (15.5 cm,0);     
         \filldraw[thick,fill=white] (16 cm,0) circle (5 mm) node [shift={(0.0,-0.5)}] {\mfn{5}};
           \filldraw[thick,fill=white] (16.5 cm, 0) -- (19.5 cm,0);  
            \filldraw[thick,fill=white] (20 cm,0) circle (5 mm) node [shift={(0.0,-0.5)}] {\mfn{6}} ;    
         \filldraw[thick,fill=white] (4cm,4cm) circle (5 mm) node [shift={(0.5,0.0)}] {\mfn{7}};     
 \filldraw[thick,fill=white] (4 cm, 4.5cm) -- (4 cm, 7.5cm);    
         \filldraw[thick,fill=white] (4cm,8cm) circle (5 mm) node [shift={(0.5,0.0)}] {\mfn{8}}; 
         \filldraw[thick,fill=white] (0cm,8) circle (5 mm) node [shift={(-0.5,0.0)}] {\mfn{\tC_1}};
                  \filldraw[thick,red,fill=white] (0 cm, 4.5cm) -- (0 cm, 7.5cm);  
                   \filldraw[thick,red,fill=white] (0 cm, 0.5cm) -- (0 cm, 6.5cm);     
                 \filldraw[thick,red,fill=white] (0cm,4) circle (5 mm) node [shift={(-0.5,0.0)}] {\mfn{\tN}};
        \filldraw[thick,fill=white] (28 cm,8) circle (5 mm) node [shift={(0.0,-0.5)}] {\mfn{\tC_2}};    
 \filldraw[thick,fill=white] (28 cm,0) circle (5 mm) node [shift={(0.0,-0.5)}] {\mfn{6'}};        
 \filldraw[thick,fill=white] (28.5 cm, 0) -- (31.5 cm,0);  
         \filldraw[thick,fill=white] (32 cm,0) circle (5 mm) node [shift={(0.0,-0.5)}] {\mfn{5'}}; 
  \filldraw[thick,fill=white] (32.5 cm, 0) -- (35.5 cm,0);       
         \filldraw[thick,fill=white] (36cm,0) circle (5 mm) node [shift={(0.0,-0.5)}] {\mfn{4'}};     
  \filldraw[thick,fill=white] (36.5 cm, 0) -- (39.5 cm,0);   
     \filldraw[thick,fill=white] (40cm,0) circle (5 mm) node [shift={(0.0,-0.5)}] {\mfn{3'}}; 
     \filldraw[thick,fill=white] (40.5 cm, 0) -- (43.5 cm,0);   
  \filldraw[thick,fill=white] (44cm,0) circle (5 mm) node [shift={(0.0,-0.5)}] {\mfn{2'}};    
   \filldraw[thick,fill=white] (44 cm, 0.5 cm) -- (44 cm, 3.5 cm);         
 \filldraw[thick,fill=white] (44cm,4cm) circle (5 mm) node [shift={(0.5,0.0)}] {\mfn{7'}};     
 \filldraw[thick,fill=white] (44 cm, 4.5cm) -- (44 cm, 7.5cm);    
         \filldraw[thick,fill=white] (44 cm,8cm) circle (5 mm) node [shift={(0.5,0.0)}] {\mfn{8'}};
     \end{tikzpicture}\\ 
       (a) & (b) \\\hline
      \begin{tikzpicture}[scale=.145]
   \filldraw[thick,fill=white] (0 cm,0) circle (5 mm) node [shift={(0.0,-0.5)}] {\mfn{1}} ;
     \filldraw[thick,fill=white] (0.5 cm, 0) -- (3.5 cm,0);  
      \filldraw[thick,fill=white] (4 cm,0) circle (5 mm) node [shift={(0.0,-0.5)}] {\mfn{2}} ;  
          \filldraw[thick,fill=white] (4 cm, 0.5 cm) -- (4 cm, 3.5 cm);  
     \filldraw[thick,fill=white] (12 cm,0) circle (5 mm) node [shift={(0.0,-0.5)}] {\mfn{4}};    
  \filldraw[thick,fill=white] (12.5 cm, 0) -- (15.5 cm,0);     
         \filldraw[thick,fill=white] (16 cm,0) circle (5 mm) node [shift={(0.0,-0.5)}] {\mfn{5}};
           \filldraw[thick,fill=white] (16.5 cm, 0) -- (19.5 cm,0);  
            \filldraw[thick,fill=white] (20 cm,0) circle (5 mm) node [shift={(0.0,-0.5)}] {\mfn{6}} ;    
         \filldraw[thick,fill=white] (4cm,4cm) circle (5 mm) node [shift={(0.5,0.0)}] {\mfn{7}};     
 \filldraw[thick,fill=white] (4 cm, 4.5cm) -- (4 cm, 7.5cm);    
         \filldraw[thick,fill=white] (4cm,8cm) circle (5 mm) node [shift={(0.5,0.0)}] {\mfn{8}}; 
         \filldraw[thick,fill=white] (0cm,8) circle (5 mm) node [shift={(-0.5,0.0)}] {\mfn{\tC_1}};
                  \filldraw[thick,red,fill=white] (0 cm, 4.5cm) -- (0 cm, 7.5cm);  
                   \filldraw[thick,red,fill=white] (0 cm, 0.5cm) -- (0 cm, 6.5cm);     
                 \filldraw[thick,red,fill=white] (0cm,4) circle (5 mm) node [shift={(-0.5,0.0)}] {\mfn{\tN}};
        \filldraw[thick,fill=white] (28 cm,8) circle (5 mm) node [shift={(0.0,-0.5)}] {\mfn{\tC_2}};
        \filldraw[thick,blue,fill=white] (32 cm,8) circle (5 mm) node [shift={(0.0,-0.5)}] {\mfn{\tN'}}; 
         \filldraw[thick,blue,fill=white] (28.5 cm, 8) -- (31.5 cm,8);  
            \filldraw[thick,blue,fill=white] (32.5 cm, 8) -- (43.5 cm,8);  
 \filldraw[thick,fill=white] (28 cm,0) circle (5 mm) node [shift={(0.0,-0.5)}] {\mfn{6'}};        
 \filldraw[thick,fill=white] (28.5 cm, 0) -- (31.5 cm,0);  
         \filldraw[thick,fill=white] (32 cm,0) circle (5 mm) node [shift={(0.0,-0.5)}] {\mfn{5'}}; 
  \filldraw[thick,fill=white] (32.5 cm, 0) -- (35.5 cm,0);       
         \filldraw[thick,fill=white] (36cm,0) circle (5 mm) node [shift={(0.0,-0.5)}] {\mfn{4'}};     
  \filldraw[thick,fill=white] (36.5 cm, 0) -- (39.5 cm,0);   
     \filldraw[thick,fill=white] (40cm,0) circle (5 mm) node [shift={(0.0,-0.5)}] {\mfn{3'}}; 
     \filldraw[thick,fill=white] (40.5 cm, 0) -- (43.5 cm,0);   
  \filldraw[thick,fill=white] (44cm,0) circle (5 mm) node [shift={(0.0,-0.5)}] {\mfn{2'}};    
   \filldraw[thick,fill=white] (44 cm, 0.5 cm) -- (44 cm, 3.5 cm);         
 \filldraw[thick,fill=white] (44cm,4cm) circle (5 mm) node [shift={(0.5,0.0)}] {\mfn{7'}};     
 \filldraw[thick,fill=white] (44 cm, 4.5cm) -- (44 cm, 7.5cm);    
         \filldraw[thick,fill=white] (44 cm,8cm) circle (5 mm) node [shift={(0.5,0.0)}] {\mfn{8'}};
     \end{tikzpicture}&
            \begin{tikzpicture}[scale=.145]
   \filldraw[thick,fill=white] (0 cm,0) circle (5 mm) node [shift={(0.0,-0.5)}] {\mfn{1}} ;
     \filldraw[thick,fill=white] (0.5 cm, 0) -- (3.5 cm,0);  
      \filldraw[thick,fill=white] (4 cm,0) circle (5 mm) node [shift={(0.0,-0.5)}] {\mfn{2}} ;  
          \filldraw[thick,fill=white] (4 cm, 0.5 cm) -- (4 cm, 3.5 cm);  
     \filldraw[thick,fill=white] (12 cm,0) circle (5 mm) node [shift={(0.0,-0.5)}] {\mfn{4}};    
  \filldraw[thick,fill=white] (12.5 cm, 0) -- (15.5 cm,0);     
         \filldraw[thick,fill=white] (16 cm,0) circle (5 mm) node [shift={(0.0,-0.5)}] {\mfn{5}};
           \filldraw[thick,fill=white] (16.5 cm, 0) -- (19.5 cm,0);  
            \filldraw[thick,fill=white] (20 cm,0) circle (5 mm) node [shift={(0.0,-0.5)}] {\mfn{6}} ;    
         \filldraw[thick,fill=white] (4cm,4cm) circle (5 mm) node [shift={(0.5,0.0)}] {\mfn{7}};     
 \filldraw[thick,fill=white] (4 cm, 4.5cm) -- (4 cm, 7.5cm);    
         \filldraw[thick,fill=white] (4cm,8cm) circle (5 mm) node [shift={(0.5,0.0)}] {\mfn{8}}; 
         \filldraw[thick,fill=white] (0cm,8) circle (5 mm) node [shift={(-0.5,0.0)}] {\mfn{\tC_1}};
                  \filldraw[thick,red,fill=white] (0 cm, 4.5cm) -- (0 cm, 7.5cm);  
                   \filldraw[thick,red,fill=white] (0 cm, 0.5cm) -- (0 cm, 6.5cm);     
                 \filldraw[thick,red,fill=white] (0cm,4) circle (5 mm) node [shift={(-0.5,0.0)}] {\mfn{\tN}};
        \filldraw[thick,fill=white] (28 cm,8) circle (5 mm) node [shift={(0.0,-0.5)}] {\mfn{\tC_2}};
        \filldraw[thick,blue,fill=white] (32 cm,8) circle (5 mm) node [shift={(0.0,-0.5)}] {\mfn{\tN'}}; 
         \filldraw[thick,blue,fill=white] (28.5 cm, 8) -- (31.5 cm,8);   
 \filldraw[thick,fill=white] (28 cm,0) circle (5 mm) node [shift={(0.0,-0.5)}] {\mfn{6'}};        
 \filldraw[thick,fill=white] (28.5 cm, 0) -- (31.5 cm,0);  
         \filldraw[thick,fill=white] (32 cm,0) circle (5 mm) node [shift={(0.0,-0.5)}] {\mfn{5'}}; 
  \filldraw[thick,fill=white] (32.5 cm, 0) -- (35.5 cm,0);       
         \filldraw[thick,fill=white] (36cm,0) circle (5 mm) node [shift={(0.0,-0.5)}] {\mfn{4'}};     
  \filldraw[thick,fill=white] (36.5 cm, 0) -- (39.5 cm,0);   
     \filldraw[thick,fill=white] (40cm,0) circle (5 mm) node [shift={(0.0,-0.5)}] {\mfn{3'}}; 
     \filldraw[thick,fill=white] (40.5 cm, 0) -- (43.5 cm,0);   
  \filldraw[thick,fill=white] (44cm,0) circle (5 mm) node [shift={(0.0,-0.5)}] {\mfn{2'}};    
   \filldraw[thick,fill=white] (44 cm, 0.5 cm) -- (44 cm, 3.5 cm);         
 \filldraw[thick,fill=white] (44cm,4cm) circle (5 mm) node [shift={(0.5,0.0)}] {\mfn{7'}};     
     \end{tikzpicture}\\ 
       (c) & (d) \\ \hline
       \begin{tikzpicture}[scale=.145]
         \filldraw[thick,green] (24 cm, 8) -- (20 cm,0);
   \filldraw[thick,fill=white] (0 cm,0) circle (5 mm) node [shift={(0.0,-0.5)}] {\mfn{1}} ;
     \filldraw[thick,fill=white] (0.5 cm, 0) -- (3.5 cm,0);  
      \filldraw[thick,fill=white] (4 cm,0) circle (5 mm) node [shift={(0.0,-0.5)}] {\mfn{2}} ;  
          \filldraw[thick,fill=white] (4 cm, 0.5 cm) -- (4 cm, 3.5 cm);  
     \filldraw[thick,fill=white] (12 cm,0) circle (5 mm) node [shift={(0.0,-0.5)}] {\mfn{4}};    
  \filldraw[thick,fill=white] (12.5 cm, 0) -- (15.5 cm,0);     
         \filldraw[thick,fill=white] (16 cm,0) circle (5 mm) node [shift={(0.0,-0.5)}] {\mfn{5}};
           \filldraw[thick,fill=white] (16.5 cm, 0) -- (19.5 cm,0);  
            \filldraw[thick,fill=white] (20 cm,0) circle (5 mm) node [shift={(0.0,-0.5)}] {\mfn{6}} ;    
         \filldraw[thick,fill=white] (4cm,4cm) circle (5 mm) node [shift={(0.5,0.0)}] {\mfn{7}};     
 \filldraw[thick,fill=white] (4 cm, 4.5cm) -- (4 cm, 7.5cm);    
         \filldraw[thick,fill=white] (4cm,8cm) circle (5 mm) node [shift={(0.5,0.0)}] {\mfn{8}}; 
         \filldraw[thick,fill=white] (0cm,8) circle (5 mm) node [shift={(-0.5,0.0)}] {\mfn{\tC_1}};
                  \filldraw[thick,red,fill=white] (0 cm, 4.5cm) -- (0 cm, 7.5cm);  
                   \filldraw[thick,red,fill=white] (0 cm, 0.5cm) -- (0 cm, 6.5cm);     
                 \filldraw[thick,red,fill=white] (0cm,4) circle (5 mm) node [shift={(-0.5,0.0)}] {\mfn{\tN}};
        \filldraw[thick,fill=white] (28 cm,8) circle (5 mm) node       [shift={(0.0,-0.5)}] {\mfn{\tC_2}};
  \filldraw[thick,green,fill=white] (24 cm,8) circle (5 mm) node [shift={(0.0,-0.5)}] {\mfn{\texttt{S}}};        
  \filldraw[thick,green] (24.5 cm, 8) -- (27.5 cm,8);   
        \filldraw[thick,blue,fill=white] (32 cm,8) circle (5 mm) node [shift={(0.0,-0.5)}] {\mfn{\tN'}};         
         \filldraw[thick,blue,fill=white] (28.5 cm, 8) -- (31.5 cm,8);  
 \filldraw[thick,fill=white] (28 cm,0) circle (5 mm) node [shift={(0.0,-0.5)}] {\mfn{6'}};        
 \filldraw[thick,fill=white] (28.5 cm, 0) -- (31.5 cm,0);  
         \filldraw[thick,fill=white] (32 cm,0) circle (5 mm) node [shift={(0.0,-0.5)}] {\mfn{5'}}; 
  \filldraw[thick,fill=white] (32.5 cm, 0) -- (35.5 cm,0);       
         \filldraw[thick,fill=white] (36cm,0) circle (5 mm) node [shift={(0.0,-0.5)}] {\mfn{4'}};     
  \filldraw[thick,fill=white] (36.5 cm, 0) -- (39.5 cm,0);   
     \filldraw[thick,fill=white] (40cm,0) circle (5 mm) node [shift={(0.0,-0.5)}] {\mfn{3'}}; 
     \filldraw[thick,fill=white] (40.5 cm, 0) -- (43.5 cm,0);   
  \filldraw[thick,fill=white] (44cm,0) circle (5 mm) node [shift={(0.0,-0.5)}] {\mfn{2'}};    
   \filldraw[thick,fill=white] (44 cm, 0.5 cm) -- (44 cm, 3.5 cm);         
 \filldraw[thick,fill=white] (44cm,4cm) circle (5 mm) node [shift={(0.5,0.0)}] {\mfn{7'}};     
     \end{tikzpicture}& \\
       (e) &  \\
\end{tabular}     
 \caption{Dynkin diagrams for the steps leading to the enhancements 
          $\mra_3+\mra_6+\mra_9$ (c) and $3\mra_6$ (e), starting from a point with $\mra_6+\mra_3+\mra_1 +\mre_8$ (a).
           Intermediate stages where the symmetry is broken are shown in (b) and (d).}   
\label{dynkin3a6}                  
\end{figure}

In conclusion, we have arrived at $\mra_3+\mra_6+\mra_9$ and $3\mra_6$. The former has Wilson lines
 $A_1= -(\frac18 w_3 \times \frac25 w'_1)$, $A_2=\frac14 w_3 \times (-\frac15 w'_1)$, and complimentary lattice $T$ with
Gram matrix $Q=[2,0,140]$.
For the latter $A_1= -\frac18 w_3 \times (-\frac12 w'_1+\frac18 w'_8)$, $A_2= \frac14 w_3 \times (-\frac14 w_8')$, and $Q=[2,1,4]$.
For both, $E=\delta_{ij}$.

\subsection{All maximal rank groups for $d=2$}
\label{ss:alld2}

From the results in \cite{SZ} we infer
that there are 359 distinct maximally enhanced heterotic models on $T^2$, some of which share the same gauge group. The number of distinct maximal rank gauge groups found is 325. Using the extended diagram formalism of section \ref{sec:shiftd2} we are able to obtain the moduli for 331 of these models. The more powerful computational methods described in sections \ref{ss:swl} and \ref{ss:neighbor} allow us to obtain the moduli for the remaining 28 models, as well as alternative moduli for the other 331. 

In Table \ref{tab:alld2}, displayed in appendix \ref{app:tablesmaximal}, we show a representative for each of the 359 models in the 
$\mre_8 \times \mre'_8$ heterotic theory. The data for each point consists of the root lattice $L$, which gives the gauge group, the isotropic subgroup $H_L$, the complementary lattice $T$, and the moduli $E_{ij}, A_1, A_2$. 
The lattice $T$ is conveyed by its Gram matrix, computed from the moduli as described in section \ref{ss:ldata}.
Once $T$ is known we can determine the order of $H_L$ using the relation \eqref{mwrel}. We can then check that the
appropriate isotropic subgroup of $A_L$ exists as in the examples worked out in section \ref{ss:d2}. In this way
we can confirm the results of \cite{SZ} for the $H_L$ corresponding to each pair $(L,T)$.

In contrast to the $d = 1$ case, we do not have an explicit form of the fundamental domain of the moduli space, which would give 
us a clear criterion for choosing the moduli. Instead, we have selected those that have the simplest form. 
In some cases we present two different sets of moduli, one in which the Wilson lines are simple but the 
$E_{ij}$ are different from the standard ones in \eqref{E_standard}, and another where the opposite happens.  
Moduli obtained with the Fixed Wilson lines algorithm of section \ref{ss:swl}, or the Neighborhood algorithm of section \ref{ss:neighbor},
are respectively distinguished by a $\dagger$ or by a $*$, next to the Wilson line $A_2$. 
The remaining $331$ moduli were obtained using the EDD method of section \ref{sec:shiftd2}.
Notice that  for the groups 18, 23 and 40, the EDD method only gives the data for one of the possible $T$ lattices. 

As expected from the general arguments of section \ref{ss:nikhet}, in all cases the $E_{ij}$ are rational numbers and the
Wilson lines are quantized in the sense of eq.~\eqref{mwrel}. Moreover, it can be shown that for every pair $(L,T)$,
it is always possible to find Wilson lines such that $E_{ij}=\delta_{ij}$. Examples of this result are \# 15 or \# 19 in 
Table \ref{tab:alld2}. 

The torus metric and the $b$-field can be easily derived from the moduli $E_{ij}$ and $A_i$ substituting in 
$g_{ij}=\frac12(E_{ij} + E_{ji} - A_i \cdot A_j)$ and $b_{ij}=\frac12(E_{ij} - E_{ji}) $.
The complex structure and K\"ahler moduli, $\tau$ and $\rho$, can then be computed from their definition in \eqref{cplx},
or alternatively from the relations to the $E_{ij}$ in \eqref{Ecplx}. Note that in most cases in Table \ref{tab:alld2},
the enhancements occur at points with $\rho=\tau$. The exception is \# 2, but as mentioned before, this group
can also be reached with $E_{ij}=\delta_{ij}$ and suitable Wilson lines.

The transformations of the moduli under the duality group are best found as explained in section \ref{ss:duality}.
We conjecture that all possible heterotic models on $T^d$ with maximal rank gauge group and 
given pair $(L,T$), are unique up to T-dualities. We know that this is true in $d = 1$, since the extended Dynkin diagram for the lattice 
$\text{II}_{1,17}$ uniquely encodes all such models within a fundamental region of the 
moduli space. Indeed, the only freedom
in the diagram in Figure \ref{dynkindiag2} corresponds to a reflection about the central node,
which is an automorphism of the  lattice $\text{II}_{1,17}$. 
For $d = 2$, the conjecture implies that Table \ref{tab:alld2} exhibits all maximally enhanced HE models up to T-dualities. 
In particular, we have checked that in cases such as \# 15, the two sets of moduli can be connected by an
element of $\rO(2,18,\ZZ)$.

For each model in Table \ref{tab:alld2}, the moduli in the $\hosp$ heterotic can be obtained by using the map described in section \ref{ss:hehomap}. We have explicitly verified that the Gram matrices of the lattices $L$ and $T$ are preserved under this map, which is to be expected from an orthogonal transformation. Some examples of these transformed HO models are given in Table \ref{tab:exHO}.

As for $d = 1$, we can compute the Weyl transformation for each simple root of the enhanced gauge group to obtain the reflexive subgroup of $\rO(2,18,\mathbb{Z})$ that fixes the corresponding moduli. However, since $\rO(2,18,\mathbb{Z})$ is not reflexive, computing the whole set of dualities which fix a given point is not generally straightforward, although a 
complete answer can be given in simpler cases. For instance, as discussed in section \ref{sec:complex}, we can restrict to
Wilson lines of the form $A_i=a_i w_6 \times 0$ and work with the complex moduli $(\tau,\rho, \beta)$, defined in \eqref{cplx}
and \eqref{betadef}. As explained in section \ref{sec:complex}, $\Omega=\left(\begin{smallmatrix} \tau & \beta \\ \beta & \rho \end{smallmatrix}\right)$
parametrizes the genus-two Siegel upper half-plane, and the fundamental region, as well as fixed points of $\mathrm{Sp}(4,\ZZ)$,
have been determined by Gottschling \cite{Gottschling1, Gottschling2, Gottschling3}. In particular, in Theorem 4, Lemma 7 in \cite{Gottschling3}, it is shown that the point 
$\Omega_G=\left(\begin{smallmatrix} \eta & \frac12(\eta-1) \\ \frac12(\eta-1)  & \eta \end{smallmatrix}\right)$, with
$\eta$ given in \eqref{pam325}, is fixed by the octahedral group ($O$) of order 24.
In fact, this point $\Omega_G$ can be shown to be precisely dual to the maximally enhanced point of entry \# 325 in 
Table \ref{tab:alld2}, which corresponds to
\be
\Omega_P = 
	\begin{pmatrix}
	\eta-1 & \frac12(\eta-1) \\ \frac12(\eta-1)  & \eta-1
	\end{pmatrix},
\quad  \eta = \frac{1}{3}(1+2i\sqrt{2})\, .	
\label{pam325}
\ee
At a generic point $\Omega$, the gauge group is $\rG=\uo^3 \times \mre_7 \times \mre'_8$, while at $\Omega_P$ (or equivalently $\Omega_G$) it is
enlarged to $\rG_P=\sug(4) \times \mre_7 \times \mre'_8$. It is natural to propose that the transformations
that leave $\Omega_P$ fixed are generated by Weyl reflections about the simple roots that extend $\rG$ to $\rG_P$.
We have checked that this is indeed the case. The simple roots of $\sug(4)$ are associated to the nodes
$\varphi_0$, $\varphi_{{\texttt{C}_1}}$ and $\varphi_{{\texttt{C}_3}}$, shown in \eqref{zero2} and \eqref{nodesc1c3}.
As expected, the group generated by the Weyl transformations is the permutation group $S_4$, which is isomorphic
to the octahedral group $O$. It is easy to verify that $\Omega_P$ is fixed by the transformations of order 3 and 4 displayed in \eqref{octa},
which are just products of Weyl reflections about $\varphi_0$, $\varphi_{{\texttt{C}_1}}$ and $\varphi_{{\texttt{C}_3}}$.

Actually, the maximal enhancements in \# 296, \#297 and \# 324 in Table \ref{tab:alld2}, which also lie in the slice
of moduli space with $\zeta=\beta w_6 \times 0$, correspond to fixed points analyzed in \cite{Gottschling2, Gottschling3}.
However, at the fixed points of cyclic subgroups there is no maximal enhancement. Similar results can be obtained in HO. 
It would be interesting to find more connections between the Narain moduli space and other kinds of moduli spaces, 
and to further study maximal enhancements as fixed points of duality transformations.

\section{Compactifications on $T^d$}
\label{sec:dd}

In heterotic compactifications on $T^d$ there are $d(d+16)$ moduli from background values of
the metric, the $b$-field and the 16-dimensional Wilson lines.  
The  $\mathrm{II}_{d,d+16}$ lattice vectors $(p_R; p_L, p^I)$, which depend on these moduli, are given in \eqref{momenta}.
The generalization of the algorithms discussed  in the preceding sections to study the enhancement of symmetries and the corresponding moduli  in  higher dimensional compactifications is straightforward. Here we briefly outline the procedures and present some examples.

In section \ref{sec:review} we worked out the transformation rules of the moduli under $\rO(d,d+16)$, for 
arbitrary dimension $d$. In particular, the Buscher rules found in 
\cite{buscher1, buscher} for the heterotic string were easily reobtained from a factorized duality as shown in eq.~\eqref{buscher}. 
We further generalized the  HE $\leftrightarrow$ HO map that was derived  for the circle in \cite{Keurentjes:2006cw} to  compactifications 
on $T^d$ with $d>1$. As an application,  we can  map   the simple cases $2\mre_8 +\tilde {\mathrm G}_d $ in the HE, or  
$\mrd_{16} + \tilde{\mathrm G}_d$  in the HO, with moduli $A_i=0$ and $E_{ij}, i,j=1,...,d$, given in \eqref{Eginsparg}, respectively to HO or HE. We find $E'_{ii}=E_{ii}=1, E'_{1j}=\frac14 E_{1j}$ for  $j>1, E'_{ij}=E_{ij}$ for $ i\ne 1$,  $A'_i=0, i>1$, whereas
$A'_1=\Lambda_{\small{\rm O}}-\frac14\Lambda_{\small{\rm E}}$ for $\text{HE}\to \text{HO}$  and
$A'_1=\Lambda_{\small{\rm E}}-\frac14\Lambda_{\small{\rm O}}$ for $\text{HO}\to \text{HE}$.

In section \ref{sec:lattices} we explained that all allowed groups \mbox{$\gr\times \uo^{d+16-r}$} can be obtained by lattice embedding techniques. We gave examples in $d=1, 2$ and 8, but the machinery can be applied to other dimensions.
An interesting observation is that for $d=8$ any semisimple ADE group of rank 24 seems to be allowed, as indicated by
the fact that the group of lowest dimension, namely $\sug(2)^{24}$, does occur as shown in section \ref{ss:d8}.

The generalization of the algorithms developed in section \ref{sec:d2} is straightforward, as they are based on general ideas that do 
not depend on $d$. 
In particular,  the method of extended diagrams described in section \ref{sss:gdd}, 
requires to find suitable values of $E_{ij}$ that can account  for the possible ways
to connect the toroidal nodes, i.e. the nodes corresponding to $\mathrm{II}_{d,d}$, to the affine diagram of $\mre_8 \times \mre'_8$,
or a subgroup of rank $16$ obtained with the shift algorithm.
Specifically, we can take the $E_{ij}$ given in \eqref{Eginsparg} in terms of the Cartan metric $\tilde g_{ij}$ of
an ADE group of rank $d$, which have the properties $|E_{ij}|$ equal to 0 or 1, and $\det E=1$.
One can then construct extended diagrams with $d+18$ nodes in arbitrary dimensions in a completely analogous way as done for 
$d=2$, also taking into account the twisting operation in \eqref{twisting}. 

A simple example of an EDD, for generic $d$, can be constructed in HE by choosing the $E_{ij}$ in \eqref{Eginsparg} 
with $\tilde g_{ij}$ equal to the Cartan matrix of $\mra_d$.
The affine nodes $\varphi_0$ and $\varphi_0'$ are taken to have $n_1 = -1$ and $n_d = -1$, respectively, with all other values for 
$w^i$ and $n_i$ set to zero. 
The resulting diagram is shown in figure \ref{edd3:1}. Note that in this construction the Wilson lines $A_i$, with $i = 2,...,d-1$ are always turned off, while $A_1 = \delta_1 \times 0$ and $A_d = 0 \times \delta_d'$. This EDD yields maximal enhancements such as $\sog(32+2d)$ and $\sug(17+d)$. More generally we can arrange the toroidal nodes into an $\mra_p \times \mra_{d-p}$ diagram, with $p = 1,...,d-1$, in order to obtain groups such as $\sog(18+2p)\times\sog(18 + 2(d - p))$ and $\sug(9+p)\times \sug(9+d-p)$.

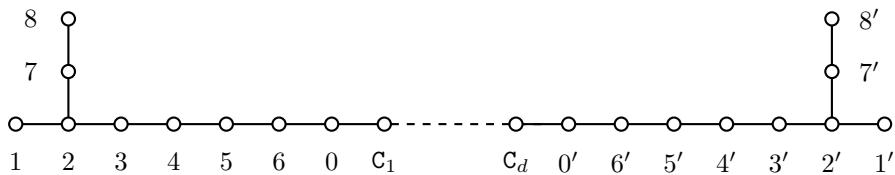
\begin{figure}[htb]
	\begin{center}
		\begin{tikzpicture}[scale=.175]
		\draw[thick] (0 cm,0) circle (5 mm) node [shift={(0.0,-0.5)}] {\mfn{1}} ;
		\draw[thick] (.5 cm, 0) -- (3.5 cm,0);    
		\draw[thick] (4 cm,0) circle (5 mm) node [shift={(0.0,-0.5)}] {\mfn{2}};
		\draw[thick] (4.5 cm, 0) -- (7.5 cm,0); 
		\draw[thick] (8 cm,0) circle (5 mm) node [shift={(0.0,-0.5)}] {\mfn{3}};
		\draw[thick] (8.5 cm, 0) -- (11.5 cm,0);     
		\draw[thick] (12 cm,0) circle (5 mm) node [shift={(0.0,-0.5)}] {\mfn{4}};
		\draw[thick] (12.5 cm, 0) -- (15.5 cm,0);     
		\draw[thick] (16 cm,0) circle (5 mm) node [shift={(0.0,-0.5)}] {\mfn{5}};
		\draw[thick] (16.5 cm, 0) -- (19.5 cm,0);     
		\draw[thick] (20cm,0) circle (5 mm) node [shift={(0.0,-0.5)}] {\mfn{6}};
		\draw[thick] (4 cm, 0.5cm) -- (4 cm, 3.5cm);     
		\draw[thick] (4cm,4cm) circle (5 mm) node [shift={(-0.5,0.0)}] {\mfn{7}};
		\draw[thick] (4 cm, 4.5cm) -- (4 cm, 7.5cm);     
		\draw[thick] (4cm,8cm) circle (5 mm) node [shift={(-0.5,0.0)}] {\mfn{8}};
		\draw[thick] (20.5 cm, 0) -- (23.5 cm,0);     
		\draw[thick, fill = white] (24cm,0) circle (5 mm) node [shift={(0.0,-0.5)}] {\mfn{0}};
		\draw[thick] (24.5 cm, 0) -- (27.5 cm,0);     
		\draw[thick, fill=white] (28cm,0) circle (5 mm) node [shift={(0.0,-0.5)}] {\footnotesize{$\texttt{C}_1$}}; 
		\draw[thick,dashed](28.5cm,0)--(40cm,0);   
		\begin{scope}[shift = {(10cm,0)}]
		\draw[thick, fill=white] (28cm,0cm) circle (5 mm) node [shift={(0.0,-0.5)}] {\footnotesize{$\texttt{C}_d$}};
		\draw[thick] (28.5 cm, 0) -- (31.5 cm,0); 
		\draw[thick, fill = white] (32 cm,0) circle (5 mm) node [shift={(0.0,-0.5)}] {\mfn{0^\prime}} ;
		\draw[thick] (32.5 cm, 0) -- (35.5 cm,0);    
		\draw[thick] (36 cm,0) circle (5 mm) node [shift={(0.0,-0.5)}] {\mfn{6^\prime}};
		\draw[thick] (36.5 cm, 0) -- (39.5 cm,0); 
		\draw[thick] (40 cm,0) circle (5 mm) node [shift={(0.0,-0.5)}] {\mfn{5^\prime}};
		\draw[thick] (40.5 cm, 0) -- (43.5 cm,0);     
		\draw[thick] (44 cm,0) circle (5 mm) node [shift={(0.0,-0.5)}] {\mfn{4^\prime}};
		\draw[thick] (44.5 cm, 0) -- (47.5 cm,0);     
		\draw[thick] (48 cm,0) circle (5 mm) node [shift={(0.0,-0.5)}] {\mfn{3^\prime}};
		\draw[thick] (48.5 cm, 0) -- (51.5 cm,0);     
		\draw[thick] (52cm,0) circle (5 mm) node [shift={(0.0,-0.5)}] {\mfn{2^\prime}};
		\draw[thick] (52.5 cm, 0) -- (55.5 cm,0);     
		\draw[thick] (56cm,0) circle (5 mm) node [shift={(0.0,-0.5)}] {\mfn{1^\prime}};
		\draw[thick] (52 cm, 0.5cm) -- (52 cm, 3.5cm);     
		\draw[thick] (52cm,4cm) circle (5 mm) node [shift={(0.5,0.0)}] {\mfn{7^\prime}};
		\draw[thick] (52 cm, 4.5cm) -- (52 cm, 7.5cm);     
		\draw[thick] (52cm,8cm) circle (5 mm) node [shift={(0.5,0.0)}] {\mfn{8^\prime}};
		\end{scope}
		\end{tikzpicture}
	\end{center}
	\caption{Extended diagram for generic values of $d$, yielding for example the maximal enhancements $\sog(32+2d)$ and $\sug(17+d)$.}\label{edd3:1}
\end{figure}

To apply the Fixed Wilson lines algorithm of section \ref{ss:swl} in $T^d$, one may  take some of the Wilson lines obtained from the previous construction, then delete the toroidal nodes in the corresponding diagram and add $d$ generic nodes to be determined by the algorithm. Finally,  the Neighborhood algorithm of  section \ref{ss:neighbor} may be implemented  starting from a point with maximal enhancement and 
$16+d$ simple roots, e.g. $2\mre_8 + d\mra_1$,  eliminate one of the simple roots 
and replace it by a generic one of $16+2d$ components to be fixed by the algorithm.
Some results obtained applying this algorithm in $d=3$ are presented in Table \ref{tab:exdim3}. 
All of them were found taking $3$ or less steps from $2\mre_8+3\mra_1$, and setting $(\lambda_1,\lambda_2)=(1,2)$
in \eqref{trunca}.

\begin{table}[H]
\renewcommand{\arraystretch}{1.05}
\setlength\tabcolsep{3pt}
\small{	
\centering
\begin{tabular}{|c|c|@{}c@{}||@{}c@{}|@{}c@{}|@{}c@{}|@{}c@{}|}\hline
$ L $&$ H_L $&$ T $&$ E $&$ A_{1} $&$ A_{2} $&$ A_{3} $\\ \hline 
$ \mra_{2}+\mra_{7}+\mrd_{10} $&$\mathbb{1}$&$ \left(
\begin{smallmatrix}
 2 & 0 & 0 \\
 0 & 2 & 0 \\
 0 & 0 & 24 \\
\end{smallmatrix}
\right) $&$ 
\left(\begin{smallmatrix}
 1 & -1 & 0 \\
 0 & 1 & 0 \\
 0 & 1 & 1 \\
\end{smallmatrix}\right)
 $&$0 \times \frac{w_{4}}{6} $&$ -\frac{w_{3}}{5} \times \frac{w_{4}}{15} $&$ -\frac{w_{3}}{5} \times  \frac{7w_{4}}{30} $\\
 [1mm] \hline
$ \mra_{1}+3 \mra_{3}+\mra_{9} $&$\mathbb{Z}_2$&$ \left(
\begin{smallmatrix}
 4 & 0 & 0 \\
 0 & 8 & 4 \\
 0 & 4 & 12 \\
\end{smallmatrix}
\right) $&$ 
\left(\begin{smallmatrix}
 1 & 0 & 0 \\
 0 & 1 & 0 \\
 0 & 0 & 1 \\
\end{smallmatrix}\right)
 $&$\left(-\frac{w_{4}}{38}- \frac{5w_{7}}{19}\right) \times  \frac{4w_{7}}{19} $&$ \left( \frac{5w_{7}}{19}- \frac{9w_{4}}{19}\right) \times   \frac{4w_{7}}{19} $&$ \left( \frac{7w_{4}}{19}- \frac{6w_{7}}{19}\right) \times \frac{w_{7}}{19}$\\ 
 [1mm]\hline
$ \mra_{10}+3 \mra_{3} $&$\mathbb{1}$&$ \left(
\begin{smallmatrix}
 4 & 0 & 0 \\
 0 & 8 & 4 \\
 0 & 4 & 24 \\
\end{smallmatrix}
\right) $&$ 
\left(\begin{smallmatrix}
 1 & 0 & 0 \\
 0 & 1 & 0 \\
 0 & 0 & 1 \\
\end{smallmatrix}\right)
 $&$  \frac{2w_{7}}{7}  \times \left(\frac{w_{3}}{4}- \frac{11w_{8}}{56}\right) $&$ -\frac{w_{7}}{7}  \times \left(-\frac{w_{3}}{2}+ \frac{27w_{8}}{28}\right) $&$ 0 \times \left(\frac{w_{3}}{4}- \frac{5w_{8}}{8}\right)$\\ 
 [1mm] \hline
$ 3 \mra_{4}+\mrd_{7} $&$\mathbb{1}$&$ \left(
\begin{smallmatrix}
 10 & 5 & 5 \\
 5 & 10 & 0 \\
 5 & 0 & 10 \\
\end{smallmatrix}
\right) $&$ 
\left(\begin{smallmatrix}
 1 & 0 & 0 \\
 0 & 1 & 0 \\
 0 & 0 & 1 \\
\end{smallmatrix}\right)
 $&$-\frac{w_{3}}{5}  \times \frac{w_{3}}{5} $&$ -\frac{w_{3}}{5}  \times -\frac{w_{3}}{5}  $&$ \left(\frac{w_{6}}{2}-\frac{w_{3}}{5}\right) \times 0$\\ 
 [1mm]\hline
$ \mra_{4}+3 \mra_{5} $&$\mathbb{1}$&$ \left(
\begin{smallmatrix}
 6 & 0 & 0 \\
 0 & 6 & 0 \\
 0 & 0 & 30 \\
\end{smallmatrix}
\right) $&$ 
\left(\begin{smallmatrix}
 1 & 0 & 0 \\
 0 & 1 & 0 \\
 0 & 0 & 1 \\
\end{smallmatrix}\right)
 $&$-\frac{w_{3}}{5}  \times 0 $&$ \left( \frac{2w_{3}}{5}-w_{8}\right) \times 0 $&$ 0 \times \frac{w_{3}}{5}$\\ 
 [1mm]\hline
$ 4 \mra_{3}+\mrd_{7} $&$ \mathbb{Z}_4 $&$ \left(
\begin{smallmatrix}
 4 & 0 & 0 \\
 0 & 4 & 0 \\
 0 & 0 & 4 \\
\end{smallmatrix}
\right) $&$ 
\left(\begin{smallmatrix}
 1 & 0 & 0 \\
 0 & 1 & 0 \\
 0 & 0 & 1 \\
\end{smallmatrix}\right)
 $&$\left(-\frac{w_{3}}{6}-\frac{w_{8}}{6}\right) \times \frac{w_{4}}{6} $&$ \left(\frac{w_{8}}{12}-\frac{w_{3}}{6}\right) \times -\frac{w_{4}}{3}  $&$ \left(\frac{w_{3}}{2}- \frac{5w_{8}}{4}\right) \times 0$\\ 
 [1mm]\hline
$ 2 \mra_{6}+\mra_{7} $&$\mathbb{1}$&$ \left(
\begin{smallmatrix}
 4 & 2 & 1 \\
 2 & 8 & 4 \\
 1 & 4 & 16 \\
\end{smallmatrix}
\right) $&$ 
\left(\begin{smallmatrix}
 1 & 0 & 0 \\
 0 & 1 & 0 \\
 0 & 0 & 1 \\
\end{smallmatrix}\right)
 $&$  \frac{2w_{7}}{7}  \times \left( \frac{2w_{1}}{7}- \frac{3w_{6}}{7}\right) $&$ -\frac{w_{7}}{7}  \times \left( \frac{8w_{6}}{7}- \frac{3w_{1}}{7}\right) $&$ 0 \times \left( \frac{3w_{1}}{7}-\frac{w_{6}}{7}\right)$\\ 
 [1mm]\hline
$ \mrd_{4}+\mrd_{9}+\mre_{6} $&$\mathbb{1}$&$ \left(
\begin{smallmatrix}
 4 & 2 & 0 \\
 2 & 4 & 0 \\
 0 & 0 & 4 \\
\end{smallmatrix}
\right) $&$ 
\left(\begin{smallmatrix}
 1 & 0 & 0 \\
 0 & 1 & 0 \\
 0 & 0 & 1 \\
\end{smallmatrix}\right)
 $&$-\frac{w_{8}}{2}  \times 0 $&$ 0 \times \frac{w_{5}}{3} $&$ 0 \times \frac{w_{5}}{3}$\\ 
 [1mm]\hline
$ \mrd_{6}+\mrd_{7}+\mre_{6} $&$\mathbb{1}$&$ \left(
\begin{smallmatrix}
 2 & 0 & 0 \\
 0 & 2 & 0 \\
 0 & 0 & 12 \\
\end{smallmatrix}
\right) $&$ 
\left(\begin{smallmatrix}
 1 & 0 & 0 \\
 0 & 1 & 0 \\
 0 & 0 & 1 \\
\end{smallmatrix}\right)
 $&$-\frac{w_{3}}{5}  \times  \frac{2w_{5}}{15} $&$ -\frac{w_{3}}{5}  \times  \frac{2w_{5}}{15} $&$ 0 \times \frac{w_{5}}{3}$\\ 
 [1mm]\hline
$ 2 \mrd_{6}+\mre_{7} $&$\mathbb{Z}_2$&$ \left(
\begin{smallmatrix}
 2 & 0 & 0 \\
 0 & 2 & 0 \\
 0 & 0 & 2 \\
\end{smallmatrix}
\right) $&$ 
\left(\begin{smallmatrix}
 1 & 0 & 0 \\
 0 & 1 & 0 \\
 0 & 0 & 1 \\
\end{smallmatrix}\right)
 $&$-\frac{w_{6}}{2}  \times \frac{w_{6}}{2} $&$ -\frac{w_{8}}{2}  \times 0 $&$ \left(\frac{w_{8}}{2}-w_{6}\right) \times 0$\\ 
 [1mm]\hline
\end{tabular}
\caption{Data for some groups of maximal rank, for the $\mre_8\times \mre'_8$ heterotic on $T^3$.}
  \label{tab:exdim3}
}
\end{table}

In this exploration of the $d = 3$ moduli space we have chosen at each step of the Neighborhood algorithm one representative model for each maximal enhancement, i.e. for each embedding $(L,T) \subset \text{II}_{3,19}$. This was also done in $d = 2$, as explained in section \ref{ss:neighbor}. This procedure would be exhaustive only if any two models corresponding to the same pair $(L,T)$ are equivalent under T-duality. Indeed, we posed this conjecture in section \ref{ss:alld2} for generic $d$. It would be interesting to understand this better, for example by using the techniques of lattice embeddings of Nikulin \cite{Nikulin80, Morrison}, or by further studying the dependence of the full heterotic spectrum on the data of $L$ and $T$.
 
We also remark that all the examples in table \ref{tab:exdim3} have $E_{ij} = \delta_{ij}$ or can be shown to be T-dual to a model satisfying this condition. Taking into account the fact that all maximal enhancements in $d = 1$ and $2$ can be constructed with $E_{ij} = \delta_{ij}$,  we expect that this fact extends to the case at hand with $d = 3$. In fact, we conjecture that this is a generic feature of all Narain moduli spaces, with arbitrary  $d$. To see the physical significance of this statement, note that the condition $E_{ij} = \delta_{ij}$ implies that the antisymmetric field $b_{ij}$ is turned off. In many cases it is also true that the Wilson lines are orthogonal, $A_i \cdot A_j = 0$, $i \neq j$, further implying that the metric $g_{ij}$ is diagonal and so $T^d = S^1 \times \cdots \times S^1$. 
However, we do not have a formal proof that this can be done for all the maximally enhanced models.

\section{Final remarks}
\label{sec:conclusions}

In this paper we have explored the rich landscape of perturbative heterotic string compactifications on $T^d$. These lead to 
non-chiral theories in $(10\msm{-}d)$ dimensions with rank $(d+16)$ gauge groups, which realize the upper bound on the rank 
arising in string constructions with 16 supercharges \cite{Kim:2019ths}. 
At special points  in moduli space, the $(d+16)$ $\uo$ symmetries can get enhanced, and we stated lattice embedding
criteria to determine whether a given gauge group is realized or not in a toroidal compactification. The use of these criteria 
was explained in several examples.

We also introduced different algorithms to systematically explore the moduli space and applied them to obtain all the semi-simple 
groups of maximal rank for $d=1$ and $d=2$, as well as the values of the corresponding background fields.  The algorithms can be
implemented in arbitrary dimension.
A few examples are provided in section \ref{sec:dd} and a more exhaustive analysis is left for a future publication. 
Specifying the moduli is important for various reasons. First of all, the vertex operators and
the full 1-loop modular invariant partition function of the theory explicitly depend on the
momenta \eqref{momenta} \cite{Narain:1985jj, narain2}. Besides, the moduli could be relevant in the study of
dualities with other constructions and in phenomenological applications (combining with additional orbifold actions).

All maximal enhancements in the heterotic compactification on $T^2$ coincide with all possible singular fibers of extremal K3 surfaces 
classified in \cite{SZ}. This gives additional evidence for the duality between compactifications of the heterotic string 
on $T^2$ and F-theory on K3, as well as relevant information for the study of extremal K3 surfaces.
Some realizations of these surfaces have been studied in detail, see \cite{K3book, K3SZ} and references therein. 
In the early days some examples were found by analyzing F-theory on orbifold limits of K3 \cite{Dasgupta:1996ij}.
Other examples have been obtained more recently by considering enhancements at special
points in the moduli space of K3 surfaces with Picard number less than 20 \cite{Kimura:2017rhk, Kimura:2018oze, afcern, Chabrol:2019bvn}.   
The identification of the moduli that give particular enhancement points supplies further ingredients  for a closer 
examination of the explicit map between heterotic and  K3 moduli. This map
was constructed in \cite{LopesCardoso:1996hq}, in the particular case when all Wilson lines vanish, hence with
two complex moduli $\tau$ and $\rho$.
A step further is the map of \cite{Malmendier:2014uka}, which includes Wilson lines that break a $\sug(2)$ in
$\mre_8$ such that the 16 complex moduli $\zeta^I$ reduce to the single complex parameter in \eqref{betadef}.
In the latter case the matching of the moduli was presented in \cite{afcern}, where also the moduli at points
of maximal enhancement were identified. 

Many other interesting questions deserve further study. For instance, we would like to identify the fundamental region in moduli space 
for $d\ge 2$. In the HE theory compactified on the circle, this region is given in Table \ref{tab:fundreghe}, and it was nicely described in \cite{Cachazo:2000ey} in terms of a chimney with side walls  at certain values of $A^I$,
and bottom bounded by a spherical wall at  $E=1$. In general it is also practical to use as moduli the Wilson lines $A_i^I$
together with the $E_{ij}$ that depend on the torus metric and the Kalb-Ramond field, cf. \eqref{genmetric}.
Our work  hints at two important features of the fundamental region. 
One is that all groups of maximal enhancement arise at $\det E=1$. This is obvious in $d=1$, as  the central node in the Extended 
Dynkin diagram of Figure \ref{dynkindiag2}, corresponding to $E=1$, cannot be deleted. In $d=2$ we have explicitly verified it, and 
for higher $d$ it seems to be always possible. 
The second observation is that all groups of maximal enhancement  arise at a single point in the fundamental domain. 
These two features  imply that any maximal enhancement point  at $\det E\neq 1$ is not in the fundamental region,  
and can be brought to $\det E =1$ by dualities. This suggests that $\det E=1$ is always a component of the boundary of the 
fundamental region (it corresponds to the bottom of the chimney for $d=1$). We conjecture that these two features are 
generic properties of the fundamental region and leave the proof for future work.

Classification of all allowed groups that can appear in compactifications of the perturbative heterotic
string on $T^d$ is an important problem, posed already in the early days \cite{Narain:1985jj} and revived recently
in the context of the swampland program \cite{Kim:2019ths}.
In this work we have stated criteria to solve this problem, and given the answer
for $d=1,2$. Actually, the solution includes not only the groups with maximal enhancement,
but also groups $\rG_r \times \uo^{d+16-r}$, with $r \le (16+d)$. For $d=1$ all possible $\rG_r$ can be deduced from the EDD,
and for $d=2$ they are listed in \cite{ShimadaK3}.
A natural question is whether different $\rG_r$ could arise in other non-chiral string constructions with 16 supercharges. 
For $d=2$, our results contain the groups with maximal enhancement found in the covariant lattice formulation \cite{Balog:1989xd}.
On the other hand, it is well known that $(10\msm{-}d)$-dimensional theories with
semisimple non-ADE groups of rank $(8+d)$, e.g. $\text{USp}(20)$ for $d=2$, can be built in the fermionic 
formulation \cite{Chaudhuri:1995fk} and as asymmetric orbifolds \cite{Chaudhuri:1995bf, Mikhailov:1998si, deBoer:2001wca}.
It would be interesting to know if some other CFT 
construction could give for instance 8-dimensional theories with 16 supercharges and an ADE gauge 
group of rank 18, such as $\mre_8 \times \sog(14) \times \sug(4)$, which is forbidden in the heterotic on $T^2$.
It would also be helpful to understand if a theory with a forbidden group could suffer from
global anomalies as discussed in \cite{8danomaly}.

Finally, we have observed that the landscape becomes less constrained as the internal torus dimension increases. 
Presumably, in $d = 8$, i.e. in two-dimensional theories, any rank 24 ADE group can appear in 
a toroidal compactification of the heterotic string.

\subsection*{Acknowledgements}
We are grateful to Gerardo Aldazabal, Martin Cederwall, Lilian Chabrol, Luis Ib\'a\~nez, Sergio Iguri, Axel Kleinschmidt, Stefano Massai, Fernando Quevedo, Stefan Theisen, 
and Cumrun Vafa  for interesting comments and valuable insights. 
Special recognition is due to Christoph Mayrhofer for earlier collaboration and continuous advice. 
We thank Daniel Allcock, Bert Schellekens, Ichiro Shimada and Ernest Vinberg for enlightening emails.
This work was partially supported
by the ERC Consolidator Grant 772408-Stringlandscape, PIP-CONICET- 11220150100559CO, UBACyT 2018-2021,  ANPCyT- PICT-2016-1358 (2017-2020) and NSF under Grant No. PHY-1748958. 
A.~Font acknowledges hospitality and support at various stages from CERN-TH, MITP, ICTP,
as well as the IFT UAM-CSIC via the Centro de Excelencia Severo Ochoa Program under Grant SEV-2016-0597 and  C.~N\'u\~nez from the Simons Foundation  and the ICTP.
M.~Gra\~na would like to thank KITP, Santa Barbara, for 
hospitality.

\vspace*{1cm}

\appendix

\section{Notation and basics concerning lattices} 
\label{ap:not}

\noindent
\leftline{\underline{$L$,  even positive definite lattice of rank $r$}}

Typically $L$ will be the sum of ADE root lattices. 
There is a basis formed by roots $\alpha_i$ with $\alpha_i^2=2$. The Gram matrix of $L$ has elements $\alpha_i \cdot \alpha_j$. 
It is equal to the Cartan matrix when $L$ is the root lattice of an ADE group. 

\medskip\noindent
\leftline{\underline{$d(L)$, discriminant of $L$}}

It is defined to be the determinant of the Gram matrix of $L$. By assumption $d(L) \not=0$.

\medskip\noindent
\leftline{\underline{$L^*$, dual lattice}}

Lattice generated by the weights $w_i$ defined by $w_i \cdot \alpha_j = \delta_{ij}$. Clearly $L \subset L^*$.

\medskip\noindent
\leftline{\underline{$A_L$, discriminant group}}

It is defined as $A_L=L^*/L$, also named $D_L$ or $G_L$ in the literature. 

It can be shown that $A_L$ is a finite Abelian group of order $d(L)$. 

Since $\mre_8$ is unimodular, its discriminant group is trivial. 
For $L=\mra_n, \mrd_{2m+1}, \mrd_{2m}, \mre_6, \mre_7$, 
\mbox{$A_L\cong\ZZ_{n+1}, \ZZ_4,  \ZZ_2 \times \ZZ_2, \ZZ_3, \ZZ_2$.}

\medskip\noindent
\leftline{\underline{$\ell(A_L)$, minimal number of generators of $A_L$}}

For example, for $L=2\mre_6 + \mra_6$, $\ell(A_L) =2$, because
$\ZZ_3 \times \ZZ_3 \times \ZZ_7 \sim  \ZZ_3 \times \ZZ_{21}$. Notice that $\ell(A_L) \le r$.

\medskip\noindent
\leftline{\underline{$ q_L$, discriminant quadratic form}}

It is  a map $q_L: A_L \to \QQ/2\ZZ, \  x+L \, \mapsto \, x^2 \,\text{mod}\, 2$.

For example for $L=\mra_n$, $A_L=\ZZ_{n+1}$ is generated by the class of the fundamental weight $[w_1]$.
Thus $q_L([w_1])=w_1^2 =\frac{n}{n+1}$, whereas $q_L([w_j])=w_j^2 =\frac{j(n+1-j)}{n+1}= \frac{j^2 n}{n+1}$,
with equalities mod 2. 

For $L=\mrd_{2m+1}$, $A_L=\ZZ_4$ is generated by the spinor class $[s]$ with $q_L([s])=\frac{2m+1}4$.

For $\mrd_{2m}$, $A_L=\ZZ_2 \times \ZZ_2$. One $\ZZ_2$ is generated by the spinor class $[s]$ with $q_L([s])=\frac{m}2$,
and the other $\ZZ_2$ by the vector class $[v]$ with $q_L([v])=1$. 

For $\mre_6$, $A_L=\ZZ_3$ is generated by  the fundamental weights of $[\bf{27}]$ with $q_L([{\bf{27}}])=\frac43$.

For $\mre_7$, $A_L=\ZZ_2$ is generated by the fundamental weights of $[{\bf{56}}]$ with $q_L([{\bf{56}}])=\frac32$.

\medskip\noindent

\leftline{\underline{$T$, even positive definite lattice of rank $d$}}

It is characterized by the Gram matrix $(Q)_{ij}=u_i \cdot u_j$, where $u_i$ are
the basis vectors. 

A generic even 1 dimensional lattice, denoted $\mra_1\langle m\rangle$, is a multiple by $m$ of the $\mra_1$ lattice.
It is generated by a vector $u_1$ with $u_1^2=2m$ and has discriminant group $\ZZ_{2m}$, in turn generated
by $(u_1^*)^2=\frac1{2m}$.

We will mostly consider $d=2$ and
as in \cite{SZ}, represent $Q$ as $[u_1^2, u_1\!\cdot\! u_2, u_2^2]$. 
For classification of even 2-dimensional lattices see chapter 15 in \cite{CS}, and section 2 in \cite{SZ} for a short account.
$Q$ can be brought to Smith normal form $\text{diag}(s_1,s_2)$, with
positive integer entries. Then $A_T\cong \ZZ_{s_1} \times \ZZ_{s_2}$.  Notice that if $s_1$ and $s_2$ are
coprimes then $A_T \cong \ZZ_{s_1 s_2}$. 
We will also need to compute the discriminant form $q_T$. From $Q^{-1}$ we can read off
$u_i^*\cdot u_j^*$, where $u_1^*$, $u_2^*$ are the basis vectors of the dual lattice $T^*$. 
Besides, $Q^{-1}$ gives the $e_i^*$ in terms of $e_i$. With this data
we can then find the generators of $A_T$ and derive $q_T$. 
For example, for $T$ with $Q=[2, 1, 4]$, $A_T\cong \ZZ_7$ and
$Q^{-1}=[\frac{4}{7}, -\frac1{7}, \frac27]$. The generator of $A_T$ can be taken to be $u_2^*$ which satisfies
$7 u_2^* = -u_1 + 2 u_2 \in T$, and has the lowest norm. Then $q_T$ takes values $\frac{2 j^2}7 \, \text{mod}\, 2$,
$j=0, \ldots, 6$.

\medskip\noindent
\leftline{\underline{$H_L$, isotropic subgroup of $A_L$}}

$H_L \subset A_L$ is isotropic if $q_L\big|_{H}=0$. 

For instance, for $L=\mra_8$, with
$A_L=\ZZ_9$, the subgroup $H_L=\ZZ_3$ generated by $w_3 \sim 3 w_1$ is isotropic because 
$q_L([w_3])=\frac{18}9=2=0 \, \text{mod}\, 2$.

Another example is $L=\mrd_8$, with $A_L=\ZZ_2\times \ZZ_2$. Now there is
an isotropic $H_L=\ZZ_2$ generated by the spinor class with $s^2=\frac84=2=0\, \text{mod} \, 2$.

An important example is $L=\mrd_{16}$ which has an isotropic group $H_L=\ZZ_2$ generated by the spinor weight
with $s^2=\frac{16}4=4=0\, \text{mod} \, 2$. 

\medskip\noindent
\leftline{\underline{Orthogonal complement}}

Given a sublattice $S$ of $\Gamma$, $S \subset \Gamma$, the orthogonal complement of $S$ in $\Gamma$
is defined to be the set \mbox{$S^\perp=\{ x \in \Gamma\,  \big| \, x.y=0 \,\, \forall y \op \in S\}$.}

\medskip\noindent
\leftline{\underline{$M$, overlattice of $L$}}

If $L \subset M$ and the index $[M:L]$ is finite then $M$ is an overlattice of $L$. 
This means that $M$ and $L$ have the same rank. In fact,  $[M:L]^2=d(L)/d(M)$. The index is also denoted by $|M/L|$.

The important Proposition 1.4.1 of  Nikulin states that the set of even overlattices of $L$ corresponds bijectively with the set of 
isotropic subgroups of $A_L$ \cite{Nikulin80}.
The overlattice corresponding to $H_L$ can be constructed as $M_H=\{x \in L^* \big| [x \, \text{mod}\, L] \in H_L\}$.
(see e.g. proposition $\alpha$ in \cite{Braun:2013yya}). This means that the elements of $M_H$ are weights that can be written
as roots plus generators in $H_L$. 
Besides, the discriminant form $q_{M_H}$ is given by the discriminant form $q_L$ restricted to 
$H_L^\perp/H_L$. Orthogonality is defined with respect to the bilinear quadratic form $b_L$ \cite{Braun:2013yya}. In practice,
$y \in H_L^\perp$ if $y \in A_L$ and $y\cdot x =\text{integer}$ for all $x\in H_L$.
To avoid cluttering we will drop the subscript in $M_H$ when $H_L$ has been specified.

As an example, take $L=\mra_8$ and $H_L=\ZZ_3$ 
so that  $M/L \cong \ZZ_3$ and $d(M)=\frac9{3^2}=1$. Then $M$ has elements $x=y + n w_3$, with $y \in L$
and $n=0,1,2$. It can be shown that this $M$ is isomorphic to $\mre_8$, which is the unique rank 8 even unimodular
lattice. 

For $L=\mrd_8$ the overlattice associated to $H_L=\ZZ_2$ has elements $x=y + n s$, with $y \in L$ and $n=0,1$.
This is nothing but  $\mre_8$, as expected since the overlattice has $d(M)=\frac{4}{2^2}=1$.

For $L=\mrd_{16}$ the overlattice corresponding to $H_L=\ZZ_2$ is the even unimodular lattice $\Gamma_{16}$
with elements $x=y + n s$, with $y \in L$ and $n=0,1$. Unimodularity follows from $M/L\cong \ZZ_2$ implying
$d(M)=\frac{4}{2^2}=1$. $\Gamma_{16}$ is the HO lattice. 

\medskip\noindent
\leftline{\underline{$M_{\text{root}}$, root sublattice of $M$}}

It is the sublattice of $M$ generated by roots, i.e. by vectors of norm 2.

For example, for the overlattice of $L=\mrd_{16}$, $M_{\text{root}}=L$.
For $L=\mrd_8$ this is not the case because the overlattice $\mre_8$ has many more
roots. This reflects the fact that for $\mrd_8$ the additional element $s$ in the overlattice has $s^2=2$.

\medskip\noindent
\leftline{\underline{Primitive embedding}}

A lattice $S$ is primitively embedded in another lattice $\Gamma$ if $S \subset \Gamma$ and
$\Gamma/S$ is torsion-free. 

For example, $\mra_8 \subset \mre_8$ but the embedding is not primitive
because $\mre_8/\mra_8\cong \ZZ_3$ as explained above. An example of primitive embedding is
$\mra_3 \subset \mre_8$. Since $\mra_3$ has rank 3 and $\mre_8$ is even unimodular, this  
follows from Theorem 1.12.4 of Nikulin \cite{Nikulin80} quoted below. It can then be shown that $\mrd_5 \subset \mre_8$ is 
primitive because $\mrd_5$ is the orthogonal complement of $\mra_3$ in $\mre_8$, and also that
$\mre_8$ is an overlattice of $\mrd_5 + \mra_3$.

\medskip\noindent
\leftline{\underline{Nikulin's Theorem 1.12.4 \cite{Nikulin80}}}

Every even lattice of signature $(t_{(+)}, t_{(-)})$ admits a primitive embedding in an even unimodular
lattice of signature $(l_{(+)}, l_{(-)})$, with $l_{(+)} - l_{(-)} \equiv 0\, \text{mod}\, 8$, if
\be
t_{(+)} \le  l_{(+)}, \qquad t_{(-)} \le l_{(-)}, \qquad t_{(+)} + t_{(-)} \le \frac12( l_{(+)} + l_{(-)}) \, .
\label{nik1124}
\ee
In particular, if $r \le (8 +d)$ then $L$ of signature $(r,0)$ admits a primitive embedding in $\mathrm{II}_{d+16,d}$. 
  
\section{Complements to section \ref{sec:lattices}}
\label{ap:extralat}

In this appendix we present some additional material for the discussion of
the lattice embedding formalism.

\subsection{Embeddings of groups with rank $r<d+16$}
\label{ap:gen}

The problem is now to embed $L$ of signature $(r,0)$, $r < d+16$, in the 
even unimodular Narain lattice $\mathrm{II}_{d+16,d}$.  
In this case there are also three criteria that read
\begin{quotation}
\noindent
{\bf {Criterion 1, from Corollary 1.12.3 \cite{Nikulin80}}} \hfill\\
\noindent
{\em If $\ell(A_L)<16+ 2d-r$  then $L$ has an embedding in $\mathrm{II}_{d+16,d}$.}
\end{quotation}
\begin{quotation}
\noindent
{\bf {Criterion 2, from Theorem 1.12.2(c) \cite{Nikulin80}} }\hfill\\
\noindent
{\em $L$ has a primitive embedding in $\mathrm{II}_{d+16,d}$ if and only if there exists a lattice $T$ of signature $(d,d+16-r)$
such that $(A_T, q_T)$ is isomorphic to $(A_L,q_L)$. } 
\end{quotation}
\noindent
\noindent
\begin{quotation}
\noindent
{\bf {Criterion 3, from Theorem 7.1 \cite{ShimadaK3}}} \hfill\\
\noindent
{\em $L$ has an embedding in $\mathrm{II}_{d+16,d}$ if and only if $L$ has an overlattice $M$ with
the following properties:
\begin{itemize}
\item[\rm{(i)}] there exists an even lattice $T$ of signature $(d,d+16-r)$ such that $(A_T, q_T)$ is isomorphic to $(A_M,q_M)$,
\item[\rm{(ii)}] the sublattice $M_{root}$ of $M$ coincides with $L$.
\end{itemize}
}
\end{quotation}
\noindent
Recall that Theorem 1.12.4 \cite{Nikulin80} further implies that when $r\le (8+d)$ there is always a primitive
embedding.
The above criteria clearly reduce to those in section \ref{s:max} setting $r=d+16$. 
The lattice $T$ now has indefinite signature so the application would be more complicated.

\subsection{More on the complementary lattice $T$ of signature $(d,0)$}
\label{ap:extraT}

In section \ref{ss:nikhet} we have argued that $T=K\langle -1 \rangle$. 
To complete the proof that $(A_M, q_M) \cong (A_K, -q_K)$ we can use the following theorem of \cite{CS}:
Let $L_1$ and $L_2$ be two sublattices of a unimodular lattice $L_3$ such that\footnote{$L \otimes \mathbb{R}$ means the set of all points obtained by real linear combinations of the basis vectors of $L$}
\begin{equation}
L_1 \oplus L_2 \subset L_3, \qquad L_1 = (L_1 \otimes \mathbb{R}) \cap L_3, \qquad L_2 = (L_2 \otimes \mathbb{R}) \cap L_3. \nn
\end{equation} 
Then the discriminant groups $L_1^* / L_1$ and $L_2^* / L_2$ are isomorphic. The isomorphism is given by 
$y_1 + L_1 \to y_2 + L_2$, where $y_1 \in  L_1^*/L_1$ and $y_2 \in L_2^*/L_2$, whenever 
$y = y_1 + y_2$ generates an isotropic subgroup of $L_1 \oplus L_2$.

To apply this theorem to our problem
we take $L_1=M$, $L_2=K$, and $L_3=\mathrm{II}_{d,d+16}$, with $K$ and $M$ given in \eqref{Kheterotic}
and \eqref{Mheterotic}. 
We have $M \otimes \mathbb{R} = \mathbb{R}^{0,d+16}$ and $K \otimes \mathbb{R} = \mathbb{R}^{d,0}$. 
Moreover, $\mathbb{R}^{0,d+16} \cap \mathrm{II}_{d,d+16} = M$ and 
$\mathbb{R}^{d,0} \cap \mathrm{II}_{d,d+16} = K$. 
It follows that $M$ and $K$ have isomorphic discriminant groups. It remains to show that they have isomorphic discriminant forms.
The Narain lattice $\mathrm{II}_{d,d+16}$ is generated by the lattice sum $M\oplus K$ together with some isotropic vectors
(glue vectors in the language of \cite{CS}).
These vectors are generically of the form $y = y_1 + y_2$, where $y_1$ and $y_2$ are non trivial vectors in the discriminant 
groups of $M$ and $K$, respectively, and are connected by the discriminant group isomorphism. Since $y$ must be even, 
we have $y^2 = 0 \mod 2$. Therefore, $y_1^2 + y_2^2 = 0 \mod 2$, because $M$ and $K$ are orthogonal.  
We thus find $y_1^2 = - y_2^2 \mod 2$.
This shows that $q_M \cong -q_K$, and so $T$ as defined is the complementary lattice of $M$.

\section{Groups of maximal enhancement in $d=1$ and $d=2$}
\label{app:tablesmaximal}

In this appendix we present the Tables containing all the groups of maximal enhancement in one and two dimensions.
The list of groups realized in $S^1$ compactifications of the heterotic string is displayed in Table \ref{tab:alld1}.
The groups realized in $T^2$ compactifications of the $\mre_8 \times \mre'_8$ heterotic string are shown
in Table \ref{tab:alld2}. To simplify notation we dropped the primes in the $\mre'_8$ weights.
In Table \ref{tab:exHO} we give the realization of some of these groups in the $\hosp$ theory.

{\footnotesize
\begin{center}
\renewcommand{\arraystretch}{1.0}
\begin{longtable}{|c|c|c|c|c||c|c||c|c|}
\hline
\# & $L$ & $A_L$ & $H_L$ & $T$  & $R^2_{\text{E}}$ & $A_{\text{E}}$ & $R^2_{\text{O}}$ & $A_{\text{O}}$\\[3pt]
\hline
1 & $2\mre_8 + \mra_1$ &  $\ZZ_2$  & $\mathbb{1}$ & $\langle 1\rangle$ & 1 & $ 0 \times 0$
& $\frac1{16}$ & $\frac14(\tw_7+\tw_9)$  \\[3pt]\hline
2& $\mre_8 +\mre_7+ \mra_2$ &  $\ZZ_6$  & $\mathbb{1}$ & $\langle 3\rangle$ & $\frac34$ & $\frac12 w_6\times 0$
& $\frac1{12}$ & $\frac16(\tw_6+2\tw_9)$   \\[3pt]\hline
3& $\mre_8 +\mre_6+ \mra_3$ &  $\ZZ_{12}$  & $\mathbb{1}$ & $\langle 6\rangle$ & $\frac23$ & $\frac13 w_5\times  0$
& $\frac3{32}$ & $\frac18(\tw_5+3\tw_9)$   \\[3pt]\hline
4 & $\mre_8 + \mrd_9 $ &  $\ZZ_4$ & $\mathbb{1}$ & $\langle 2\rangle$ & $\frac12$ & $ 0\times  \frac12w_8$
& $\frac1{8}$ & $\frac12 \tw_7$    \\[3pt]\hline
5 & $\mre_8 +\mrd_5+ \mra_4$ &  $\ZZ_{20}$  & $\mathbb{1}$ & $\langle 10\rangle$ & $\frac58$ & 
$\frac14 w_4\times  0$
& $\frac1{10}$ & $\frac1{10}(\tw_4+4\tw_9)$   \\[3pt]\hline
6 & $\mre_8 + \mra_9 $ &  $\ZZ_{10}$ & $\mathbb{1}$ & $\langle 5\rangle$ & $\frac59$ & $0\times\frac13 w_1$
& $\frac5{49}$ & $\frac17(3\tw_7+\tw_{15})$    \\[3pt]\hline
7 & $\mre_8 + \mra_8 + \mra_1 $ &  $\ZZ_{18}$ & $\mathbb{1}$ & $\langle 9\rangle$ & $\frac9{16}$ & 
$\frac14 w_7\times  0$
& $\frac1{9}$ & $\frac49\tw_9$    \\
\end{longtable}
\end{center}
}

\begin{tiny}
\begin{center}
\def\arraystretch{1.0}
\begin{longtable}{|c|c|c|c||@{}c@{}c@{}c@{}c@{}|@{}c@{}|@{}c@{}|} 
\hline
$ \# $ & $ L $ & $ H_L $ & $ T $ & $ E_{11} $ & $ E_{21} $ & $ E_{22} $ & $ E_{12} $ & $ A_1 $ & $ A_2 $ \\ \hline 
$ 1 $ & $ 6 \mra_{3} $ & $ \mathbb{Z}_4 \times \mathbb{Z}_4  $ & $ [4,0,4] $ & $ 1 $ & $ 0 $ & $ 1 $ & $ \frac{1}{2} $ & $  \frac{w_{6}}{2}\times\frac{w_{6}}{2}  $ & $ \left (\frac{w_{2}}{4}-\frac{3 w_{6}}{4}\right) \times \left(\frac{w_{2}}{4}-\frac{3 w_{6}}{4}\right ) \dagger$ \\  \hline
$ 2 $ & $ 2 \mra_{1}+4 \mra_{4} $ & $ \mathbb{Z}_5  $ & $ [10,0,10] $ & $ 1 $ & $ \frac{1}{5} $ & $ 1 $ & $ -\frac{1}{5} $ & $  \frac{w_{3}}{5}\times\frac{w_{3}}{5}  $ & $ \left (w_{1}-\frac{3 w_{3}}{5}\right) \times \left(w_{1}-\frac{3 w_{3}}{5}\right ) \dagger$ \\ \hline
$ 3 $ & $ 2 \mra_{2}+2 \mra_{3}+2 \mra_{4} $ & $ \mathbb{1} $ & $ [60,0,60] $ & $ 1 $ & $ 0 $ & $ \frac{11}{12} $ & $ \frac{5}{12} $ & $  \frac{w_{5}}{3}\times\frac{w_{6}}{2}  $ & $ \left (\frac{w_{2}}{3}-\frac{2 w_{5}}{3}\right) \times \left(\frac{w_{2}}{4}-\frac{3 w_{6}}{4}\right ) \dagger$ \\ \hline
$ 4 $ & $ 3 \mra_{1}+3 \mra_{5} $ & $ \mathbb{Z}_2 \times \mathbb{Z}_6  $ & $ [2,0,6] $ & $ 1 $ & $ 0 $ & $ 1 $ & $ 0 $ & $  \frac{w_{6}}{2}\times\frac{w_{8}}{2}  $ & $ \left (\frac{3 w_{6}}{4}-\frac{w_{2}}{4}\right) \times \left(\frac{w_{2}}{2}-\frac{3 w_{4}}{4}\right ) *$ \\ \hline
$ 5 $ & $ 4 \mra_{2}+2 \mra_{5} $ & $ \mathbb{Z}_3 \times \mathbb{Z}_3  $ & $ [6,0,6] $ & $ 1 $ & $ 0 $ & $ 1 $ & $ 0 $ & $  \frac{w_{5}}{3}\times\frac{w_{5}}{3}  $ & $ \left (\frac{w_{2}}{3}-\frac{2 w_{5}}{3}\right) \times \left(\frac{w_{2}}{3}-\frac{2 w_{5}}{3}\right ) $ \\ \hline
$ 6 $ & $ \mra_{3}+3 \mra_{5} $ & $ \mathbb{Z}_2 \times \mathbb{Z}_3  $ & $ [4,0,6] $ & $ 1 $ & $ 0 $ & $ 1 $ & $ 0 $ & $  \frac{w_{5}}{3}\times\frac{w_{5}}{3}  $ & $ \left (\frac{w_{1}}{2}-\frac{w_{5}}{2}\right) \times \left(\frac{w_{1}}{2}-\frac{w_{5}}{2}\right ) $ \\
\end{longtable}
\end{center}
\end{tiny}

\begin{table}[H]\begin{center}
\renewcommand{\arraystretch}{1.0}
{\footnotesize
\begin{tabular}{|c|c||cccc|c|c|}\hline
$ \# $ & $ L $ & $ E_{11} $ & $ E_{21} $ & $ E_{22} $ & $ E_{12} $ & $ A_1 $ & $ A_2 $\\ \hline
$ 1 $&$ 6  \mra_{3} $&$ 1 $&$ 0 $&$ \frac{3}{2} $&$ \frac{1}{4} $&$ \frac14(\tw_6+\tw_{10}) $&$\frac14(\tw_2-\tw_6-\tw_{10}+\tw_{14})  $\\\hline
$ 2 $&$ 2  \mra_{1}+4  \mra_{4} $&$ 1 $&$ 0 $&$ 1 $&$ -\frac{1}{4} $&$\frac14(\tw_3+\tw_{13})  $&$ \tw_1+\tw_{15}+\frac15(\tw_8-3\tw_3-3\tw_{13}) $\\\hline
$ 13 $&$ 3  \mra_{6} $&$ 1 $&$ 0 $&$ 1 $&$ 0 $&$\frac12\tw_6  $&$\frac13(2\tw_6-\tw_9-\tw_{15})  $\\\hline
$ 21 $&$ 2  \mra_{1}+2  \mra_{5}+ \mra_{6} $&$ 1 $&$ 0 $&$ 1 $&$ 0 $&$\frac1{14}(3\tw_4+4\tw_{11})  $&
$ \frac1{28}(14\tw_2-15 \tw_4-6 \tw_{11} +14 \tw_{15}) $\\\hline
$65$ & $\mra_3+\mra_6+\mra_9$ & $1$ & $0$ & $1$ & $0$ & $\frac12\tw_6$ & $\frac1{30}(3\tw_6-16\tw_{15})$\\\hline
$177$ & $2\mrd_5+\mrd_8$ & $1$ & $0$ & $1$ & $0$ & $\frac12(\tw_4+\tw_{12})$ & $\frac14(\tw_4-\tw_{12})-\tw_9-\tw_{11}$\\\hline
$ 196 $&$  \mra_{2}+ \mrd_{10}+ \mrd_{6} $&$ 1 $&$ -\frac{1}{2} $&$ 1 $&$ 0 $&$\frac12\tw_6  $&$0  $\\\hline
$ 219 $&$ 3  \mre_{6} $&$ 1 $&$ 0 $&$ 1 $&$ -\frac{3}{4} $&$\frac14(\tw_{11}+\tw_5)  $&$0  $\\\hline
$ 297 $&$  \mra_{2}+2  \mre_{8} $&$ 1 $&$ 0 $&$ 1 $&$ -\frac{1}{4} $&$ \frac14(\tw_7+\tw_9) $&$0  $\\\hline
$ 319 $&$  \mra_{1}+ \mrd_{9}+ \mre_{8} $&$ \frac{5}{4} $&$ 0 $&$ 1 $&$ 0 $&$ \frac12\tw_9 $&$0  $\\\hline
\end{tabular}
}
\caption{Data for some groups of maximal rank, for the $\text{Spin}(32)/{\mathbb Z}_2$ heterotic on $T^2$.}
  \label{tab:exHO}\end{center}\end{table}

%%%%%%%%%%%%%%%%%%%%%%%%%%%%%%%%%%%%%%%%%%%%%%%%%%%%%%%%%%%%%%%%%%%%%%%%%

%%%%%%%%%%%%%%%%%%%%%%%%%%%%%%%%%%%%%%%%%%%%%%%%%%%%%%%%%%%%%%%%%%%%%%%%%

\end{document}